\documentclass[a4paper,10pt]{article}
\usepackage[utf8]{inputenc}
\usepackage{authblk}

\usepackage{graphicx}
\usepackage{epstopdf}
\usepackage{todonotes}
\usepackage{lineno,hyperref}
\usepackage{upgreek}
\usepackage{units}
\usepackage{multirow}
\usepackage{array}
\usepackage{amssymb}	
\usepackage{amsmath}	
\usepackage{textcomp}
\usepackage{pdfpages}

\usepackage{caption}
\usepackage{subcaption}

\title{The ABC130 barrel module prototyping programme for the \mbox{ATLAS} strip tracker}

\author[1]{Luise Poley}
\author[2]{Craig Sawyer}
\author[3]{Sagar Addepalli}
\author[4]{Anthony Affolder}
\author[5]{Bruno Allongue}
\author[6]{Phil Allport}
\author[1]{Eric Anderssen}
\author[5]{Francis Anghinolfi}
\author[7]{Jean-François Arguin}
\author[8]{Jan-Hendrik Arling}
% ...

\affil[1]{Lawrence Berkeley National Laboratory, Cyclotron Road, Berkeley, USA}
\affil[2]{Particle Physics Department, STFC Rutherford Appleton Laboratory, Harwell Science and Innovation Campus, Didcot, United Kingdom}
\affil[3]{Martin A. Fisher School of Physics, Brandeis University, Waltham, United States of America}
\affil[4]{Santa Cruz Institute of Particle Physics, University of California, High Street, Santa Cruz, United States of America}
\affil[5]{Experimental Physics Department, CERN, Geneva, Switzerland}
\affil[6]{School of Physics and Astronomy, University of Birmingham, Edgabston, Birmingham, United Kingdom}
\affil[7]{Group of Particle Physics, University of Montreal, Rue University, Montreal, Canada}
\affil[8]{Deutsches Elektronen-Synchrotron, Notkestra\ss{}e, Hamburg, Germany}
% ...

\begin{document}

\maketitle

\begin{abstract}

For the Phase-II Upgrade of the \mbox{ATLAS} Detector~\cite{ATLAS}, its Inner
Detector, consisting of silicon pixel, silicon strip and transition
radiation sub-detectors, will be replaced with an all new \unit[100]{\%} silicon tracker, composed of a pixel tracker at inner radii and a strip tracker at outer radii. The future
\mbox{ATLAS} strip tracker will include 11,000 silicon sensor modules in the central
region (barrel) and 7,000 modules in the forward region (end-caps),
which are foreseen to be constructed over a period of
\unit[3.5]{years}. The construction of each module consists of a
series of assembly and quality control steps, which were engineered to
be identical for all production sites. In order to develop the tooling and procedures for assembly and testing of these modules, two series of major prototyping programs were conducted: an early program using
readout chips designed using a \unit[250]{nm}
fabrication process (ABCN-25)~\cite{Petaletpaper}~\cite{stave} and a subsequent program using a follow-up chip set made using \unit[130]{nm} processing (ABC130 and
HCC130 chips). This second generation of readout chips was used for an
extensive prototyping program that produced around 100
barrel-type modules and contributed significantly to the development
of the final module layout. This paper gives an overview of the
components used in ABC130 barrel modules, their assembly procedure and
findings resulting from their tests.

\end{abstract}

\tableofcontents

\section{Introduction} 

For the High-Luminosity Upgrade of the Large Hadron Collider, the
\mbox{ATLAS}~\cite{ATLAS} Inner Detector will be replaced with a new, all
silicon Inner Tracker (ITk), composed of a pixel tracker~\cite{TDRp}
and a strip tracker~\cite{TDRs}. 

\begin{figure}
 \centering
\begin{subfigure}{\textwidth}
 \centering
 \includegraphics[width=\linewidth]{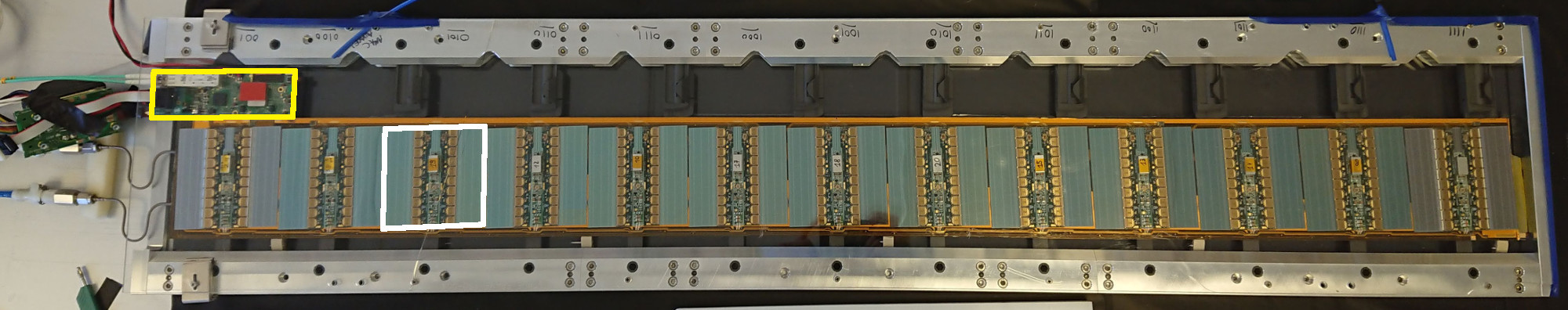}
 \caption{Stave for the ITk strip tracker barrel: thirteen modules are
   arranged in one row}
 \label{fig:intro_stave}
 \end{subfigure}
 
 \begin{subfigure}{\textwidth}
 \centering
 \includegraphics[width=0.3\linewidth]{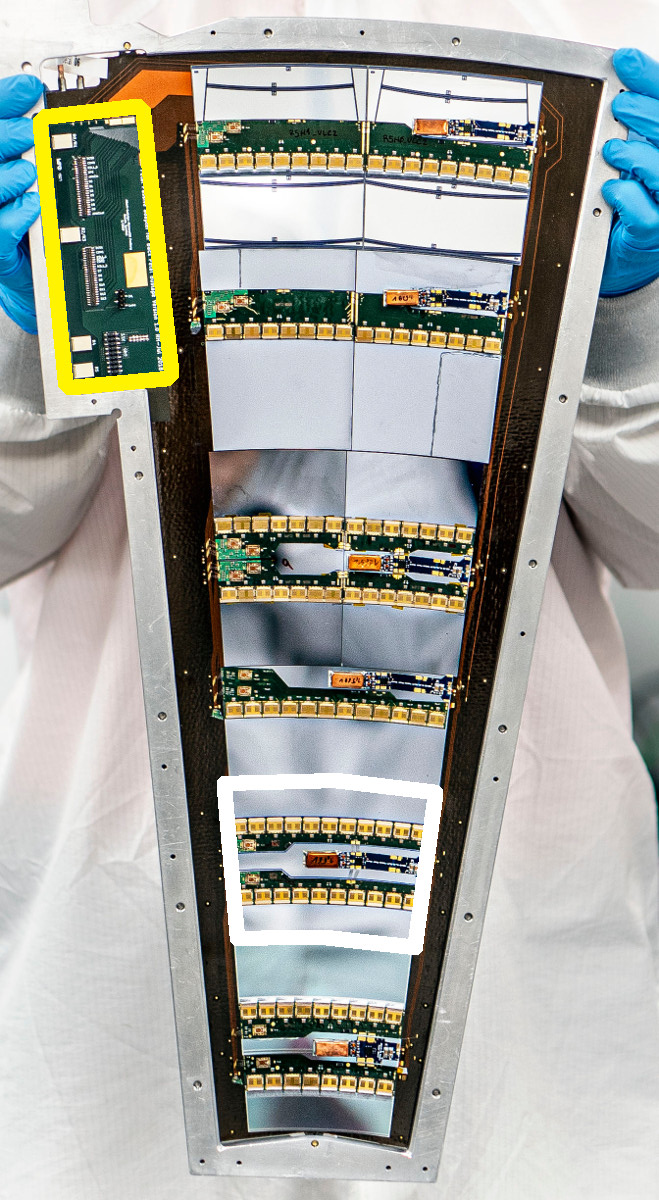}
 \caption{Petal for the ITk strip tracker end-caps: six modules are
   arranged in six rings}
 \label{fig:intro_petal}
\end{subfigure}
\caption{Support structures with modules for the ITk strip tracker
  barrel (composed of staves) and end-caps (consisting of petals). Sensor strips on staves are aligned parallel to the beam axis modulo a \unit[26]{mrad} stereo angle on either side of the stave whilst strips on petals are arranged perpendicular to the beam axis with a \unit[20]{mrad} stereo angle implemented into the sensors themselves. An individual module is indicated in white, with sensor strip implants oriented perpendicular to the hybrids on each module segment. The end-of-substructure card (see~\cite{TDRs}) of each
  structure is indicated in yellow. In the outer three rings of the end-cap so-called split modules are implemented due to the limited area of \unit[6]{inch} silicon wafers so that each ring module contains two silicon strip modules.}
\label{fig:intro_structures}
\end{figure}

The main component of the ITk strip tracker is the module, comprising
a silicon strip sensor, multiple custom readout chips mounted on a
electronic circuit, called a hybrid, and a powerboard. In the central
region of the ITk strip detector, the four barrel layers comprise
11,000 modules mounted on staves such that the sensors are arranged
parallel to the beam axis (see figure~\ref{fig:intro_stave}). The two end-caps in the
forward region are constructed from six disks supporting a total of 7,000 modules
mounted on petals such that the sensors are arranged orthogonal to the
beam axis (see figure~\ref{fig:intro_petal}).

Modules in the strip tracker barrel and end-caps were designed to
contain the same materials and components, which have the same
functionality, but different geometries. Only two types of sensors are
used in the barrel region, whereas six sensor geometries are
required for hermetic coverage of the end-cap. Here, only modules
designed for the barrel region are presented.

An extensive prototyping program was conducted in preparation for the
production of 11,000 barrel modules at ten construction sites in the
US, UK and China. The aim of the prototyping programme was to
develop realistic tests of the concepts for tooling and assembly,
readout software and testing procedures, hence, the prototype modules use
readout chips, sensors and other components similar to those foreseen
to be used in production.

\section{Components}
\label{sec:comp}

\begin{figure}
\begin{subfigure}{.47\textwidth}
 \centering
 \includegraphics[width=\linewidth]{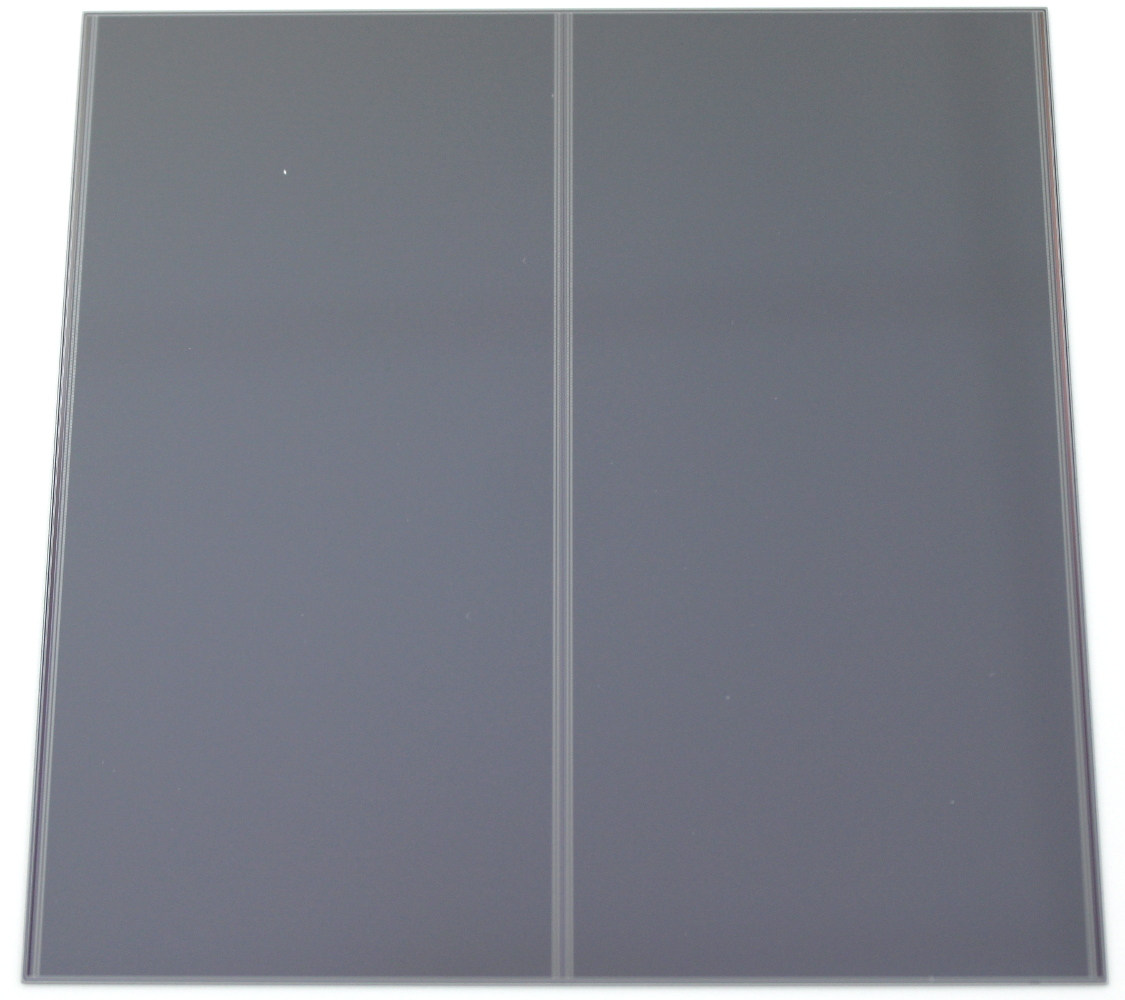}
 \caption{\mbox{ATLAS17LS} sensor with two segments containing long
   strips with a length of about {\unit[5]{cm}} each and two rows of bond pads per segment.}
 \label{fig:sensor_LS}
\end{subfigure}
\begin{subfigure}{.05\textwidth}
\hfill
\end{subfigure}
\begin{subfigure}{.47\textwidth}
 \centering
 \includegraphics[width=\linewidth]{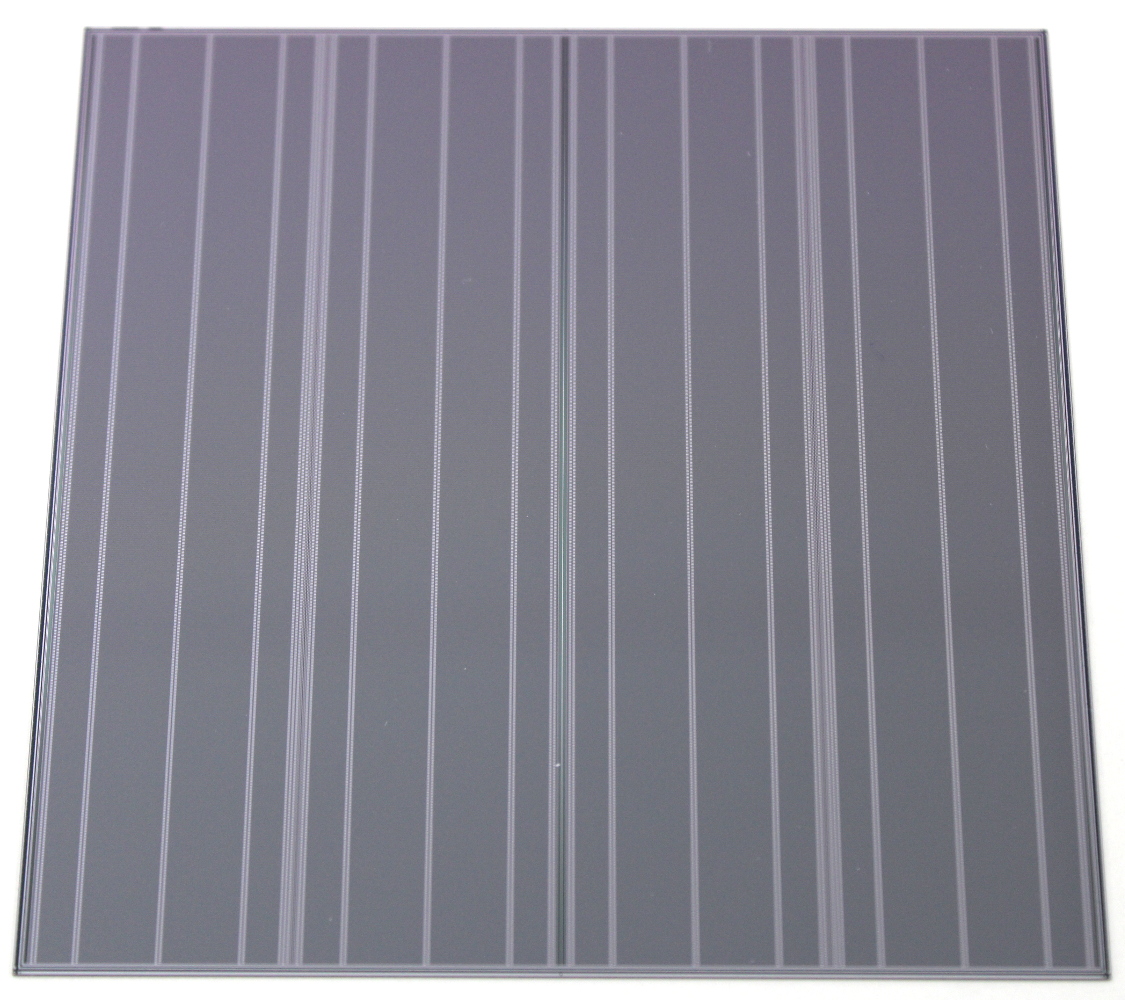}
 \caption{\mbox{ATLAS12SS} sensor with four short strip segments (strip
   length about {\unit[2.5]{cm}}) and five rows of bond pads per
   segment.}
 \label{fig:sensor_SS}
\end{subfigure}
\caption{\mbox{ATLAS} barrel long strip and short strip sensors used for the
  construction of ABC130 barrel modules. Sensor strips are oriented
  horizontally, with each segment comprising 1282 sensor
  strips. The vertical lines seen here are rows of the bond pads, the only large-scale feature in the strip area discernible by eye.}
\end{figure}

In the central region of the ITk strip tracker (barrel), two versions of
modules are used:
\begin{itemize}
 \item short strip (SS) modules in the inner two barrel layers, where
   each sensor strip has a length of about \unit[2.5]{cm}
 \item long strip (LS) modules in the outer two barrel layers, with
   sensor strip lengths of about \unit[5]{cm}
\end{itemize}
Despite their different strip lengths, both module types have similar
sizes, which are determined by the size of the silicon strip
sensor (about \unit[$10\times10$]{cm$^2$} each). Therefore, strips are
arranged in two rows on LS sensors and in four rows on SS sensors (the terms row and segment are used interchangeably throughout this manuscript),
where each row consists of 1280 signal strips and two unconnected edge strips (see
figures~\ref{fig:sensor_LS} and~\ref{fig:sensor_SS}). Accordingly, LS modules require 2560 readout channels (corresponding to 10 ABC130
readout chips with 256 channels each) and SS modules 5120
(corresponding to 20 ABC130 readout chips).

\begin{figure}
\begin{subfigure}{.47\textwidth}
 \centering
 \includegraphics[width=\linewidth]{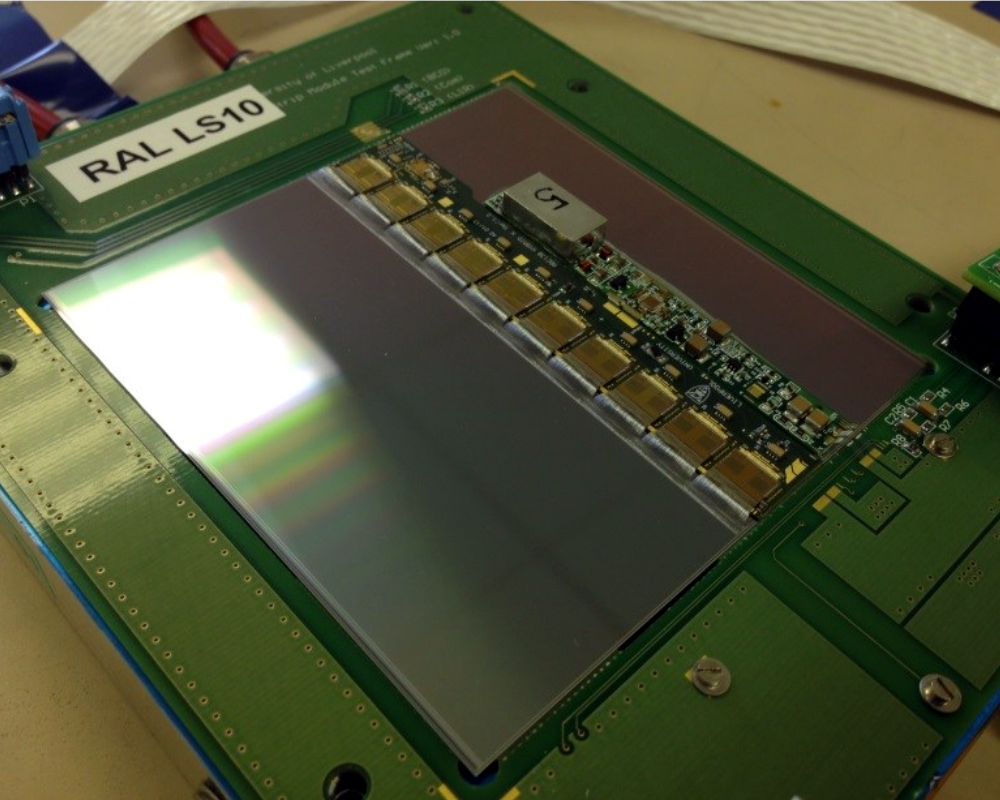}
 \caption{ABC130 LS barrel module on an LS test frame: one X-type
   hybrid is mounted at the border between LS strip segments
   with the powerboard mounted on the same segment.}
 \label{fig:module_LS}
\end{subfigure}
\begin{subfigure}{.05\textwidth}
\hfill
\end{subfigure}
\begin{subfigure}{.47\textwidth}
 \centering
 \includegraphics[width=\linewidth]{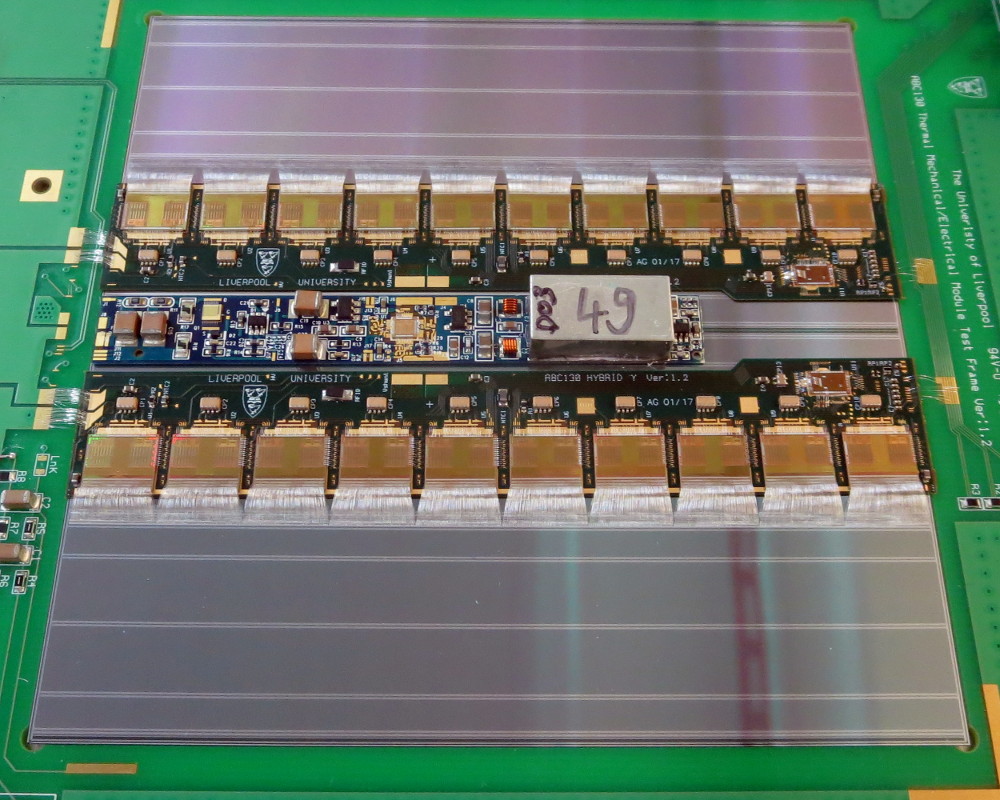}
 \caption{ABC130 SS module on an SS test frame: one X-type hybrid and
   a Y-type are mounted at the borders between two short strip
   segments with one powerboard between them.}
 \label{fig:module_SS}
\end{subfigure}
\caption{ABC130 long-strip and short-strip modules.}
\end{figure}

Despite requiring different numbers of readout channels and chips,
electronic components for barrel modules were designed to be compatible with both
sensor geometries. Flexible circuit boards supporting ABC130
readout chips, called hybrids (section~\ref{comp:hyb}), were designed,
with one hybrid required per two strip segments. An SS module uses two such hybrids,
an X-type and a Y-type version, whereas
an LS module uses only one X-type hybrid (see
figures~\ref{fig:module_LS} and~\ref{fig:module_SS}). Flex circuit boards called powerboards (section~\ref{comp:pb}), which support a
DCDC power converter, high voltage switch and a monitoring chip, match
both LS and SS module layouts, thereby minimising the number of
components to be designed, tested and qualified for production.

\subsection{Sensors}
\label{subsec:comp_sensors}

Short strip modules for the ABC130 barrel module program were
constructed using \mbox{ATLAS12} barrel sensors~\cite{ATLAS12}, a prototype
version of the sensors to be used in the \mbox{ATLAS} ITk developed from the
predecessor \mbox{ATLAS07} sensors~\cite{ATLAS07}. The sensors are fabricated from 6-inch floatzone wafers in a single-sided process.

The sensors have a nominal thickness of $\unit[310\pm20]{\upmu\text{m}}$ with a maximum thickness variation of \unit[10]{$\upmu$m} across the sensor area. 
After dicing, \mbox{ATLAS12} sensors have a size of $\unit[96.7\times96.6]{\text{mm}^2}$.
Compared to \mbox{ATLAS07} sensors, the dead space in periphery of the sensor was reduced from approximately \unit[1]{mm} to \unit[500]{$\upmu$m} per edge.

Each ATLAS12SS sensor consists of four segments with 1282 strip implants each, where the first and last strip serve as field shaping strips. The strips have a length of $\unit[23.9]{\text{mm}}$ and a strip pitch of \unit[74.5]{$\upmu$m}. 

\begin{figure}
\centering
\includegraphics[width=0.8\linewidth]{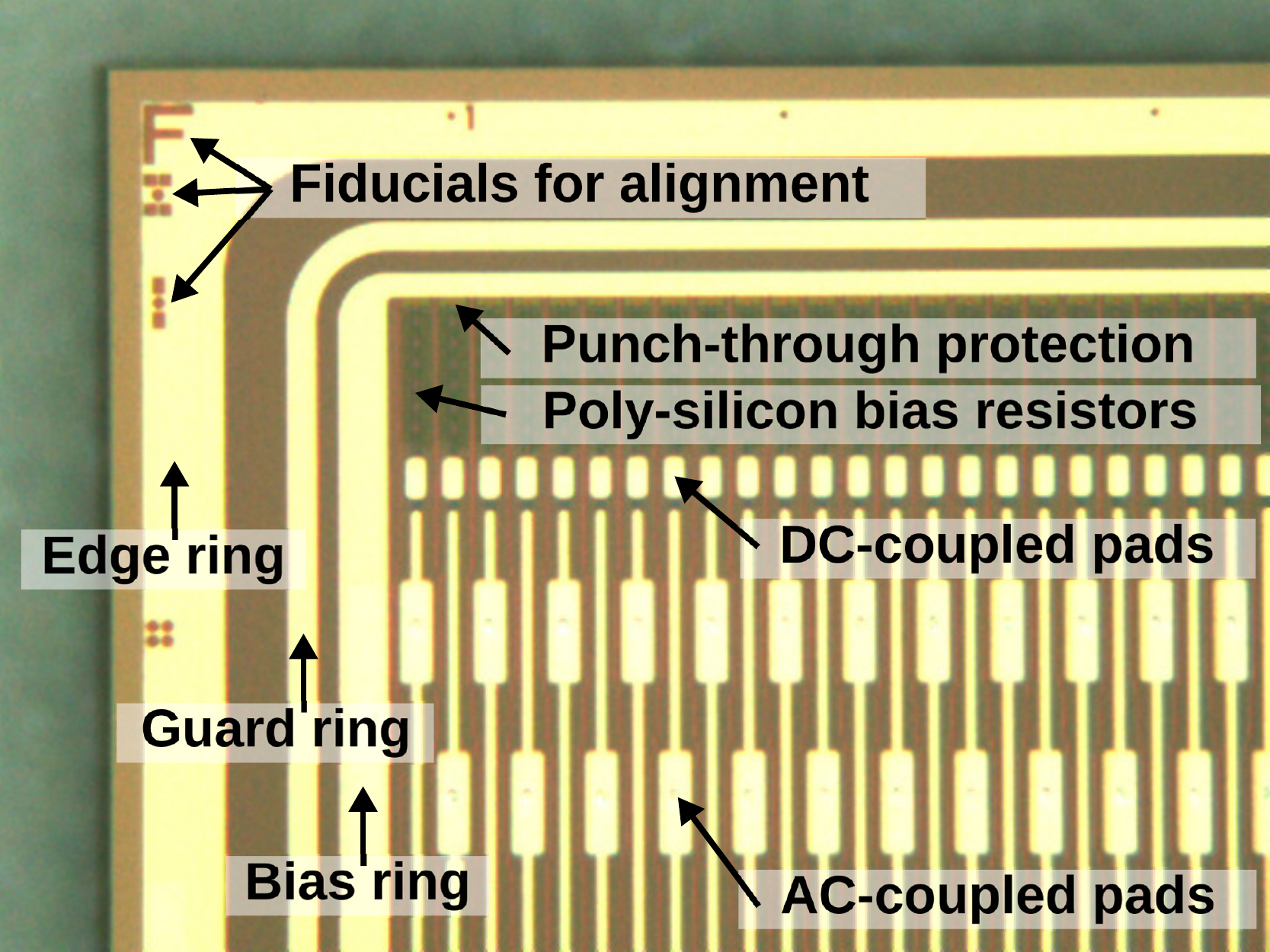}
\caption{Detail image of an ATLAS barrel sensor and its design features}
\label{fig:sensorlayout}
\end{figure}

In order to cope with the high-radiation environment of the ITk, strip sensors are made from p-doped bulk material with n$^{+}$-doped strip implants. The bulk remains as p-doped after radiation damage, therefore the sensor depletion zone grows from the strip implant side towards the backside, allowing for a significant signal collection even when operated underdepleted due to radiation damage at the end-of life fluence.
Each n$^{+}$-doped strip implant is connected to an n-doped implant ring surrounding all strip implants (bias ring) to
hold all strip implants at the same potential during operation. The
bias ring is surrounded by another n-doped implant ring (guard ring)
and a p-doped implant ring (edge ring) laid-out next to
the dicing edge. Figure~\ref{fig:sensorlayout} shows an overview of the different sensor design features. This edge ring prevents the depleted region,
evolving from the bias ring, from extending to and along the dicing edge between the edge ring and the p-doped backplane (held at high voltage), and is needed to prevent an early breakdown~\cite{Diodes}. Detailed studies of the electrical properties of the \mbox{ATLAS12} can be found in~\cite{HOMMELS}.

With increasing radiation damage, the Si/SiO$_2$ passivation layer on the sensor surface 
will experience a build-up of defects and suffer from the surface damage of the ionising dose, which could lead to a short circuiting of the n$^{+}$-strip implants. This is prevented using an inter-strip isolation technique based on p-stop traces, which was chosen out of several options tested on \mbox{ATLAS07} devices.
The sensor design also includes a protection structure for the AC coupling of the strips against beam splashes, a so called gated Punch Trough Protection~\cite{PTP}.
The gated PTP design of the \mbox{ATLAS12} sensors extends the strip below the bias resistor, leaving a \unit[20]{$\upmu$m} gap to the bias rail and the gap covered by the extended sheet of the bias rail, which results in a hard breakdown across this gap in case of an excessive potential.

Another novelty of the \mbox{ATLAS12} is the staggered design of the bond pads to match the four rows of bond pads on ABC130 ASICS (see section~\ref{sec:ABC}) to facilitate wire bonding. The bond pad design in the sensor mirrors the bond pad arrangement of the ABC13 ASICS, which results in a four row bonding process, where each subsequent row increases in height. 

During the development of \mbox{ATLAS12} sensors, it was discovered that they were sensitive to humidity~\cite{humidity}. This meant
sensor breakdown, indicated by high leakage current, at reverse bias voltages below the nominal operating voltage of \unit[-500]{V} was observed at ambient humidity
levels. Therefore, a protocol was established over the course of the
ABC130 barrel module program, which required a minimisation of sensor exposure to higher humidity levels to prevent early breakdowns:
\begin{itemize}
 \item storage of sensors at modules at a maximum of \unit[10]{\%} humidity
 \item sensor tests to be performed at maximum relative humidity of \unit[20]{\%}
 \item minimisation of time sensors spent outside of dry storage, e.g. assembly
\end{itemize}

In addition to the construction of short strip modules, several long strip modules were
constructed using \mbox{ATLAS17LS} sensors~\cite{ATLAS17LS}, which were developed after \mbox{ATLAS12} sensors to prototype the long strip geometry. In contrast to the \mbox{ATLAS12}, the \mbox{ATLAS17LS} sensors are slightly larger with dimensions 
of $\unit[98.0\times97.6]{\text{mm}^2}$, utilising the full usable area out of 6-inch wafer. ATLAS17LS sensors have a strip pitch of \unit[75.5]{$\upmu$m} and two long strip segments with 1280 \unit[4.83]{mm} strips each. 
Additionally, the wafer layout included new test structures and updated fiducial marks for spatial referencing for the \mbox{ATLAS17LS} design. The fabrication of \mbox{ATLAS17LS} sensors used split batches to test options for alternative passivation and a non-standard active depth~\cite{ATLAS17LS}.

\subsection{Readout chips}
\label{subsec:comp_chips}

\subsubsection{ABC130}
\label{sec:ABC}

Each ABC130 chip \cite{ABC130Spec} provides the initial data acquisition and readout chain for up to 256
sensor strips. Submitted in June 2013, it is the second generation of the \mbox{ATLAS} Strips readout family of
custom Application Specific Integrated Circuits (ASICs) since the ABCD~\cite{DabrowskiABCD}, which was used for the SemiConductor Tracker (SCT) readout. The ABC130 follows the
ABCN-25~\cite{DabrowskiABCN25}, which implemented ABCD in a new process, with some improvements, but kept a similar architecture. The ABC130 is the next member of this ``\mbox{ATLAS} Binary Chip'' family, and its
suffix is from its implementation in IBM's (now GLOBALFOUNDRIES') CMOS8RF\_DM 130nm technology. The die
has a size of \unit[$6.8\times7.9$]{mm$^2$} with the wide side meant to be oriented orthogonally to the
direction of the sensor strips and along the edge of the hybrid circuit board. With these dimensions, it allows for bonding of the input pads to the sensor strip pitch, while still allowing space for decoupling capacitors to be
placed between chips.

\begin{figure}
\begin{subfigure}{.50\textwidth}
 \centering
 \includegraphics[width=\linewidth]{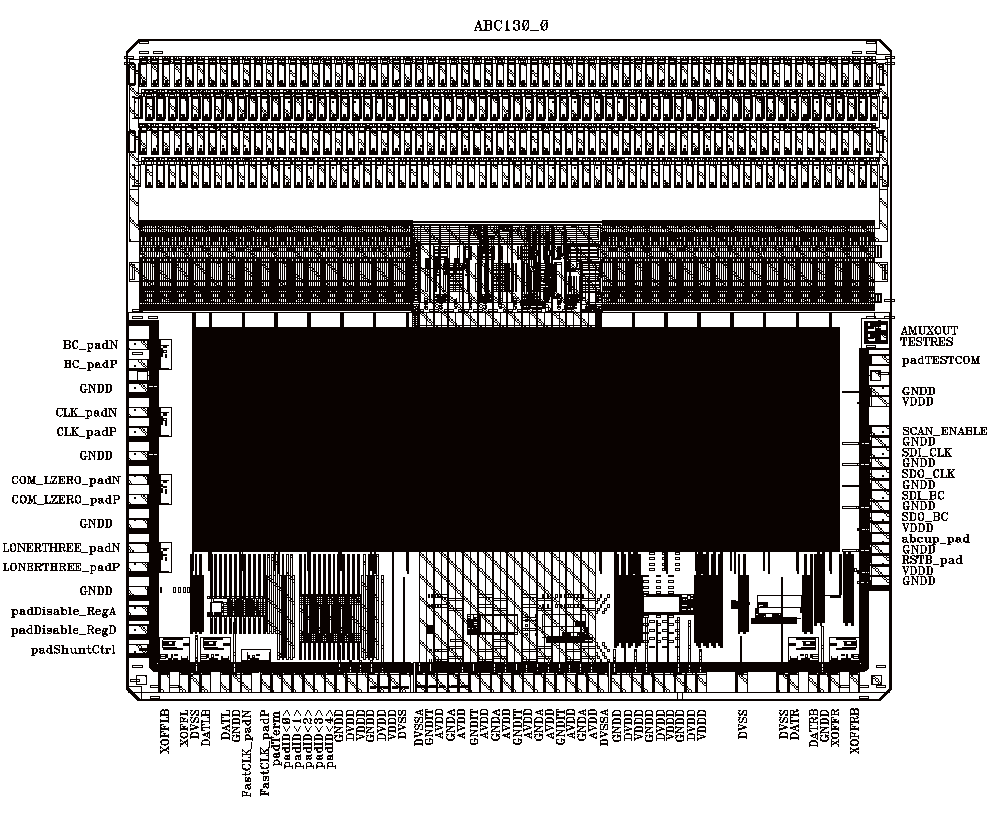}
 \caption{Pad layout of ABC130\_0 die (the ABC130\_1 die is a superset, see section~\ref{para:abc130_dig_io}).}
 \label{fig:abc130_pads}
\end{subfigure}
\begin{subfigure}{.02\textwidth}
\hfill
\end{subfigure}
\begin{subfigure}{.46\textwidth}
 \centering
 \includegraphics[width=\linewidth]{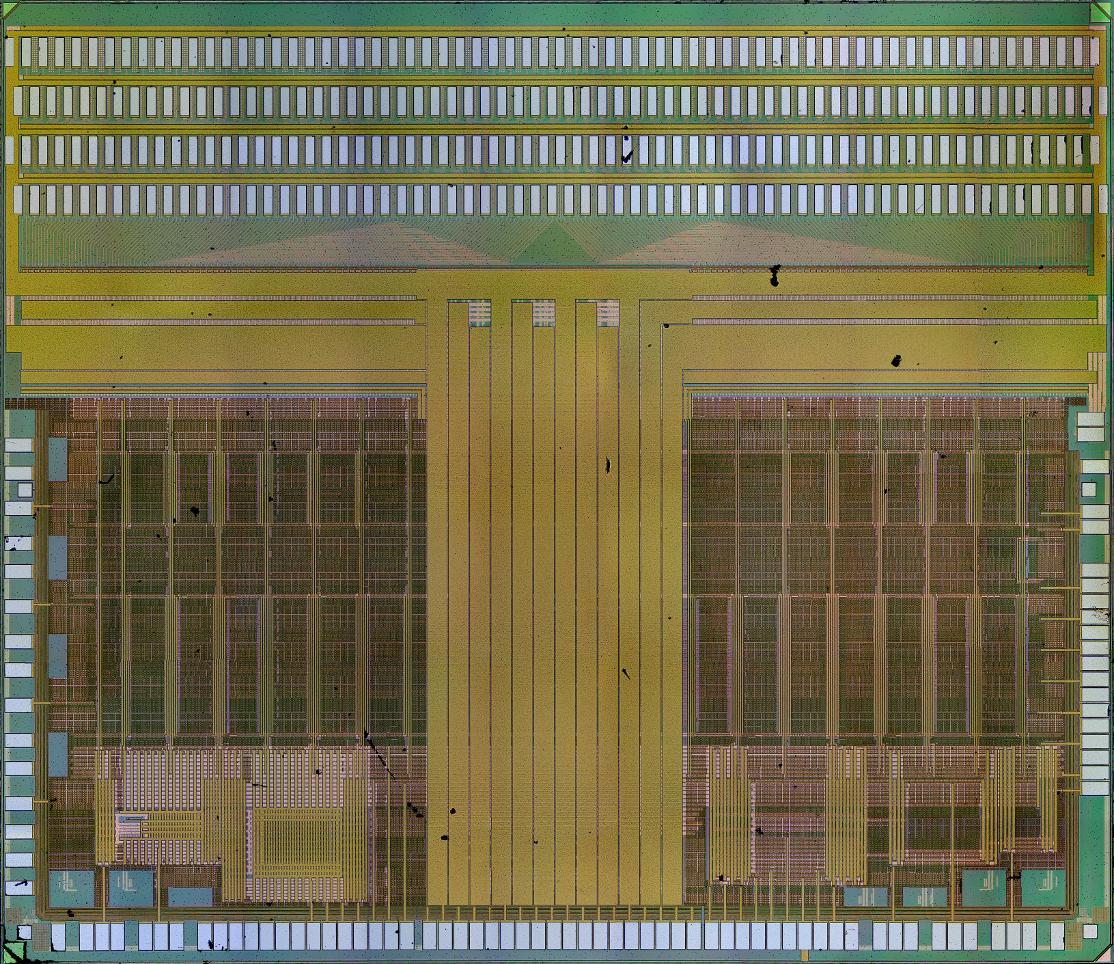}
 \caption{Photograph of ABC130 die showing I/O and four rows of front-end pads}
 \label{fig:abc130_photo}
\end{subfigure}
\caption{ABC130 die layout}
\end{figure}

The first significant change from ABCN-25 is that the smaller feature size allowed a doubling of the number of readout channels per chip. The front-end input pads are
arranged in a novel configuration of four staggered
rows of 64 pads each (see figure \ref{fig:abc130_photo}) for wire bonding to
the AC sensor pads (see section~\ref{subsec:MA}). Ground pads at either end of
each row provide for a sensor ground reference (HV decoupling and guard
ring). The pitch of \unit[119]{$\upmu$m} is chosen to allow direct bond connection from
the available pad sizes to the sensor pitch.
These pads are arranged so that one ASIC can be connected to two rows of
strips on the sensor, with the edge of the ABC130 placed close to the
boundary. The connections from both strip rows to the ASIC amplifier channels are interleaved, which provides a powerful performance cross-check in case of problems.
These two rows are referred to as odd (running away from the ASIC) and even
(running under the ASIC).
The odd strips are also connected by long bonds that reach over the 
top of those for the odd strips (see figure \ref{fig:sensorwires}).
Power and signal connections are restricted to the other three sides of the die
and are wire-bonded to the hybrid circuit board (see section~\ref{subsec:assem_hyb}).

Another substantial change over the ABCN-25 is in the readout system. It was substantially updated and new trigger levels have been added in order to raise the trigger rate from \unit[100]{kHz} to \unit[500]{kHz}, and to include the regional readout concept~\cite{TDRs}. The new architecture has three main stages:
\begin{itemize}
  \item First, the inputs are sampled from the front-ends on every cycle of the
LHC Bunch Crossing (BC) clock (\unit[40.079]{MHz}), and put into a synchronous
pipeline (the ``L0 buffer''). This allows an external process (the L0 trigger)
up to \unit[6]$\upmu$\text{s} to choose which crossings to read out, with \unit[1]{BC} = \unit[25]{ns};
  \item When the L0 accept (L0A) arrives at the ABC130 (a fixed period from
the original BC), the appropriate data is copied to the ``L1 buffer'';
  \item The final readout command (either R3 or L1A, described below) can then be received up to
\unit[512]$\upmu$\text{s} later, and refers to a specific location in the buffer.
\end{itemize}
This architecture allows collection of data into the L1 buffer at a higher
rate than the output bandwidth allows, as not all data might be selected for
read out. The Regional Readout Request (R3) trigger is designed to be
acted on by a small proportion of modules, selectable at the HCC-level (see section~\ref{sec:HCC}), based on
where that module is in the detector, and provides fast readout of data that
can provide input to the L1 Trigger system~\cite{TDRs}. This proportion is
expected to be no more than \unit[10]{\%} of the strip tracker on average.
The L1A is then used to read out full information for the required BCs.

\begin{figure}
\centering
\includegraphics[width=0.7\linewidth]{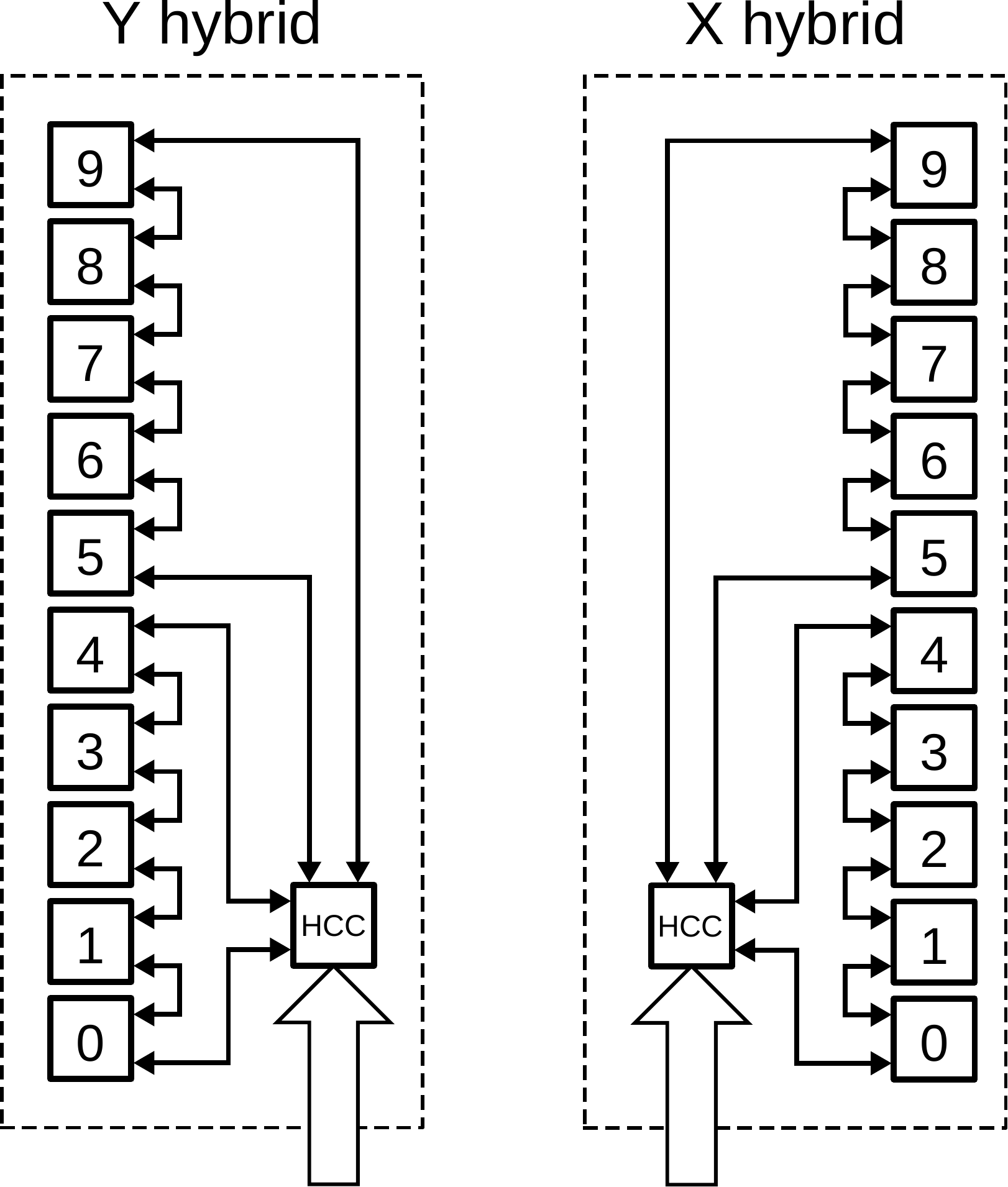}
\caption{ABC130s are connected in daisy-chains of 5 ASICs, with each end of the chain connected to an HCC130 to allow bidirectional access to the chain in the event one of the ABC130s fails~\cite{TDRs}.}
\label{fig:abc130_daisychain}
\end{figure}

All digital signalling is carried out using SLVS\cite{SLVS} differential I/O
between ABC130s and an HCC130 (see section~\ref{sec:HCC}), in a bi-directional daisy-chain fashion (see
figure~\ref{fig:abc130_daisychain}). This allows for the failure of individual
ASICs as the readout direction from downstream ASICs in the chain can be reversed.

\paragraph{Analogue Front-End}

The ABC130 analogue front-end block consists of 256 independent input channels
accepting negative signals from the AC-coupled n-type strips of the ITk
strips sensor. The architecture and performance of the individual front-end
channels are detailed in~\cite{Kaplon2012FrontEE}. Each channel's preamplifier
is designed around a single-ended buffered telescopic cascode with an NMOS
input transistor. The input transistor is enclosed with active feedback built
with a PMOS transistor biased in saturation. This feedback scheme allows for
full control of the DC potential at the preamplifier output and permits the
use of a very power-efficient shaper stage with a single-ended input. This
particular configuration of the input stage was originally designed for the
p$^{+}$\,on\,n sensors intended for the \mbox{ATLAS} tracker upgrade and was later
modified for the negative-going signal of the current n$^{+}$\,on\,p sensor.
On the ABC130, the input current from the sensor strip to the input channel
modulates the transconductance of the feedback transistor and causes
degradation of the noise performance and gain of that stage. Figure~\ref{fig:NoisevsCint} shows a comparison of the noise performance for negative and positive signals. These issues, as well as degradation of the
noise performance after irradiation, have been addressed in the new design
implemented for the ABCStar~\cite{ABCStarSpec}.

The preamplifier input stage has been optimised for the value of \unit[5]{pF}, covering the expected range of input capacitance of the short strip (SS) sensors, but also operates effectively with the higher capacitance of long strip sensors (LS) as well as the different end-cap sensor configurations. This includes the expected parameter variation to maximum lifetime irradiation of the modules. The inputs are protected with non-silicided NMOS thin oxide devices, whose width was chosen as a tradeoff between ESD protection and the parasitic capacitance it added to each channel (\unit[$\sim0.4$]{pF}). The circuit was designed to withstand a \unit[1.5]{kV} Human Body Model event and a \unit[0.6]{A} Transmission Line Pulse (\unit[5]{ns} rise time, \unit[100]{ns} duration). Therefore, a series \unit[22]{$\Omega$} resistor is used to improve response to the Charge Device Model, a tradeoff between protection and added noise. Inputs can be left unconnected without affecting the performance of any other channel.

The effective channel gain is \unit[80]{mV/fC} at nominal bias currents and process parameters with a signal response peaking time of around \unit[22]{ns}. The Full Width at Half Maximum (FWHM) of the response is around \unit[35]{ns} and the overall shaping function is close to a second order CR-RC filter. In terms of frequency response, if the AC-coupling between booster and shaper as well as the limitation of the preamplifier bandwidth are neglected, the front-end channel can be approximated by a bandpass filter with a center frequency of about \unit[15]{MHz} and a roll-off of around \unit[20 or 40]{dB/decade} for frequencies below or above, respectively. The time walk of the discriminator is below \unit[16]{ns} for nominal threshold settings and \unit[50]{\%} of a minimum ionising particle’s (MIP) signal after the total expected radiation dose. 

This timing performance guarantees correct data association to a given BC for the worst
case of signal charge sharing, where the charge is shared equally between neighbouring
strips, and at the end of lifetime of the experiment. The response linearity
is better than \unit[5]{\%} for signal charges from 0 to \unit[-4]{fC}, and
better than \unit[15]{\%} from 0 to \unit[-8]{fC}. Expected noise is
\unit[850]{e$^-$} for SS sensors, and \unit[1150]{e$^-$} for LS sensors.
Double-pulse resolution for a \unit[-3.5]{fC} signal followed by a
\unit[-3.5]{fC} signal is \unit[$\leq75$]{ns}, and maximum recovery time for
a \unit[-80]{fC} signal followed by a \unit[-3.5]{fC} signal is \unit[200]{ns}.
For a \unit[-1]{fC} signal, the gain sensitivity to power supply voltage is
\unit[$<1$]{\%} per \unit[100]{mV}. The chip features a low dropout (LDO) linear
voltage regulator that, in addition to improving the rejection of power supply noise at the front-end and providing an accurate voltage to the chip independent of any voltage drops on the hybrid, allows the analogue core operating voltage to be set
to within \unit[$\pm20$]{mV} of the target voltage of \unit[1.2]{V}. In
addition, the preamplifier input transistor and feedback bias currents can
be tuned to compensate for process variation with internal 5-bit Digital-to-Analogue Converters (DACs) referenced to an internal bandgap circuit (\unit[$592\pm40$]{mV}).

A common threshold level is set with an on-chip 8-bit DAC and is distributed to
the 256 input channel discriminators by means of current mirrors. Due to 
process variation, there is about \unit[20]{mV} rms of threshold variance
between channels, so a 5-bit TrimDAC is provided for each channel. The
magnitude (range of the DAC) of this tuning can also be adjusted.
In this way, the
inter-channel threshold variance can be reduced into the single millivolt
range (see section~\ref{chartests}).

The output of the threshold comparators is then sampled on the rising
edge of the BC clock and shifted into the L0 Buffer (FIFO). A mask register is
supplied to force a zero into the pipeline and allow skipping of any noisy
channels.

Each channel includes the ability, selectable by a calibration enable bit, to inject a tunable
calibration pulse to simulate a strip ``hit''. 
This capability can be used to calibrate the full tracker or a module performance on a per-strip level or as a Built-In Self-Test (BIST) function for the inputs during testing of
wafers. 
Each channel receiving a calibration pulse is connected to a \unit[60]{fF} \unit[$\pm1$]{\%} capacitor
(\unit[$\pm$10]{\%} over full production skew) through a CMOS switch. The injected charge is defined by setting a defined voltage using an 8-bit calibration DAC. A fixed-width calibration pulse (8BC $\approx$
\unit[200]{ns}), generated by a chip-control command, activates a chopper circuit that applies the voltage to provide a controlled amount of charge (0 to \unit[-9]{fC}) to the input of each channel where the calibration pulse is enabled. The polarity of the
calibration pulse is also controllable by a bit in the control registers. The relative phase of the
calibration pulse can be varied using a programmable strobe delay circuit from 0 to \unit[80]{ns} so the
position of the pulse relative to the BC can be tuned for optimal results.

\paragraph{Power and Ground}

The ABC130 has independent digital and analogue power domains, each with its own power (DVDD and AVDD) and ground (GNDD and GNDA) pad connections, and each has its own on-chip programmable Low-DropOut (LDO) regulator that can be used to provide the required regulated \unit[$+1.2$]{V} core voltage. Options were also included on the chip to allow for the application of the core voltages using external connections. By providing a sufficient number of power pads connected to the outputs of the LDOs (VDDD and VDDA) that are normally connected to decoupling capacitors, these could also safely be used as power inputs if the LDO's input pads are not being powered. Furthermore, a voltage-controlled high current shunt circuit was included to allow for series powering of the chips. All of these modes of powering the ABC130 were tested, and it was decided to use the LDOs as voltage regulators to provide core voltages for both the analogue and digital portions of the chip when used on modules. 

As part of the front-end pads array, there are four ground pads on each end to provide a ground reference for the sensor's HV decoupling and guard ring. Furthermore, there are three special sets of ground pads: analogue ground pads specifically for the front-end (GNDIT), and one pad each for the digital and analogue ESD circuit returns. On modules, all of these are wire-bonded to the respective digital or analogue ground planes of the hybrid.

The LDOs can be controlled by programming registers: each has its own Control
Enable Bit and a register field that allows them to be tuned to
\unit[$1.20\pm0.02$]{V} in 16 steps. If the Control
Enable Bit is not set, then
the output voltage of the LDOs (VDDD and VDDA) applied to the chip's core are
the voltages applied to their inputs (DVDD and AVDD) minus the minimal drop
across the LDOs. The chip is fully functional in this state; however, the LDOs
should be tuned to \unit[1.2]{V} for proper operation during data taking. In
addition to the programmable control, the chip has dedicated pads that can be
used to disable the LDOs in case the chip is to be powered by externally
provided core voltages or using the shunt circuit. The default state on power
on with no connection to the pads is to have the LDOs disabled but controllable.

The nominal pre-irradiation current at \unit[$+1.2$]{V} is \unit[40]{mA} for the digital portion of the chip, and \unit[70]{mA} for the analogue circuitry. However, due to the Total Integrated Dose current ``bump'' (TID bump, see section~\ref{sec:TIDBUMP}) experienced by this CMOS technology, the amount of digital current drawn by the chip will increase by a factor $\mathcal{O}(\unit[100]{\%})$
with increasing radiation dose before falling back to near pre-irradiation levels as the dose moves out of the TID bump range (around \unit[1]{Mrad}). To allow data taking to be consistent before, through, and beyond the TID bump operating region, the shunt circuit can be used to draw the difference between the expected maximum TID Bump current, and the current being drawn by the chip at the current TID. As the current increases through the TID Bump, the shunt current can be reduced so the overall current remains constant. Similarly, the shunt current can be increased again as the TID bump current begins to decrease, again maintaining a constant operating temperature and current draw as the TID increases, and helping to ensure comparable results for measurements taken throughout the irradiated operating regions of this CMOS technology. For the next chip generation, the ABCStar, a procedure for its pre-irradiation has been developed to pass the TID bump before the ASICs are assembled into modules. The shunt circuit is disabled by tying the Shunt Control analogue input to ground.

\paragraph{Digital Input and Output}
\label{para:abc130_dig_io}

There are two types of digital I/O pads on the ABC130: low-voltage
single-ended CMOS I/Os for low speed signals (LVCMOS~\cite{LVCMOS}), and
high-speed differential SLVS I/O with a nominal \unit[600]{mV} common-mode
voltage and \unit[400]{mV} differential voltage (SLVS~\cite{SLVS}).

Generally, static I/O uses the low-voltage CMOS single-ended signalling, and
all clocks and data I/O use high-speed differential SLVS signalling. Most
LVCMOS I/O is left without wire-bonds when assembled onto a module, with the
exception of the RSTB, and a
5-bit Chip ID (see section~\ref{subsec:assem_hyb}). Clock and command lines are implemented in a common-bus multi-drop configuration on the hybrids. Any command communications to the ABC130s contain a Chip ID and are only acted on
by the chip whose ID matches the one in the command (with the exception that
ID = 31 is a broadcast address and all ABC130s must respond to those
commands). Similarly, all packets output by an ABC130 include its Chip ID to
allow any packets it generates to be associated with that particular chip.
Chip IDs only need be unique within a group of ABC130 ASICs read out by the
same HCC.

In addition to the LVCMOS signals used during operation on a module, a number
of other pads were provided as experimental features, as risk mitigation, or
to assist in testing die before dicing the manufactured wafer of dice. These
include:
\begin{itemize}
 \item pads to disable the digital and/or analogue LDOs (active high with CMOS pull-downs)
 \item a Termination Enable pad (active high with a CMOS pull-down) that can be used to provide on-chip \unit[75]{$\Omega$} (\unit[82]{$\Omega$} max.) termination for the SLVS receivers
 \item the ``abc up'' pad (active high with CMOS pull-down) that can be used to invert the sense of the internal reset tree
 \item 5 pads to implement a JTAG~\cite{JTAG} test interface (Scan\_Enable, SDI\_CLK, SDI\_BC, SDO\_CLK, and SDO\_BC).
\end{itemize}
The ABC130 will operate properly with any or all of these pads left unconnected.

All dynamic operations of the chip use high-speed differential SLVS I/O (each logical signal has both a positive and negative pad to provide differential input, output, or I/O as appropriate):
\begin{itemize}
 \item two clock inputs, BC and RCLK (Readout CLocK)
 \item two command and trigger inputs, COM\_LO and L1\_R3
 \item a set of bi-directional data and flow-control signals: one set for the ``left'' side of the chip, DATAL and XOFFL; and one set for the ``right'' side of the chip, DATAR and XOFFR
\end{itemize}

The \unit[40]{MHz} nominal differential BC clock is provided to all the ABC130s
on a module via the HCC130 and is used to trigger sampling of the
front-end inputs.
The BC is also used as the clock for both the COM\_L0 and L1\_R3
differential Dual-Data Rate (DDR) inputs with effective input data rates of 2
times BC (\unit[80]{Mbps} nominal). Each input is split into two \unit[40]{Mbps}
signals. On the rising edge of BC, the COM\_L0 is latched as the command data
stream to the ABC130s; and on the falling edge of BC, that signal is latched
as the L0 trigger. Similarly, the L1 trigger
data stream on L1\_R3 is latched on the rising edge of BC, and the R3 trigger
data stream is latched on the falling edge of BC. Finally, the HCC130
provides the differential RCLK signal at up to four times the rate of the BC
(\unit[160]{MHz} nominal) that is used to clock data on the DATAL/R digital
readout pads and the XOFFL/R flow-control pad signals.

The bi-directional signals are configured in pairs, so that when DATAL is an
output, DATAR is an input and vice versa. Similarly when DATAL is an output,
XOFFL is an input and vice versa as in table~\ref{tab:abc130_bidirections}.
A single configuration register bit determines whether the ABC130 is
operating in a ``right to left'' mode or in a ``left to right'' mode.
When configured as outputs, the differential output current of the drivers are
programmable between \unit[1]{mA} and \unit[7]{mA}, in 8 steps.

\begin{table}[htbp]
\centering
\begin{tabular}{l|l|l|l}
\textbf{Signal} & \textbf{Side of ASIC} & \textbf{I/O in L-R} & \textbf{I/O in R-L} \\
\hline
DATAL   & Left  & Input  & Output    \\
DATAR   & Right & Output & Input     \\
XOFFL   & Left  & Output & Input     \\
XOFFR   & Right & Input  & Output    \\
\end{tabular}
\caption{Bidirectional signals, which side of the ASIC they are positioned, and whether they are inputs or outputs}
\label{tab:abc130_bidirections}
\end{table}

When in ``right to left'' mode, DATAR is an input and forwards data received
from its neighbour to the right through to DATAL.
XOFFL is an input (recieving flow-control signalling from its neighbour to the
left) and XOFFR is an output (providing flow-control to its neighbour to the
right). When in ``left to right'' mode, the data, flow-control, and I/O
directions are reversed.
The ABC130s on a module are
connected in a daisy-chain with the DATAR and XOFFR on one chip connected to
the DATAL and XOFFL of the next chip to its ``right''.
The farthest ``left'' and farthest ``right'' ends of the daisy-chain are
connected to the HCC130 (see section~\ref{sec:HCC}), which can
receive data and/or provide flow-control signals from either of the ends of
the daisy-chain. This architecture allows part of the daisy-chain to be
configured as ``left to right'' and the other part as ``right to left'' to
handle the case of a single failed ABC130 anywhere in the daisy-chain. All
chips to the ``left'' of a faulty ABC130, if any, are configured in the
``right to left'' direction; and all chips to its ``right'', if any, are
configured in the ``left to right'' direction. This way, maximal physics data
can be read out from a partially operational module (see
figure~\ref{fig:abc130_daisychain}).

The ABC130 was manufactured on a Multi-Project Wafer (MPW) run along with an unrelated chip. There were two
different versions of the ABC130 in each reticle on the wafer: the ABC130\_0 and the ABC130\_1, which
were identical in function except the ABC130\_1 additionally had experimental circuitry for a Fast
Cluster Finder (FCF) and additional pads to support that functionality. The FCF was designed to provide
prompt, BC-synchronous, cluster position data to an external device that could be used to correlate
clusters between tracking layers and select high $p_{\textrm{T}}$ (transverse momentum) coincidences to a trigger processing
unit. Ultimately, this functionality was not used on the modules built with ABC130 chips, and was not
tested during wafer testing either. The operation of this circuitry is beyond the scope of
this article, but can be found in the ABC130 Specification~\cite{ABC130Spec}.

\paragraph{Chip operation}
\label{sec:ABCop}

For normal operation, after the chip is reset, the registers on each ABC130 are initialised using the command stream of the COM\_L0 DDR input with the values that have been determined to provide it with nominally tuned settings, and to set all relevant mode bits necessary to put the chip into the desired operational configuration. These settings include:
\begin{itemize}
 \item the LDO tuning value required to provide \unit[1.2]{V} core voltages to the analogue and digital circuit domains
 \item all of the front-end control DACs and TrimDACs to correct for process and inter-channel variation
 \item the channel mask registers to disable any known faulty input channels
 \item the required SLVS driver currents
 \item and the threshold value.
\end{itemize}
Setting the threshold to an optimal value for each ABC130 - which is distributed to all of its 256 input channels, each of which is fine-tuned by the per-channel TrimDACs - is critical, as it determines the front-end's sensitivity to signals from the sensor strips it is reading out, and its susceptibility to noise from the sensor and the front-end circuitry. The threshold can be set based on the requirements of a particular data-taking session and determines the ``hit'' rate (which can include both signal and noise), and thus the maximum data transmission bandwidth from the module during operation. There are features that can limit the maximum data rate of the ABC130s, but using these results in the discarding of potential hits (see below).

The decision to record a hit or no-hit is taken on the rising edge of the BC
clock. The state of all 256 input channels of every ABC130 on a module will be
sampled into its L0 Buffer, a \unit[256]{bit}-deep FIFO. The state is formed
by the logical AND of the input comparators reading out the sensor strip it is
wire-bonded to, and the inverse of the associated Mask Registers bits (a 1 bit
in a Mask Register will force its associated channel to always read as the 0,
or ``no hit'', state). Each of these 256-bit input vectors is pushed from the
front-end onto the L0 Buffer FIFO along with the value of an 8-bit, command
resettable, BC counter (BCID).
Because this process is continuous, a sample will
remain in the L0 Buffer to be read out for a maximum of \unit[6.387]{$\upmu$s}
before falling off the far end of the FIFO.

\begin{figure}
\centering
\includegraphics[width=0.9\linewidth]{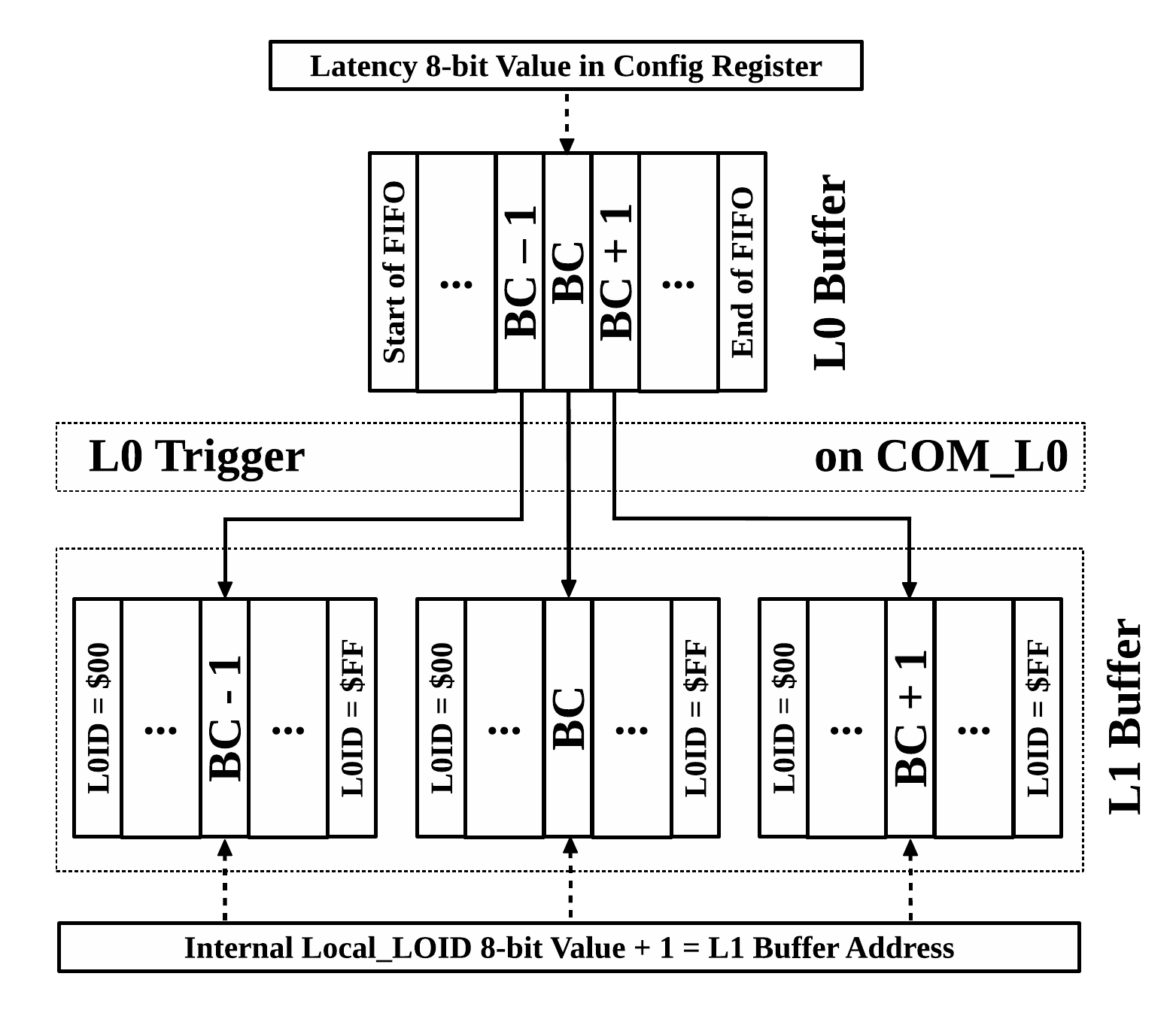}
\caption{Transfer of event from L0 Buffer to L1 Buffer on receipt of L0 Trigger on COM\_L0 input.}
\label{fig:abc130_event_xfer}
\end{figure}

The next step is to capture the data from the L0 buffer into the L1 buffer.
To save input vectors for possible later readout, an L0 (first level) trigger
accept needs to be issued. This L0A is actioned by logic-level 1 on the
COM\_L0 DDR input.
When an L0A is received, one ``event'' from the L0 Buffer is transferred to the
L1 Buffer.  The Latency is the fixed number of BCs between when the front-end is sampled
and when an L0A is received by the module from the trigger system to
store that sample. This is configurable by
the setting of the 8-bit Latency value in the chip's control register set, and
specifies the address in the L0 Buffer of the centre of a 3-BC long ``event''.
As shown in figure~\ref{fig:abc130_event_xfer}, an entry in the L1 Buffer consists of three 256-bit
memory blocks, which will be used to store the L0 Buffer entries for:
the previous bunch crossing, the bunch crossing of interest, and the next
bunch crossing. All three of the
values copied to the L1 Buffer (both the 256-bit input vector and
the associated 8-bit BCID) will further be tagged with an
8-bit Local L0 IDentifier Counter (Local\_L0ID Counter) value. This whole event
is stored at the address in the L1 Buffer specified by the Local\_L0ID Counter
after it is incremented by one. Like the BC Counter, the Local\_L0ID Counter
is settable to a known initial value (usually \$FF) when data taking begins
so both the trigger system and the ABC130s are in synchronization and the
first L0 Trigger writes into location 0 of the L1 Buffer. Since the L1
Buffer can store 256 entries, for a \unit[500]{kHz} L0 Trigger rate, for
example, it can hold the data for an event for \unit[512]{$\upmu$s} before it
is overwritten with another event.

The final stage is to read out the event data from the L1 buffer.
To read out the physics data of an event, an L1 or R3 Trigger is issued by the trigger system via the
L1\_R3 DDR input (through the HCC130). Each of the trigger commands consists
of a three-bit header - 110 for L1 and 101 for R3 - followed by the 8-bit L0ID value of the event to read
out. The L0ID value sent with the trigger is simply the number of L0As sent
since the Local\_L0ID counter was reset, and corresponds to the memory
address of the L1 Buffer that will be read out.
Depending on whether an L1 or R3 Trigger was issued, the event is sent to the
L1-DCL (Data Compression Logic), or the R3-DCL, which generate a sequence of
fixed length data packets.

The R3-DCL produces only a single output packet, and is intended to provide a
quick snapshot of whether or not any clusters were detected for a particular
bunch crossing event. Conversely, the L1-DCL produces a comprehensive output of all cluster data for the relevant event where the hit data matches a specified
pattern, and can result in many packets being queued for transmission.
Both DCLs find clusters of hits, by searching the channels first in one
set of 128 channels followed by the other 128 channels. Thus clusters are
found only between strips in the same sensor region. Also, it does not
``wrap'' from one set to another.
These are recorded in the output packet as 0-127 and 128-255.

\begin{figure}
\centering
\includegraphics[width=0.9\linewidth]{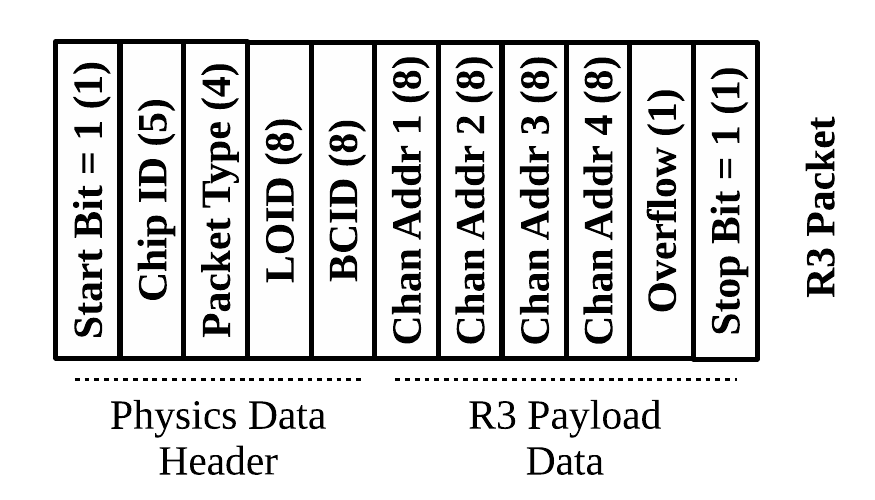}
\caption{Format of Regional Readout Request (R3) ABC130 Output Packet, the number in parenthesis indicates the corresponding number of bits.}
\label{fig:abc130_r3_packet}
\end{figure}

For an R3 Trigger, the R3-DCL will generate a single output packet flagging
whether there are no hits, some hits (1-4), or many hits (more than 4).
The R3-DCL can be configured through the
``EN\_01'' bit in the configuration registers to either define a ``hit'' by looking for a hit only in the L1 Buffer block corresponding to the selected BC
(level mode), or to look for a level change from 0 to 1 between the previous
BC and the selected input vector (edge mode).
The R3-DCL only registers clusters that have hits in a maximum of three
channels, larger clusters are ignored. The location of the first hit
is reported for clusters with width of 1 or 2, and the location of the central
strip when the cluster width is 3.
The R3-DCL will report the locations of a maximum of 4
hit clusters per event, and will set an overflow bit in the output packet if
there are more than 4 valid
clusters~\cite{ABC130Spec}. The format of the R3 packet is detailed in
figure~\ref{fig:abc130_r3_packet}.

\begin{figure}
\centering
\includegraphics[width=0.9\linewidth]{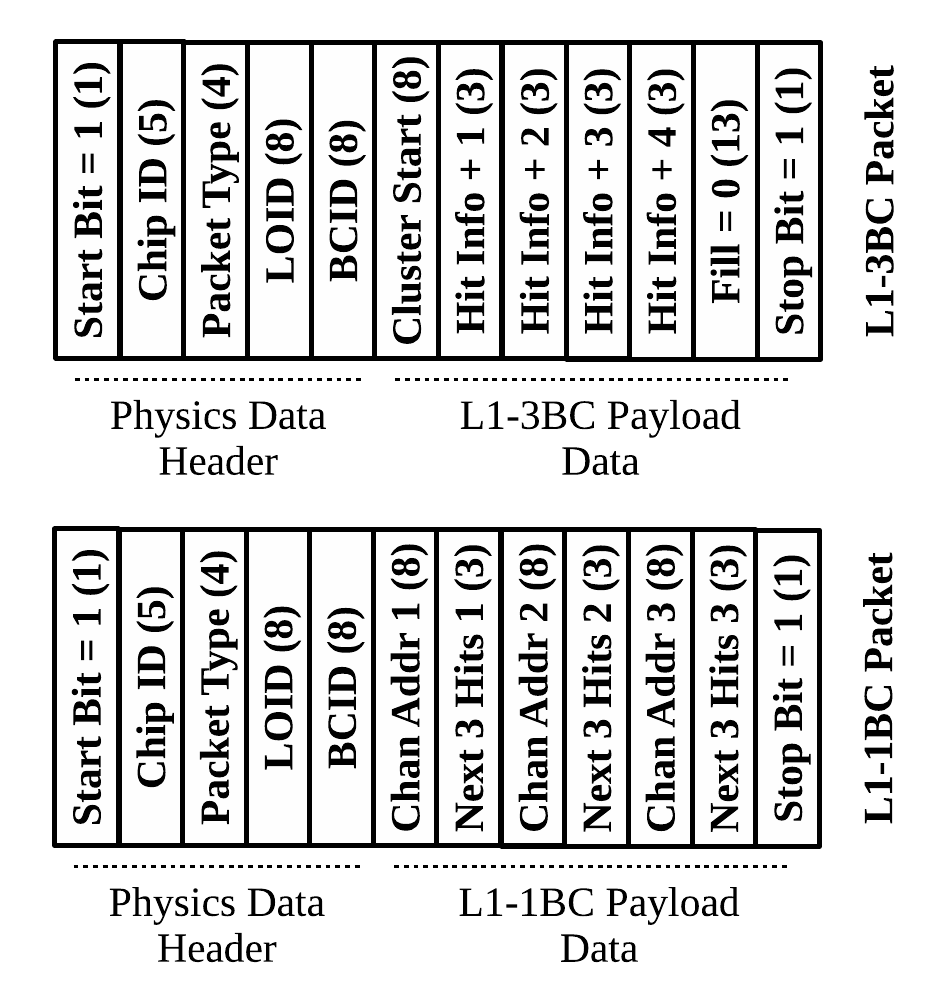}
\caption{Format of Second Level (L1) Trigger ABC130 Output Packets [number in parenthesis is number of bits].}
\label{fig:abc130_l1_packets}
\end{figure}

For an L1 Trigger, the L1-DCL reports clusters in one of two formats as selected by the ``mcluster'' bit
in the configuration registers: either L1-3BC mode where information on all 3 recorded BCs are reported; or L1-1BC mode, where only cluster patterns are reported (see figure~\ref{fig:abc130_l1_packets}).
For both modes, a compression mode can be configured to choose which clusters
to report based on the pattern of bits in the 3 recorded BCs. There are
two patterns for use during normal data taking, X1X (level) and 01X (edge), where the X indicates ``don't care''.
A further ``any hit'' mode to be used for detector alignment matches
(1XX or X1X or XX1). A final XXX mode is intended only for chip testing.
Clusters are scanned for in the two sets of 128 channels based on this mode
and clusters formed.
When a hit is found that matches the selected pattern, that bit forms
the first bit of that cluster 
and its location is used to report the
start of the cluster in the packet.
For the L1-3BC
mode, that location is used as the channel address reported in the packet. The address is followed by the
3 bits for the hit on that channel (from the 3 recorded BCs), and the next
three channels(whether or not they have any hits in them).
The DCL then moves to the following channel to check for a new cluster start.
Because 4 channels are reported per L1-3BC packet, a total of 64 packets
could potentially be created if at least every 4th channel had a hit to cause a cluster to be reported.
For the L1-1BC mode, up to three clusters can be
reported: each one comprised of the 8-bit cluster start location, and the
one-bit hit status of the following three channels (3 bits) based on whether that channel matched the hit
criteria. Like the L1-3BC mode, the next cluster is searched for after the last bit of the previously
reported 4 channels of cluster data. Because 3 clusters can be reported in each packet, up to 22 L1-1BC
packets could be generated if every 4th channel had a hit to cause a cluster to be reported.
It
should be noted that due to the potentially large number of packets generated for an L1 Trigger event
with many hits, the possibility of saturating the front-end circuitry with noise or hits from the sensors
when the thresholds are set low needs to be managed carefully. If a large number of hits are expected,
the L1 Trigger rate needs to be controlled carefully to ensure no loss of data in those situations. A
further feature is provided in the ABC130 that allows the number of packets generated to be capped at
some specified number less than 64 for L1-3BC mode or 32 for L1-1BC mode. While this mode could result in
data loss, the assumption would be that the high-occupancy data is not useful.

\begin{figure}
\centering
\includegraphics[scale=0.7]{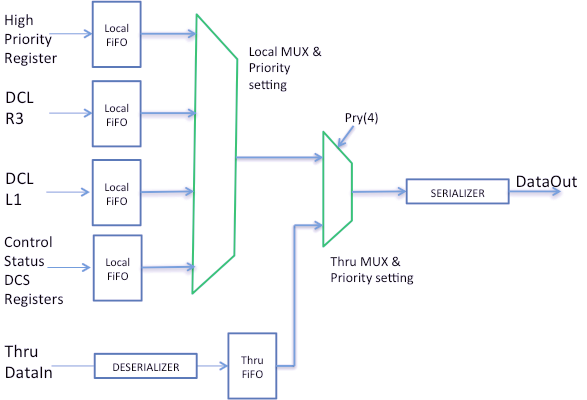}
\caption{ABC130 Packet Output Multiplexer with Priority Control}
\label{fig:abc130_outmux}
\end{figure}

As packets are created by the DCL, they are pushed onto the appropriate
flow-controlled FIFOs and are then pulled from the FIFOs and serialized
according to
their relative priority. In addition to FIFOs for the L1-DCL and R3-DCL,
there is a
FIFO to queue configuration register reads, and a separate FIFO to output the
reading of a special high-priority status register (Register \$3F). These are
output in order of highest to lowest priority: high-priority register reads,
R3-DCL, L1-DCL, and then regular register reads at the lowest priority.
Furthermore, packets that are being transmitted through the ABC130 from an
adjoining ABC130 or HCC130 are interleaved into the output data stream based on
the setting of 4 Pry (priority) bits in the configuration registers. If there
are packets from the internal data sources to send, Pry sets the number of
through-packets that might be forwarded before one internally generated packet
needs to be sent.
Thus, if Pry is set to 0, then a through-packet will
always be sent before any internally generated packets. If Pry is set to 8, for
example, then 8 through packets (if present) will be forwarded before sending
the next internally generated packet. If the through-packet FIFO (which
is 4 deep) is about to be filled, the XOFF signal is asserted to the
upstream chip to prevent the FIFO from overfilling and for
through-packets to be lost that way. Similarly, if the internal FIFOs are about
to fill up, the blocks that are sending data to them receive internal flow-control
signalling and must stop operation until FIFO space is available for them to
continue sending (see figure~\ref{fig:abc130_outmux} on the FIFO and priority
structure).

\paragraph{Wafer testing}

Since \mbox{ATLAS} ITk strip tracker modules have up to 12 ABC-style chips connected to one HCC, if even one chip were to fail due to a manufacturing defect during module testing, it would require a risky and complex re-work effort to attempt to recover the module. While the data-flow architecture allows for the routing of data around a failed ABC130, 256-channels of data would still be lost. If that repair process failed or damaged the module, that one failed chip could result in the costly loss of an entire hybrid or module. As such, each ABC130 die undergoes an extensive testing and characterization process while still on the wafer using specialized wafer probing equipment and custom software. The ABC130s on the wafer are categorized into good (grade A), usable in the lab or in the event of a parts shortage (grade B), and bad dice (which can be used as mechanical samples for assembly testing and tooling development).

ABC130 wafer testing was conceptualised, developed, and implemented using a commercial semi-automated wafer-probing system, a commercially manufactured custom probe card, a custom interface card and a commercial off-the-shelf (COTS) FPGA development board. The use of COTS components resulted in a much faster test system development compared to the custom electronics used for wafer testing during the construction of the \mbox{ATLAS} Semiconductor Tracker~\cite{wafer_testing}.
The wafer test software was integrated with the module test software suite: the ITk Strips DAQ (ITSDAQ, formerly SCTDAQ), see section~\ref{sec:software}. In the tests run on each ABC130 die on every wafer, the wafer is manually placed on the wafer-probing station's platen and positioned using the probe-station's microscope's digital camera. Once the wafer is aligned, the wafer test software can step automatically between each of the ABC130s on the wafer and run all necessary tests on it.

The tests begin with basic integrity testing looking for gross failures of the die in terms of power supply currents and to ensure proper contact between the probe card and all of the die's pads. The tests then conduct a number of further sanity checks including:
\begin{itemize}
 \item setting registers to default values and verifying that the power supply currents change appropriately
 \item tuning the chip's LDOs and front-end DACs
 \item scanning all other DACs
 \item a series of digital tests to check the functioning of the digital portion of the chip
 \item a series of complex functional tests are conducted to verify the chip's proper operation from front-end to data output
\end{itemize}
These final tests include tuning the chip's strobe delay value to provide optimal stimulus using the built-in calibration pulses, and then running a comprehensive 3-point gain test where each channel's front-end response to three different calibration pulse charges (\unit[0.5]{fC}, \unit[1.0]{fC}, and \unit[1.5]{fC}) is plotted to determine the gain response and noise level of all the input channels (see section~\ref{chartests}). These tests also verify the functioning of triggering blocks, the L0 and L1 Buffers, the cluster finding and sparsification blocks, and the data transmission I/O blocks. The data is analysed in real time by ITSDAQ and the parts undergo a preliminary categorisation at that phase. Further analysis is performed offline on the data produced by the wafer probing routines where comparative analyses are also done between dice and between wafers, and the die categorisation can be updated as needed at that phase.

In preparation for the next generation chip, the ABCStar, and to ramp up towards full production testing, a second wafer test site was established. Whereas previously a system primarily developed in-house had been used for wafer testing, the second test site was established at an outside company to use their commercial, fully automated, wafer-probing stations and associated industrial test software infrastructure.

\subsubsection{HCC130}
\label{sec:HCC}

\begin{figure}
 \centering
 \includegraphics[width=1.0\linewidth]{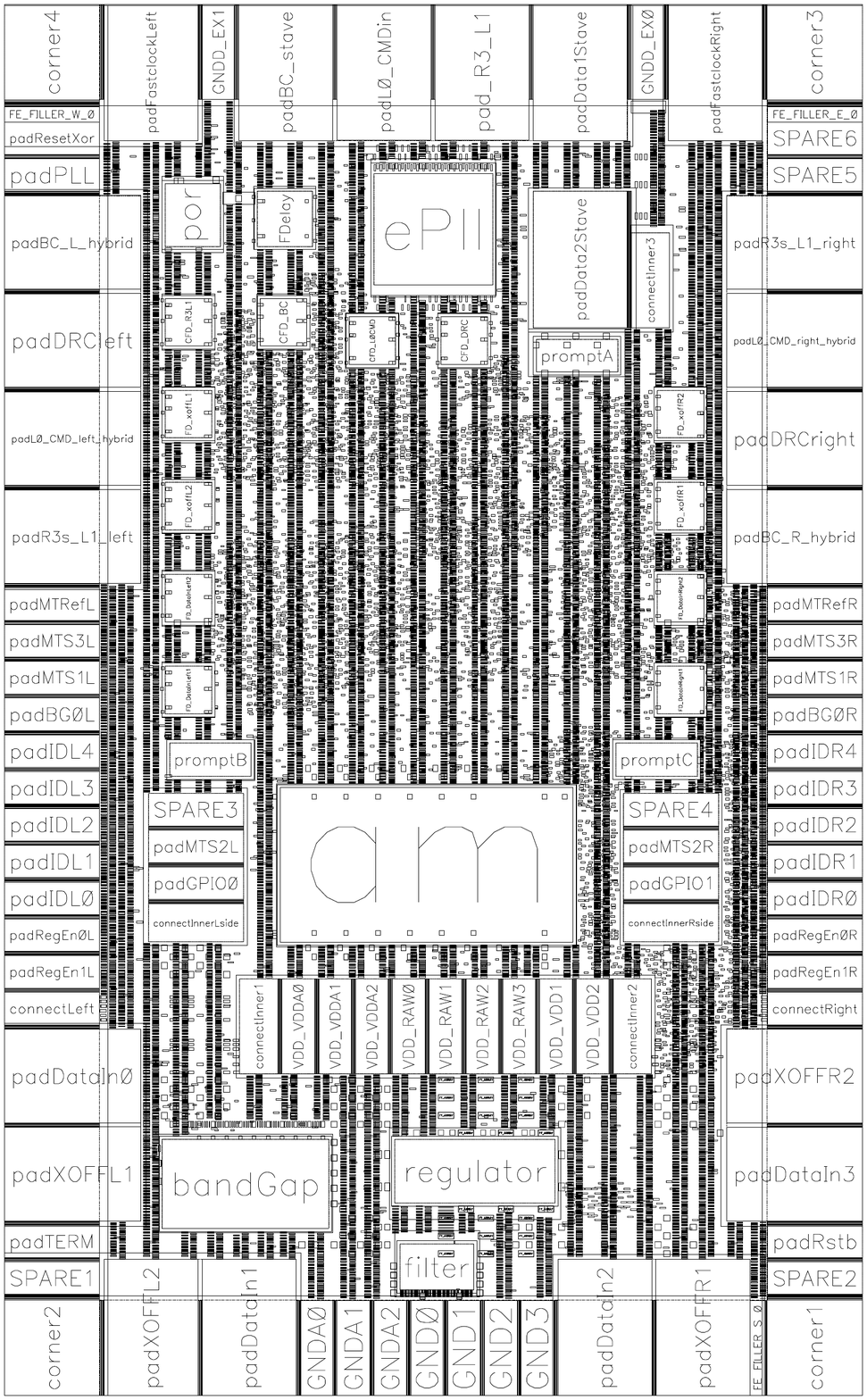}
 \caption{HCC130 Pad Ring and floorplan}
 \label{fig:HCC130PadFrame}
\end{figure}

The Hybrid Chip Controller, HCC130, the first \mbox{ATLAS} strips prototype
chip controller, was submitted for fabrication in August 2014. The
99-pad \unit[$4.7\times2.96$]{mm$^2$} ASIC (see
figure~\ref{fig:HCC130PadFrame}) was designed to provide the
interface between the hybrid-mounted front-end ABC130 and the off
detector electronics through the GBTx~\cite{GBTX} using LVDS-like low
voltage differential drivers and receivers. It also contained an
early version of the Autonomous Monitor that was functionally validated
and eventually moved to the AMAC ASIC (see section~\ref{sec:AMAC}). HCC130 receives the
\unit[40]{MHz} bunch crossing (BC) clock and two, custom protocol control
signals from multi-drop buses driven by the
GBTx. Both of these control signals are encoded with two independent
logical streams time multiplexed into one. 
Data of all types sent from each module are transmitted point to
point by the HCC130 to the GBTx at 160 or \unit[320]{Mbps}.

The L0\_COM physical control signal encodes L0, a
beam synchronous trigger that stores the ABC130 pipeline delayed data
into a 256-word deep data buffer from which data are requested for
readout. The second logical channel of L0\_COM signal is a
priority based variable length Command (COM) protocol that provides
control to set operational modes in both the HCC130 and ABC130 ASICs
and initiates requests for data from internal registers in the HCC130
and ABC130 ASICs. The difference in naming from the ABC130 indicates a
subtle difference in that the HCC130 modifies the COM signal before
forwarding to the ABC130, to mask out commands intended for ABC130 attached
to other HCC130 on the same bus.

The R3s\_L1 physical signal provides two triggers used to request readout of
data from the ABC130. The L1 is broadcast directly to the ABC130s, requesting
readout of data from one of the 256 memory locations of the ABC130s L1 buffer.
In addition to the L0ID identifying the memory location to read out, the R3s
signal contains an extra 14 bits, used to propagate the signal only to 
HCC130s with matching addresses. In this mode only addresses 2-29 are
available. If the address matches, the mask bits are stripped and the remainder
broadcast to the ABC130s.

\begin{figure}
 \centering
 \includegraphics[width=1.0\linewidth]{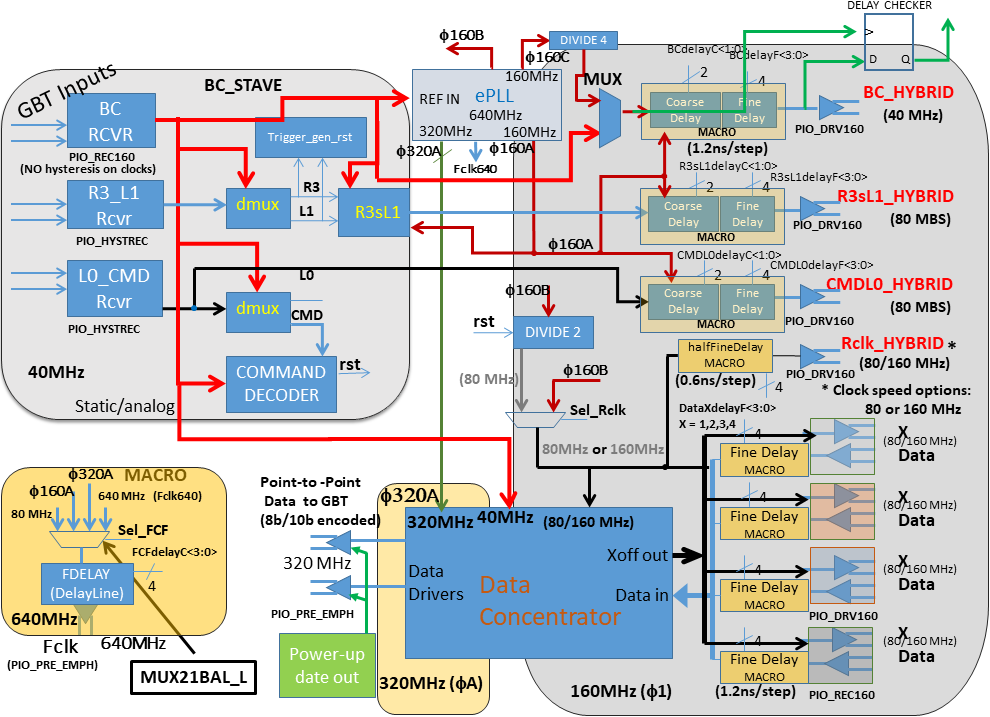}
 \caption{HCC130 block diagram}
 \label{fig:HCC130BlockDiag}
\end{figure}

The HCC130 utilises a copy of the CERN ePll block designed
for the GBTx~\cite{GBTX} to generate low jitter, 40, 80, 160 and \unit[320]{MHz}
clocks using the incoming multi-drop \unit[40]{MHz} BC as a
reference. The ePll is used internally on the HCC130 and provides the
hybrid ABC130s with a regenerated, programmable, phase delayed
\unit[40]{MHz} clock for phasing event data properly within the beam
crossings and related pipeline control. It also provides the ABC130s
on the hybrid with a selectable 80 or \unit[160]{MHz} data clock to
drive the serial loops used for data readout. HCC130 can collect data
from the hybrid through any of its four serial receivers attached
to either end of the two hybrid readout loops. Corresponding XOFF signals
allow for flow-control to the ABC130s. This readout technique
provides contingency for single and multiple chip failures in either
of the two ABC130 serial data loops. 
Once on the HCC130, a priority encoder ensures an even flow of data from
the two ends of each loop and that R3 data are sent to the DAQ system
with the highest priority. A data concentrator merges data from the
two loops. A detailed HCC130 functional block description is shown in
figure~\ref{fig:HCC130BlockDiag}.

\subsubsection{AMACv1a}
\label{sec:AMAC}

\begin{figure}
 \centering
 \includegraphics[width=0.8\linewidth]{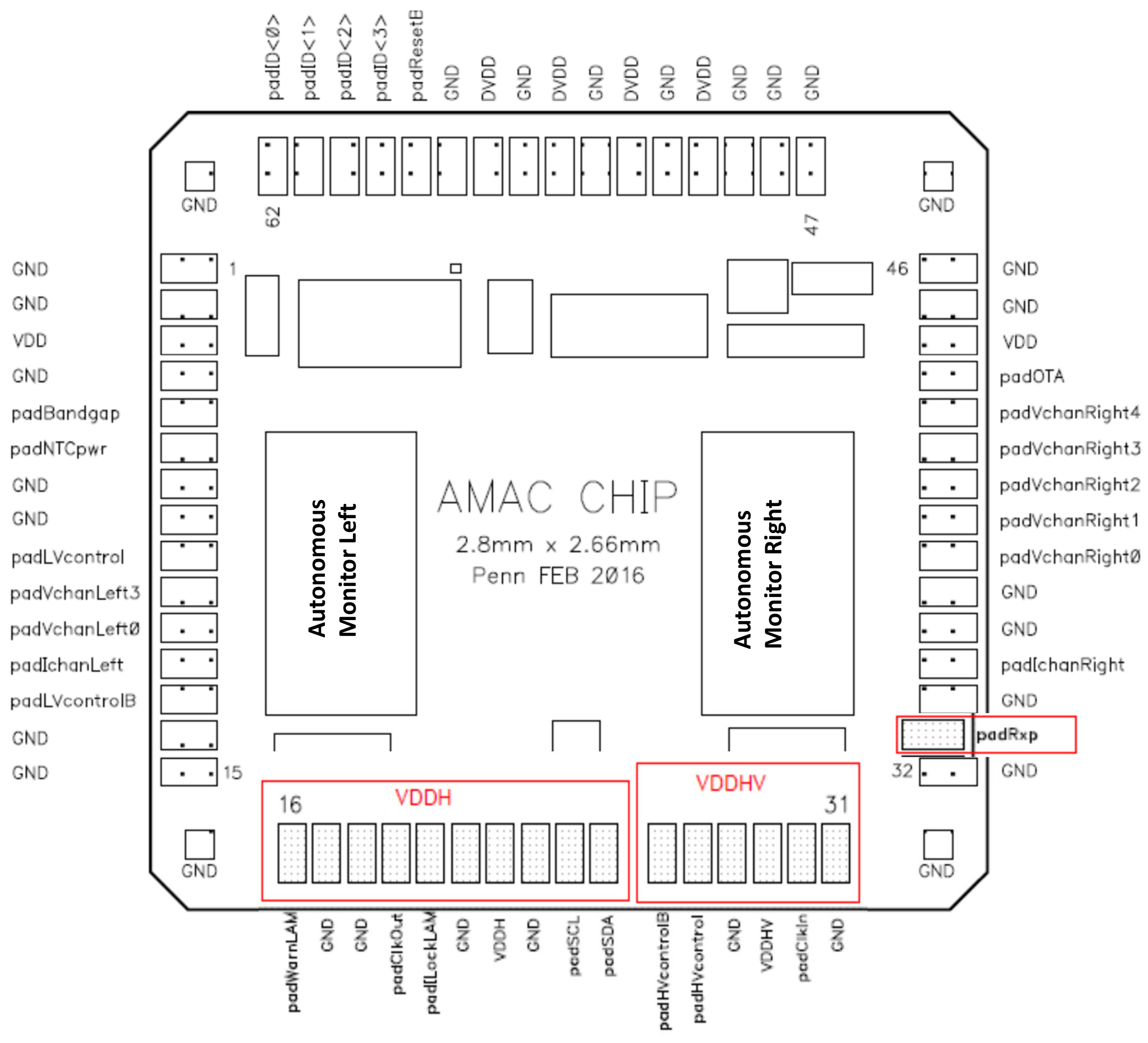}
 \caption{AMAC v1a Pad Ring and floorplan}
 \label{fig:AMACv1aDie}
\end{figure}

The Autonomous Monitoring And Control ASIC (AMAC) was submitted for
fabrication in August 2016. It was the first successful prototype of
the radiation tolerant, ten bit precision analogue monitor ASIC and was
constructed using two identical seven channel Autonomous Monitor
blocks originally housed in the HCC130. The AMACV1a pad frame is
shown in figure~\ref{fig:AMACv1aDie}: it has 62 bond pads and a die size
of \unit[$2.7\times2.8$]{mm$^2$}. An internal ring oscillator
provides a near \unit[40]{MHz} clock to control the autonomous monitor
functions and I2C protocol is used for control and readout. AMACV1a
monitors 14 independent module level parameters: voltages,
temperatures and sensor leakage current. A clock driven state machine
controls a switched capacitor stepped integration ramp to create a
common reference for the Wilkinson style ADC. Each integration step
increases the reference by \unit[1]{mV} and increments a ten bit counter reset
to 0, each 1023 step ramp cycle. Each of the 14 monitored parameters
is translated into a voltage between 0 and \unit[1]{V} and compared
with the ramp that covers the same range. When the reference ramp
exceeds the value of the measured parameter for two consecutive ramp
steps the counter value is recorded in a register and compared with
pre-programmed upper and lower limits. Out of limit values are
flagged and - if enabled - can switch the state of four logic outputs
outputs that may be wired to LV or HV supply controls. Measured
values are updated and stored locally once per millisecond. They may
be readout remotely through the I2C interface.

\subsection{Hybrids}
\label{comp:hyb}

Readout chips for ABC130 barrel modules are mounted on flexible, radiation hard and low mass Polyimide circuits (called hybrids), which were developed to carry ten
ABC130 readout chips and one HCC130 readout chip each (see section~\ref{sec:comp}).

Hybrids provide the following electrical functionalities:
\begin{itemize}
 \item Multi-drop external clock and control are connected to the hybrid
 \item On-hybrid internal clock and control are distributed to the ABC130 chips
 \item All ASICs on the hybrid are connected to a common ground and power domain
 \item Hybrid front-end data is returned to the high-level readout via the end-of-substructure card (see figure~\ref{fig:intro_stave})
\end{itemize}
The hybrid readout topology groups the ten ABC130 readout chips
into two daisy-chains of five chips each. Each chain can be read
out in either direction by the HCC.

ABC130 barrel modules are assembled by gluing hybrids with readout
chips directly onto the surface of silicon strip sensors (see
section~\ref{subsec:MA}). The circuit layout has been optimised to
minimise electrical interference into the sensitive analogue front-end electronics or sensor strips. This has been achieved by the use of a single power
and return domain with partioning of analogue and digital circuitry to
mitigate common impedance coupling of the analogue and digital
signalling. Furthermore, fast digital signalling within close
proximity of the analogue front-end are routed as differential
strip-lines to take advantage of the shielding effect of the return
planes.

In addition to ABC130 and HCC130 readout chips, hybrids are equipped
with two NTC thermistors, of which one is used to monitor the hybrid
temperature during operation. The second one is part of a
temperature interlock system required during the burn-in of
hybrids as part of their quality control.

In order to ensure a maximum yield, hybrids were designed to utilise
standard manufacturing processes with long-term
reliability. Tracks and gap sizes are about
$\unit[100]{\upmu\text{m}}$, vias (plated laser drilled holes) have a
hole diameter of about $\unit[150]{\upmu\text{m}}$ and lands of
$\unit[350]{\upmu\text{m}}$, which ensures uniform plating and
therefore reliable contacts through vias. Barrel hybrids comprise three copper layers ($\unit[18-35]{\upmu\text{m}}$
thickness) between polyimide dielectrics ($\unit[50]{\upmu\text{m}}$
layer thickness), resulting in a total hybrid thickness of
approximately $\unit[300]{\upmu\text{m}}$.

\begin{figure}
 \centering
 \includegraphics[width=0.7\linewidth]{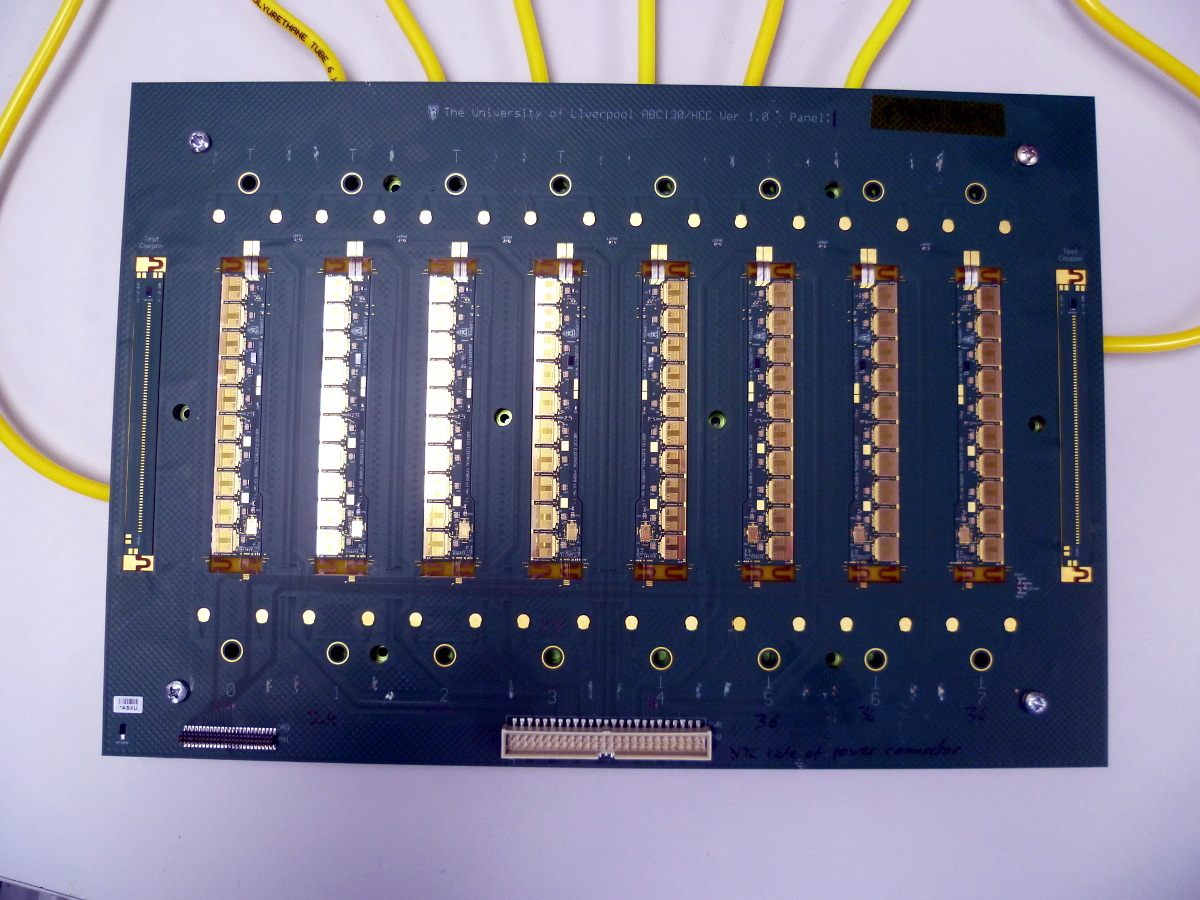}
 \caption{ABC130 barrel hybrid panel with four X-type hybrids, four
   Y-type hybrids and two test coupons.}
 \label{fig:panel}
\end{figure}
Hybrids are produced on panels (glass-reinforced epoxy laminate
sheets) which hold four X- and four Y-type hybrids (see
figure~\ref{fig:panel}) as well as two test coupons per panel, which
are used for hybrid manufacturing quality control:
\begin{itemize}
 \item testing of via reliability by via chain resistance measurements
 \item monitoring trace etching quality by testing DC resistance of test traces
 \item testing quality of surfaces for wire bonding by performing wire bonding pull tests
\end{itemize}
Hybrid panels are equipped with vacuum holes and landing pads for
assembling hybrids and readout chips (see
section~\ref{subsec:assem_hyb}) as well as traces and connectors for
the electrical testing of fully assembled hybrids (see
section~\ref{subsec:test_hyb}).

\subsection{Powerboard}
\label{comp:pb}
The powerboard fulfils three purposes on the ITk Strip Module:
\begin{enumerate}
\item DCDC regulation of the \unit[11]{V} input to \unit[1.5]{V} to supply the ASICs on the hybrid
\item High-voltage switching of up to \unit[-500]{V} to reverse bias the sensor
\item Control of low-voltage and high-voltage supply, as well as monitoring of voltages, currents, and temperatures
\end{enumerate}

The DCDC regulation is achieved by the FEAST chip~\cite{feast}, a
radiation hard custom ASIC developed by the CERN electronics group for
various experiments and their detectors. The FEAST employs a
buck-converter style switching regulator, which requires an external
inductance. For the powerboard, this inductance is an air-core solenoid
coil with a nominal inductance of \unit[545]{nH} and DC resistance of
\unit[35]{m$\Omega$}. It is required to be an air-core coil as the
detector will be placed in a \unit[2]{T} solenoid field, which precludes the application of ferrite cores.

Due to the shape and characteristics of an air-core solenoid, during operation the coil emits RF noise, which could be picked up by the silicon strips underneath and around the powerboard. Therefore, the whole DCDC circuit is enclosed by
a shield formed by a specific copper layer in the PCB underneath and a
\unit[100]{$\upmu$m} thick aluminium shield-box soldered on top with continuous seams.

Switching control of the high-voltage is gained via a GaNFET
transistor switch, by routing the high-voltage supplied to a module
onto the powerboard through the switch. This routing also allows a
low-pass RC-filter to be placed at the output of the high-voltage line,
which is connected to the silicon sensor. To switch the GaNFET, a
voltage of more than \unit[2]{V} between gate and source is needed, as the
source of the transistor after closing the circuit is at high voltage. An AC signal at a frequency of \unit[100]{kHz} and amplitude of
\unit[3.3]{V} is AC-coupled into the high-voltage domain. It is then amplified
and rectified via a quadruple charge pump circuit, which generates the
necessary gate voltage with respect to the current source potential.\\

Both the low-voltage and high-voltage domain are controlled by the
AMAC chip, which has been designed
specifically for the usage on the powerboard. It can generate an
enable signal for the FEAST chip to turn on the power to the
\unit[1.5]{V} domain and also generates the AC signal to switch the
high-voltage on or off. Furthermore, the AMAC chip features
multi-channel ADCs to measure multiple operation critical values: 
\begin{itemize}
 \item input and output voltage and current
 \item FEAST internal temperature (via an internal PTAT circuit)
 \item powerboard temperature (via NTC thermistors)
 \item hybrid temperature (via NTC thermistors, for later powerboard versions)
 \item and silicon sensor leakage current
\end{itemize}
It also contains logic to set an upper and lower boundary on
the monitored values and if these limits are violated it can interlock
the low-voltage or high-voltage of the module.

\begin{figure}
\centering
\includegraphics[width=0.9\linewidth]{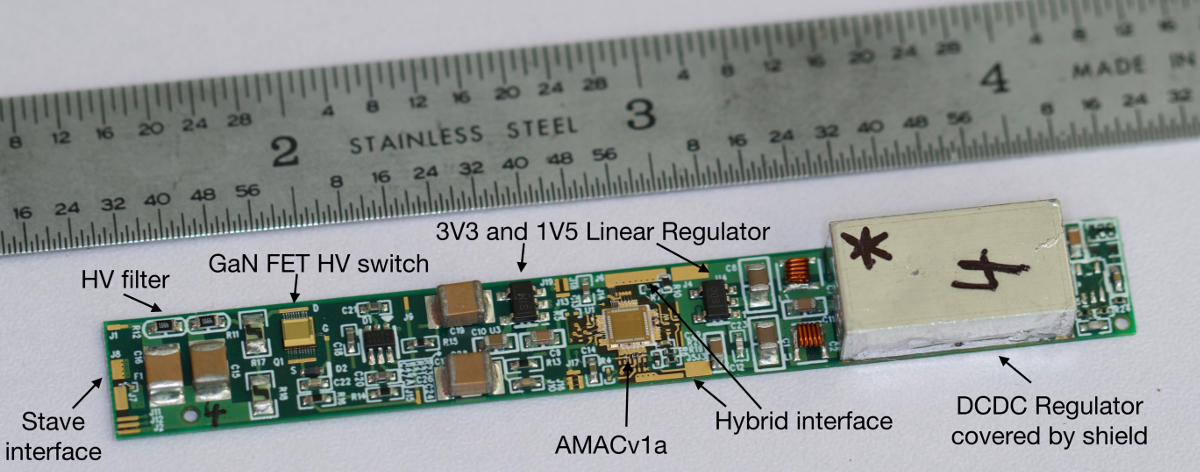}
\caption{Photograph of a fully assembled powerboard v2. The main components of design are pointed out as well as the interface.}
\label{fig:powerboard_v2}
\end{figure}

A picture of powerboard v2 can be seen in
figure~\ref{fig:powerboard_v2}, which shows the main components of the
powerboard. On this specific version of the powerboard the AMAC is
powered via two commercial linear regulators, in the next version of
the powerboard these will be replaced by a rad-hard linear regulator,
the LinPOL12V, which will also be used in the final production version
of the powerboard.

\section{Module construction}

Each ABC130 barrel module consists of one sensor (see
section~\ref{subsec:comp_sensors}), one or two hybrids (see
section~\ref{comp:hyb}) with ten ABC130 chips and one HCC130 chip each
(see section~\ref{subsec:comp_chips}) and one powerboard (see
section~\ref{comp:pb}). Modules are assembled in a defined series of
steps optimised for early defect detection to avoid the use of
low-grade electronics on good quality sensors:
\begin{enumerate}
 \item gluing readout chips to hybrids and powerboards (see
   sections~\ref{subsec:assem_hyb} and~\ref{subsec:assem_pb})
 \item electrical connection of readout chips to hybrids and powerboards using an automated wire bonding process
 \item tests of electrical functionalities of hybrids and powerboards (see sections~\ref{subsec:test_hyb} and~\ref{subsec:test_pb})
 \item gluing tested hybrid(s) to sensor (see section~\ref{subsec:MA})
 \item electrical connection of hybrid(s) and sensor using an automated wire bonding process
 \item electrical tests of module (see section~\ref{subsec:test_mod})
 \item gluing tested powerboard to sensor (see section~\ref{subsec:MA})
 \item electrical tests of module (see section~\ref{subsec:test_mod})
\end{enumerate}

Components are mechanically and thermally connected using adhesives,
which achieves the low material budget required in the \mbox{ATLAS}
tracker~\cite{TDRs}. Each hybrid and module is assembled in a manual
process using custom designed precision tooling (see section~\ref{subsec:assem_hyb}) including a stencil to ensure a reliable glue coverage and thickness between components. After each gluing step, the glue thickness between components is checked by performing metrology measurements. Since the stenciling process ensured a consistent glue volume, glue
layer thicknesses outside the specified range were
found to lead to lower quality wire bonds:
\begin{itemize}
 \item thick glue layers, corresponding to low glue coverage under components, resulted in
insufficiently supported bond pads and therefore weak wire bonds
\item thin
glue layers led to glue covering wire bond pads and prevented
electrical connections between bond pads and attached wire bonds and
thereby caused electrical failures
\end{itemize}
Additionally, glue layers with insufficient height between hybrids or powerboards and sensors led to glue spreading towards the sensor guard ring area, which was found to result in early sensor breakdowns~\cite{Cole}.

\subsection{Hybrid assembly}
\label{subsec:assem_hyb}

For the construction of hybrids, ten ABC130 readout chips and one HCC130
readout chip are glued onto an X- or Y-type flex in a series of manual
steps that use precision tooling for positioning.

Prior to assembly, the involved components were tested for electrical functionality:
\begin{itemize}
 \item circuits on the flex were tested by the manufacturer
 \item ABC130 ASICs were probed on a full wafer of chips (see section~\ref{sec:ABC})
 \item most HCC130 ASICs were probed, but after a high success rate
   during initial tests (\unit[97]{\%}) and technical difficulties
   with the test setup, tests of individual ASICs were eventually
   stopped
\end{itemize}
All components were handled in a cleanroom environment using vacuum
tools to avoid contaminations prior to wirebonding (see below).

\begin{figure}
\centering
\includegraphics[width=0.9\linewidth]{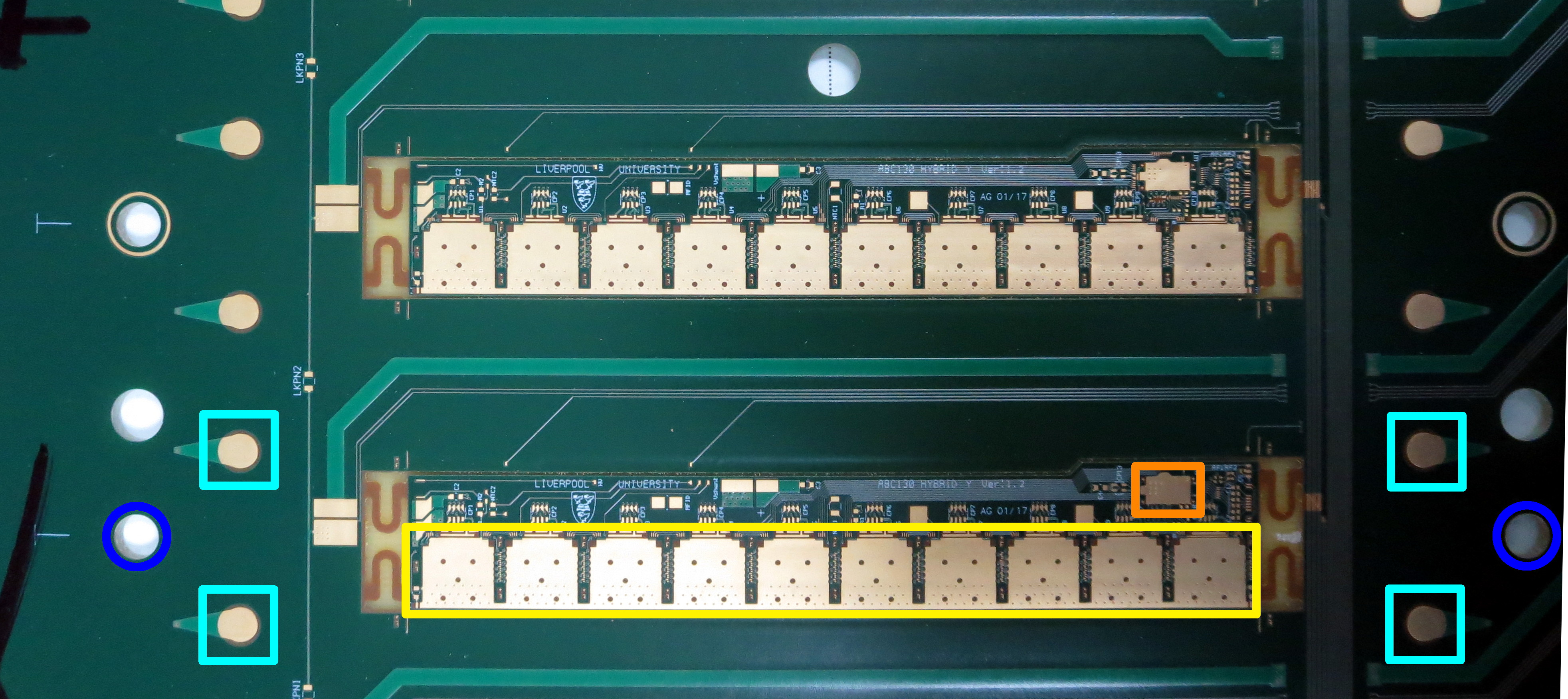}
\caption{Two hybrid flexes on a panel with pads for ABC130 chips
  (yellow) and one HCC130 (orange). Precision cut holes (blue) are used
  to position tools for population with ASICs. Landing pads (cyan) are
  used as height reference during ASIC population.}
\label{fig:hyb_panel}
\end{figure}

\begin{figure}
\centering
\includegraphics[width=0.9\linewidth]{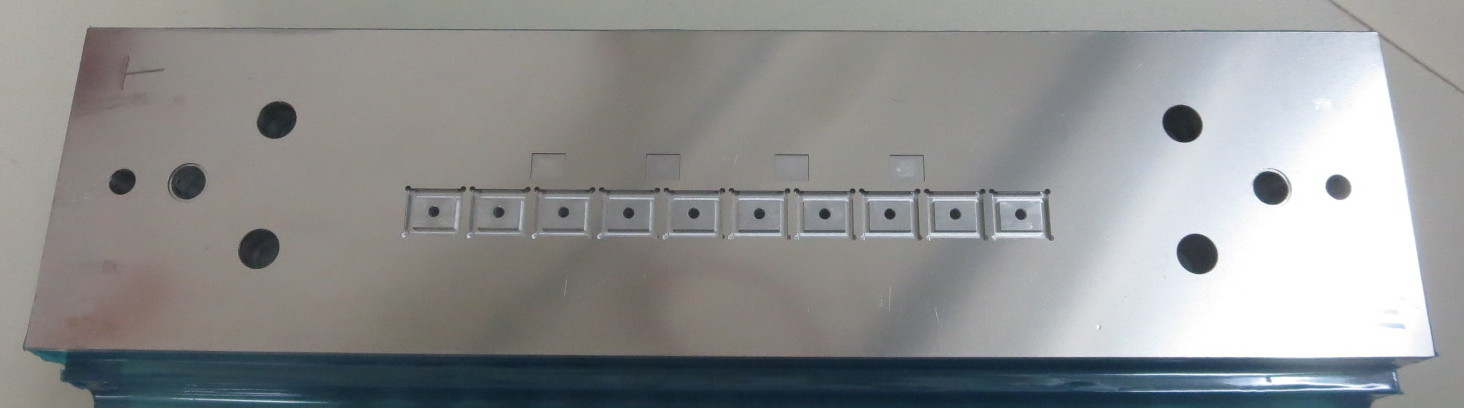}
\caption{Chip tray to align ABC130 chips in positions matching ASIC
  positions on hybrids. Alignment holes matching the alignment holes on
  panels are used to position a vacuum pick-up tool over the chips.}
\label{fig:chiptray}
\end{figure}

\begin{figure}
\centering
\includegraphics[width=0.9\linewidth]{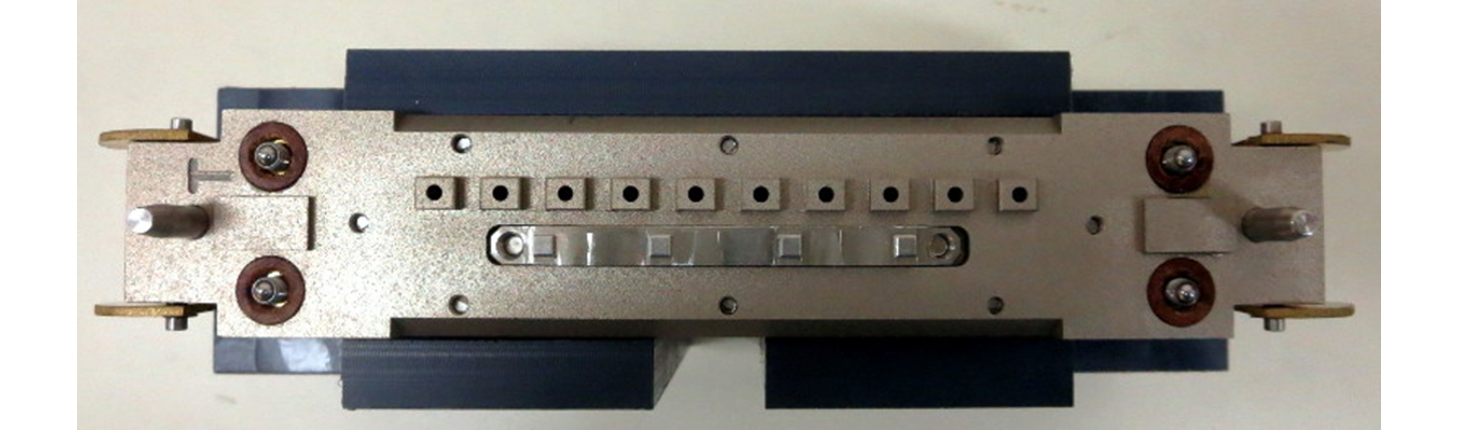}
\caption{Vacuum pick-up tool to pick up ABC130 chips to be mounted on
  a hybrid: pedestals matching ABC130 chip positions hold ASICs in
  place and are positioned on a hybrid using alignment pins matching
  alignment holes in corresponding tools. Adjustable landing feet are
  used to set the correct glue height between ASICs and
  hybrid. Landing feet are electrically insulated from the tool so
  that a contact between landing feet and pads can be checked by
  testing the corresponding resistance.}
\label{fig:pickuptool}
\end{figure}

In order to be populated with ASICs, a panel with hybrid flexes was
positioned on a vacuum chuck with vacuum holes under each hybrid
flex. Vacuum was applied in order to flatten hybrid flexes and provide
controlled glue heights between ASICs and hybrids. Positions of ASICs
on hybrids were controlled using matching positions of precision holes
and locating pins in the assembly tooling (see
figure~\ref{fig:hyb_panel}).
ASICs were first positioned in a dedicated chip tray with cutouts for each
ABC130 ASIC (see figure~\ref{fig:chiptray}), which aligned each ASIC
with respect to locating holes matching those in hybrid panels. ASICs
were picked up using a vacuum pick-up tool (see
figure~\ref{fig:pickuptool}) with individual vacuum pedestals for each
ABC130 chip and locating pins to align the tool in the chip tray.
Barrel pick-up tools were designed to be height-adjustable: each landing foot consisted of a precision metal sphere glued into a fine thread screw which could be adjusted to increase or reduce the height of the pick-up tool over a hybrid or sensor. Contact measurements of adjusted pick-up tools showed that landing feet achieved height settings with respect to ASIC pickup areas of up to \unit[10-20]{$\upmu$m}, in addition to a required ASIC pickup area flatness of $\unit[\pm5]{\upmu\text{m}}$.

In order to implement a continuous assembly and testing process for hybrids, a UV cure epoxy (Loctite 3525) was chosen after a study of several alternatives \cite{glue-paper}. It replaced the previously used silver epoxy (Tra-Duct 2902). The UV cure epoxy was dispensed on each landing pad on the hybrid using an automated glue dispenser: a combined volume of
\unit[2.0]{mg} was dispensed in a five-dot pattern matching position
indicators on each landing pad (see figure~\ref{fig:hyb_panel}) with
\unit[0.4]{mg} per glue dot. After dispensing the glue, the pick-up
tool holding ASICs was placed on top of the hybrid. 

The intended glue height between flexes and ASICs was achieved by constructing dedicated
landing feet on the pick-up tool to a height that, when placed on the landing pads of the hybrid panel, would ensure the required gap between ASICs and hybrid surface. The target glue thickness for this gluing step was \unit[80 $\pm$ 40]{$\upmu$m} at the beginning of the project, but was later increased to \unit[120 $\pm$ 40]{$\upmu$m}, which was found to produce more reliable results.

A brass weight was placed on top of
the pick-up tool to ensure a good contact of the landing feet on the
panel. Afterwards, UV LEDs (Edison Opto Federal 3535 UV Series~\cite{UVLEDs}) were placed next to the hybrid to shine
UV light into the gap between ASICs and hybrid for \unit[10]{min},
which was found to be sufficient to fully cure the glue underneath each
ASIC.

For historic reasons (the HCC chip equivalent from the ABCN-25 chip set was not mounted on hybrids~\cite{stave}, but on module test frames), ASIC pick-up tools for the ABC130 chip set did not include a vacuum pick-up area for the HCC130 chip. Therefore, HCC130 ASICs
were glued onto hybrid flexes without dedicated tooling and placed by
hand. For this gluing step, a silver-filled epoxy glue (Tra-Duct 2902)
was used, as its higher viscosity facilitated the manual assembly
process.

\begin{figure}
\centering
\includegraphics[width=0.8\linewidth]{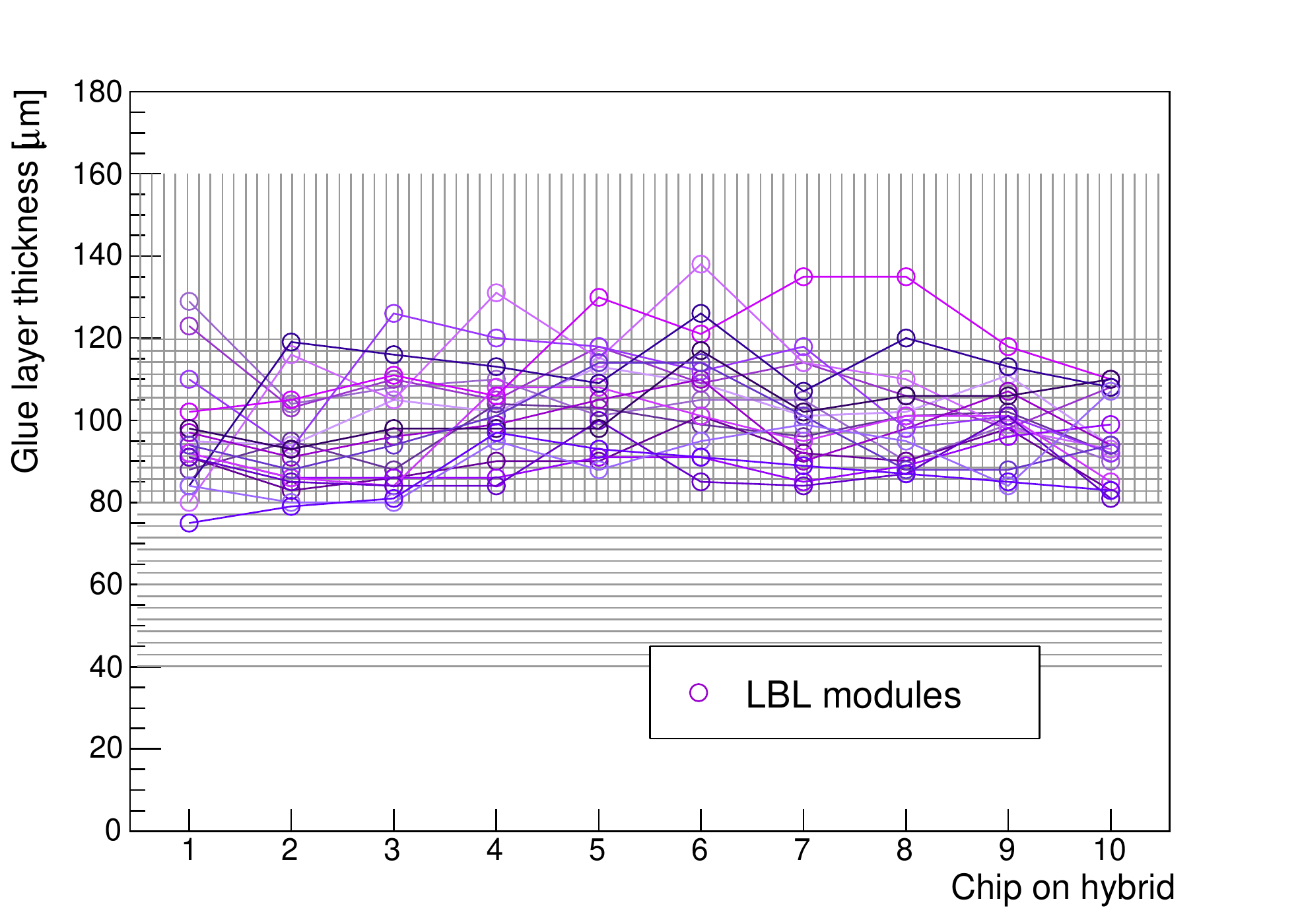}
\caption{Thicknesses of glue layers between ASICs and hybrid flexes
  measured on 16 hybrids. Over the course of the project, the target
  glue thickness was increased from {$\unit[80\pm40]{\upmu\text{m}}$}
  (indicated by horizontal stripes) to
  {$\unit[120\pm40]{\upmu\text{m}}$} (indicated by vertical stripes).}
\label{fig:hyb_thick}
\end{figure}

After each hybrid assembly, the height of each ASIC was measured to
track the achieved glue height and reliability of the gluing
process. Figure~\ref{fig:hyb_thick} shows height measurement results
for a range of hybrids: all hybrids were found to have glue thicknesses within the specified allowed range, well within the allowed uncertainties.

\begin{figure}
\centering
\includegraphics[width=0.6\linewidth]{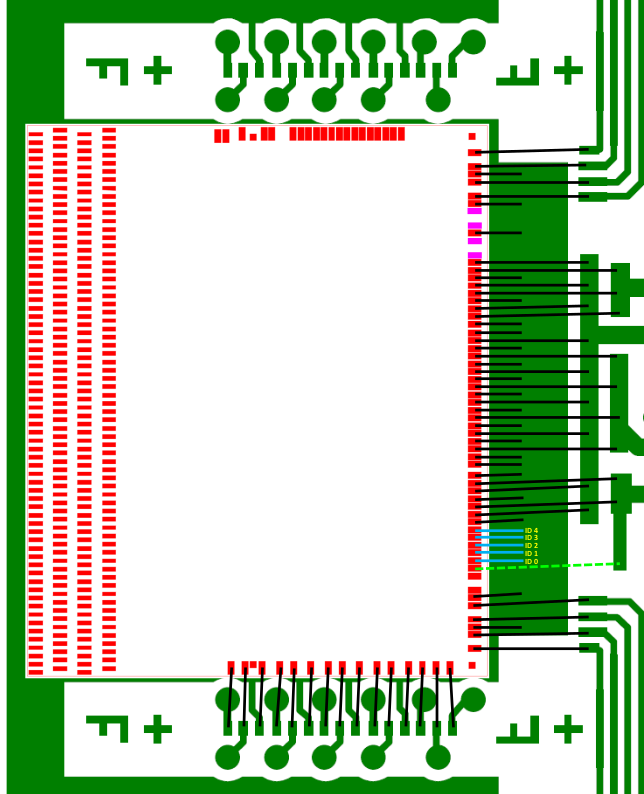}
\caption{Wire bonding scheme for electrical connections between ABC130
  ASICs and hybrid (back-end bonds). Fiducials (F- and +-shaped)
  around the ASIC corners are used for the automated alignment of a
  pre-programmed wire bonding routine to an ASIC on a hybrid. After
  attaching hybrids to a sensor, ASICs are connected to sensor strips
  using four rows of bond pads (left ASIC side).}
\label{fig:bond_ABC}
\end{figure}

After populating a hybrid with ASICs, each ASIC is electrically
connected to the hybrid using wire bonding: a \unit[25]{$\upmu$m}
aluminium wire (with \unit[1]{\%} silicon content) is fed through a
bond wedge, pressed down on the ASIC pad with a force $\mathcal{O}(\unit[10]{\text{cN}})$ while simultaneously applying ultrasonic vibrations and thereby welding the wire to the metal pad underneath.
Using the same process,
the bond wire is afterwards attached to an electroless nickel immersion gold (ENIG) plated pad on the
hybrid side and cut off. Figure~\ref{fig:bond_ABC} shows the wire
bonding scheme for wire bonds between ASIC and hybrid (back-end
bonds).
Wire bonds are placed using a program that contains position
information, loop shapes and bonding parameters for all wire bonds on
the hybrid. Prior to using the program on a hybrid, it is aligned to
the positions on hybrid and chips using fiducials (see
figure~\ref{fig:bond_ABC}).

\begin{figure}
\centering
\includegraphics[width=0.6\linewidth]{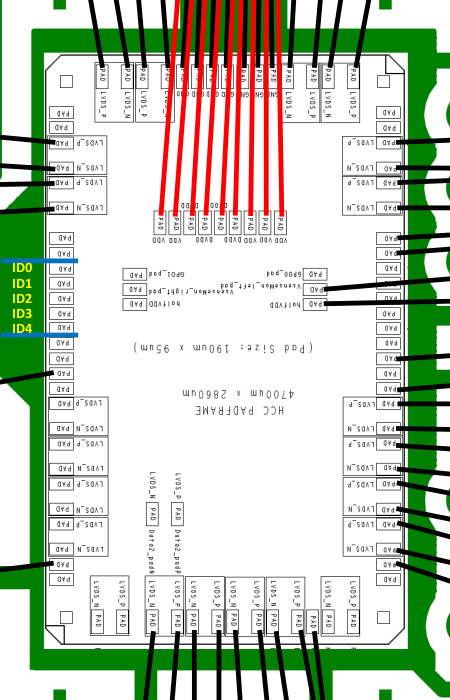}
\caption{Wire bonding scheme for an HCC130 chip on an X-type hybrid
  (mirrorred for Y-type hybrids). The address field supports five wire
  bonds in order to assign ASIC ID numbers from 0 to 31.}
\label{fig:HCC_ID}
\end{figure}

In order to read out ten identical ABC130 chips per HCC130 and two HCC130
chips on a short strip module, each ASIC is assigned an individual ID
number using bond pads on dedicated address
fields. Figure~\ref{fig:HCC_ID} shows the wire bonding scheme for an
HCC130 ASIC with its address field.
In order to assign an ID number to a chip, bond pads on the address
field are connected to a hybrid ground pad using wire bonds. Address
fields consist of several numbered fields (e.g. ID0 to ID4 on an HCC),
where each field corresponds to a binary digit (i.e. $2^0$ to $2^4$ on an
HCC). If a bond wire is attached to an address field bond pad, the
corresponding number is set to 0. The five ID pads on an HCC130 therefore
allow to assign ID numbers from 0 to 31. Ten ABC130 chips on the same
hybrid are assigned IDs 16 to 25 sequentially, while HCC130 ID numbers
can be assigned randomly, as long as HCC130 IDs on groups of 13 or 14
modules, i.e. up to 28 HCC130 chips, read out together are unique to each HCC130.

The wire bonding process is performed with the hybrid being located in
the panel where it was manufactured. During wire bonding, the hybrids
are held in place and flattened by applying vacuum to the hybrid
backplane.

\subsection{Electrical tests of hybrids and modules}
\label{subsec:test_elec}


\subsubsection{Physical DAQ system}

ABC130 objects are controlled and read out using ITSDAQ. The ITSDAQ system comprises a PC running a software
component and FPGA-based hardware to handle the digital logic
interfaces and time-critical functions. Commercial hardware and
standards-based protocols are used wherever possible; this reduces
cost, and also complexity in some cases. The connection from the PC to
the FPGA is via standard Ethernet. Custom interface boards are
manufactured to match the various ASIC and module connector
configurations, and these are plugged into the FPGA-board via
(standards based) connectors provided; most commonly an FPGA
Mezzanine Connecter (FMC) and Digilent's VHDCI based standard: VMOD.

For the FPGA hardware, commercial educator-focused ``development''
boards have been selected as these are relatively low cost, widely
available and have a product lifetime of many years. The Digilent
Atlys~\cite{DigilentAtlys} and Digilent Nexys Video~\cite{DigilentNV}
have been used widely.
The custom electronics needed to connect the development boards to
ABC130-based objects take the form of ``Interface-Boards''.  By using a
range of Interface-Boards a common FPGA-board is adapted to the wide
range objects under test.

\begin{figure} 
\centering
\includegraphics[width=0.9\linewidth]{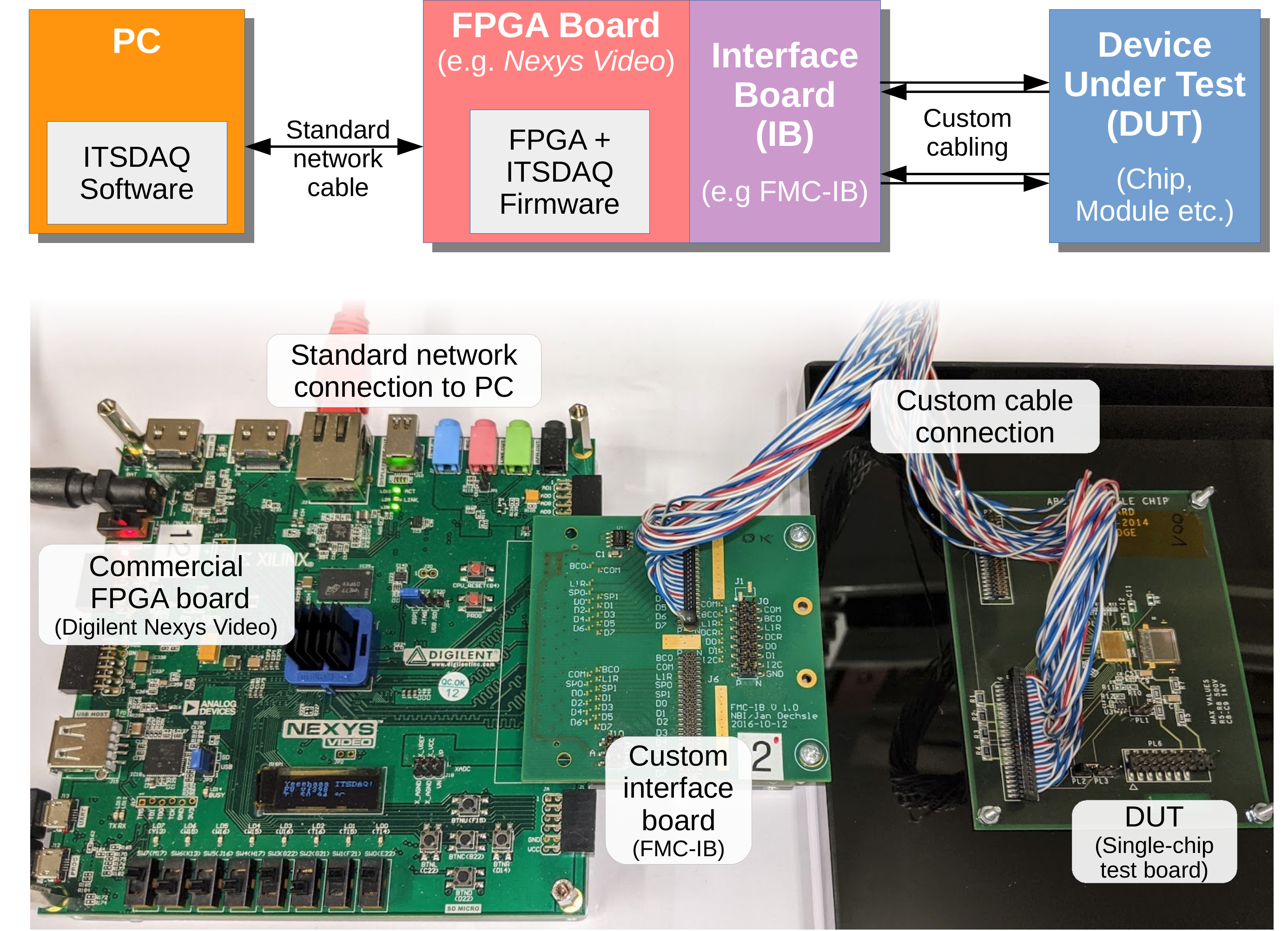}
\caption{The main components of the ITSDAQ system - as a block diagram
  and a photo of an example setup showing a Nexys Video with FMC
  Interface Board connected to a Single-chip Board.}
\label{fig:itsdaq_overview_structure_incl_hw}
\end{figure}

Figure~\ref{fig:itsdaq_overview_structure_incl_hw} shows a block
diagram of the DAQ functional components, along with an example of a
real setup (without the PC).

\subsubsection{PC/firmware interface}
\label{subsub:phy_daq_sys}

Communication with ITSDAQ hardware is via blocks of data transferred
in network packets. Initially raw Ethernet was used, but this has been
superceded by the UDP/IP protocol as it allows easier management of
the interface.
It has been found to be sufficiently reliable on point-to-point links.
The same packet format is used in both directions.

ITSDAQ network packets contain one or more opcode-blocks (``opcodes'')
formatted using a custom protocol. Table~\ref{tab:itsdaq_net_pkt_fmt}
shows the opcode wrapper protocol that forms the payload portion of
UDP packets. In the raw-Ethernet case, the first field (magic number)
is used as the standard Ethernet Type value (effectively defining a
new type of packet) and the rest is the payload. A sequence number is
provided to track packet loss. The Opcode sub-system is detailed in section~\ref{subsub:opcodes}.

\begin{table}[htbp]
\begin{tabular}{l|l|l}
\textbf{Field}  & \textbf{Size}  & \textbf{Description}                         \\
\hline
Magic number    & \unit[16]{bit} & 0x876n. In raw-Ethernet mode                 \\
                &                & ~this becomes the Type field                 \\
Sequence number & \unit[16]{bit} & Defined at source. Software can use anything,\\
                &                & ~firmware uses a counter                     \\
Packet length   & \unit[16]{bit} & Length in bytes of entire packet,            \\
                &                & ~including trailer (CRC)                     \\
Opcode count    & \unit[16]{bit} & Number of opcodes in the packet              \\
Opcode 0        &                &                                              \\
 ...            &                &                                              \\
Opcode n        &                &                                              \\
 ...            &                &                                              \\
Trailer         & \unit[16]{bit} & CRC                                          \\
\end{tabular}
\caption{ITSDAQ network packet format}
\label{tab:itsdaq_net_pkt_fmt}
\end{table}

Debugging communication (and operation) can be aided using Wireshark
software, especially if using the ITSDAQ protocol
dissector provided with ITSDAQ software.

\subsubsection{Firmware structure}
\label{subsec:firmware_struct}

\begin{figure} 
\centering
\includegraphics[width=0.9\linewidth]{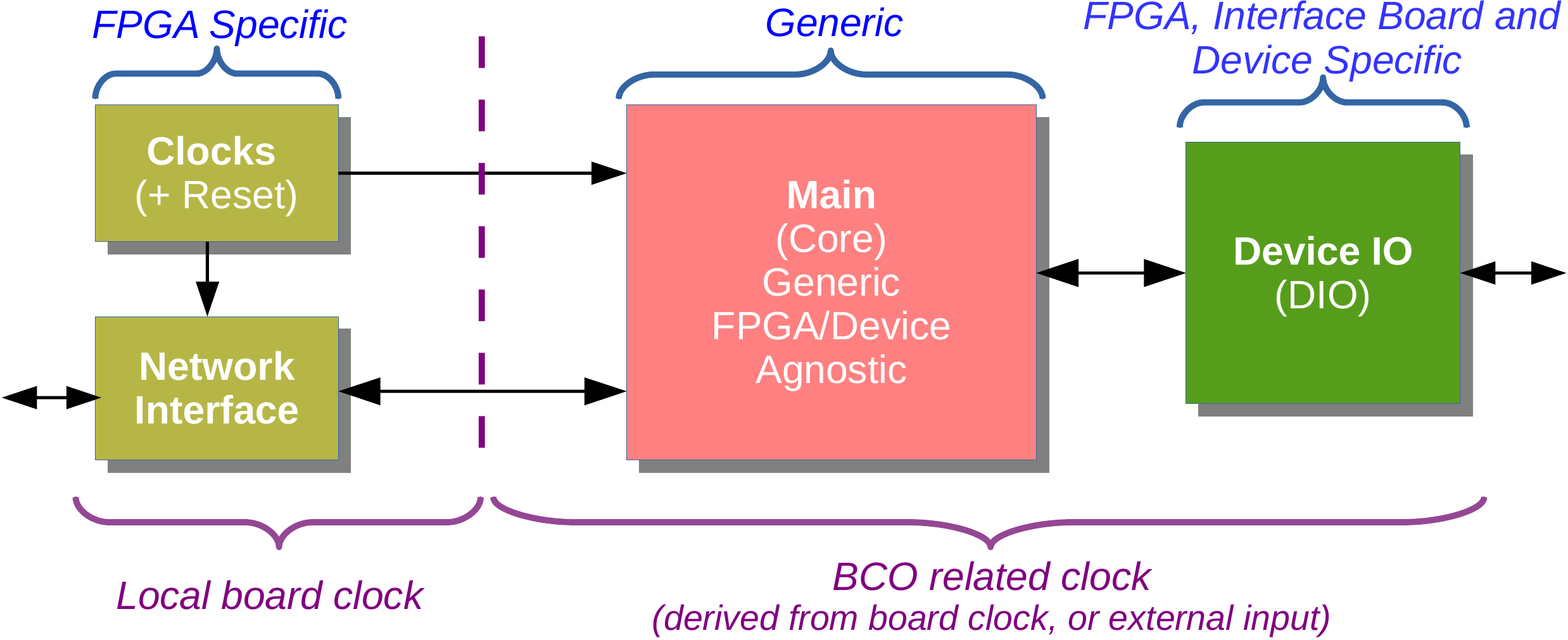}
\caption{Overview of the ITSDAQ firmware structure showing generic and
  device specific parts, and clock domains.}
\label{fig:itsdaq_firmware_overview_struct}
\end{figure}

The firmware needs to be compatible with multiple FPGAs, networking
chip sets, board-clock frequencies and devices under test. The firmware
is structured to cope with variations in board layout, FPGA family and
the devices under test. To aid this, the firmware is split into 3
distinct parts (see also figure
\ref{fig:itsdaq_firmware_overview_struct}):

\begin{itemize}
\item \textbf{Network + Clocks:} FPGA board specific; Interfaces to whatever
   network chip set is supplied on the board, and generates \unit[40]{MHz} and related clocks from the local oscillator.
\item \textbf{Main/Core:} Generic; FPGA and device agnostic, the same code is used for
all builds (but does have compile time options). 
\item \textbf{DIO:} specific to FPGA, Interface Board and device under test (DUT); Handles physical connections (pinout of
interface board, physical IO types – LVDS, CMOS, pullups etc. and FPGA
primitives – ISERDES, IODELAY2 etc.)
\end{itemize}

The core firmware provides many functions, including: control signal
encode, front-end data-format decoder, histogrammer, sequencer,
trigger generation (oscillator, random, structured bursts), I2C and
other slow interface controllers.

Firmware configuration and control signals are distributed as needed
around the FPGA using the ``Opcode Bus'' (see figure~\ref{fig:itsdaq_firmware_overview_struct}). Some of these are
responsible for sending serial streams to the DUT. Data returned from
the DUT is often multiplexed as a pair of streams sharing onto a
single line, implying 2 logical streams per physical link. Each stream
is allocated a dedicated ``Readout Unit'', which has a packet detector,
decoder, histogrammer and FIFO. A large multiplexor funnels all the
data received into a single connection to the network interface block.

Note that \unit[640]{Mb} deserialisers are used in the firmware regardless of
data rate. This allows for both simpler coarse delay setting, and
non-clock rate dependent decoding of multiple software selectable
rates.

\begin{figure} 
\centering
\includegraphics[width=0.9\linewidth]{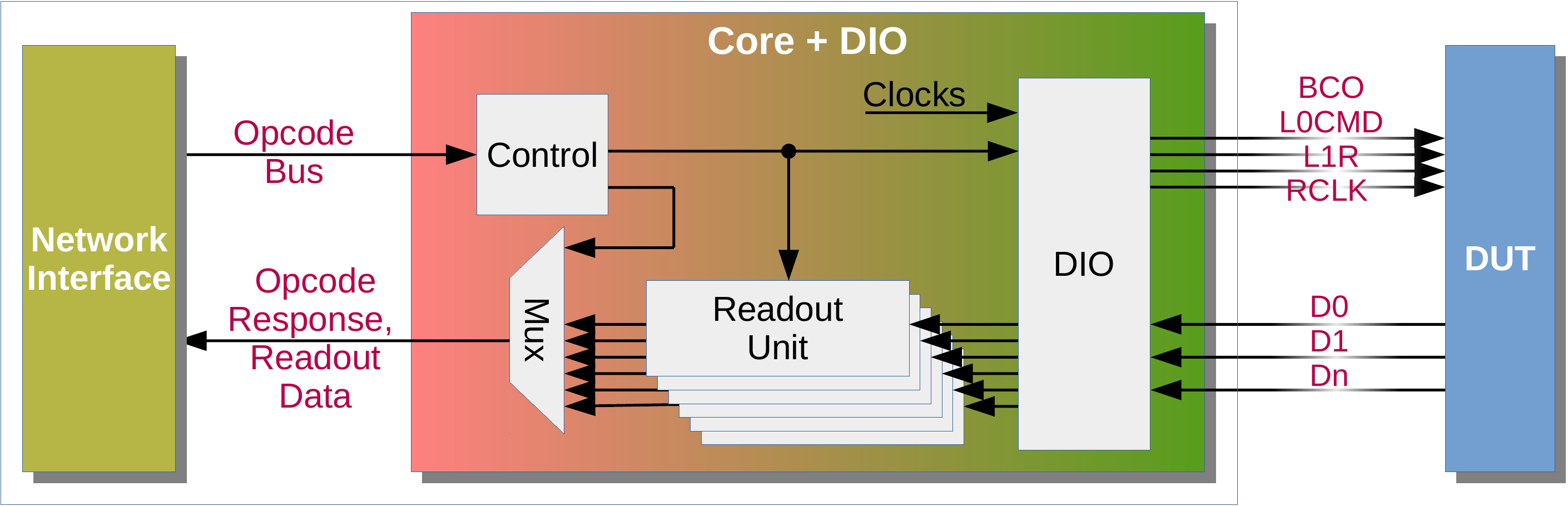}
\caption{Control and readout firmware structure showing multiple
  readout units.}
\label{fig:itsdaq_firmware_control_readout_struct}
\end{figure}

\subsubsection{Opcode sub-system}
\label{subsub:opcodes}

Opcodes are the means of communicating directly with the various
functional blocks inside the firmware (and the firmware with the
software). They allow transfers of blocks of data to firmware
``handlers''.  All handlers are connected to a common ``opcode bus''
and pull blocks of data addressed to them.  All opcode handlers will
send a response when addressed - any data that may have been
requested, or just an acknowledge to signal operation complete.

Examples of opcode handlers are: register block (allows writing to a
traditional register space), status block (returns a block of status
words), two-wire (interface to various slow control protocols) and
raw-Signal (send payload contents as a serial stream).

An ``opcode'' consists of an opcode-ID, a sequence-number and
a payload-size field, along with an (optional) data payload - see
table~\ref{tab:itsdaq_opcode_fmt}. 

\begin{table}[htbp]
\begin{tabular}{l|l|l}
\textbf{Field}  & \textbf{Size}        & \textbf{Description}                      \\
\hline
OpcodeID        & \unit[16]{bit}       & Specifies which opcode-handler is         \\
                &                      & ~being addressed                          \\
OC-Sequence No. & \unit[16]{bit}       & Generated at source                       \\
                &                      & ~Replies have same as initiating opcode   \\
Payload Size    & \unit[16]{bit}       & Payload-data length in bytes (0 is valid) \\
Payload Data    & 0-725x\unit[16]{bit} & Composition format Unique to each         \\
                &                      & opcode-type (opcode-id)                   \\
\end{tabular}
\caption{ITSDAQ opcode format}
\label{tab:itsdaq_opcode_fmt}
\end{table}

\subsubsection{Software}
\label{sec:software}

The software part of ITSDAQ is primarily developed to run on PCs
running Linux (for example CentOS 7). A Microsoft Windows version was
also been maintained for most of the period in question.  It relies on
ROOT~\cite{ROOT} for histogramming and fitting for
analysis and for the graphical user interface.

The software is used to collect data from the ASICs in various
conditions, which usually involves scanning over particular settings
of the ASIC registers and recording the data response.

A basic test thus involves:

\begin{enumerate}
 \item Load full configuration to ASICs
 \item Set parameter under test
 \item Send trigger pattern
 \item Record data response
 \item Repeat from 3 until number of triggers complete
 \item Repeat from 2 until all parameter values scanned
\end{enumerate}

A wide variety of trigger patterns is available so that (for instance)
charge can be injected into the front-end with particular timing. A
non-exhaustive list includes the addition of different reset commands,
sending multiple triggers (and potentially recording data from only
one), sending register read commands and sending arbitrary patterns (see section~\ref{sec:ABCop}).

Recording the response data normally involves decoding the pattern of
hit strips and building a hit map accumulated for all trigger patterns
with the same parameter setting. Additional modes include recording
the raw bit stream sent by the ASICs, and extracting particular parts
of responses (for instance chip ID, or the address or value in a
register read).

 \begin{figure}
\centering
\includegraphics[width=1.0\linewidth]{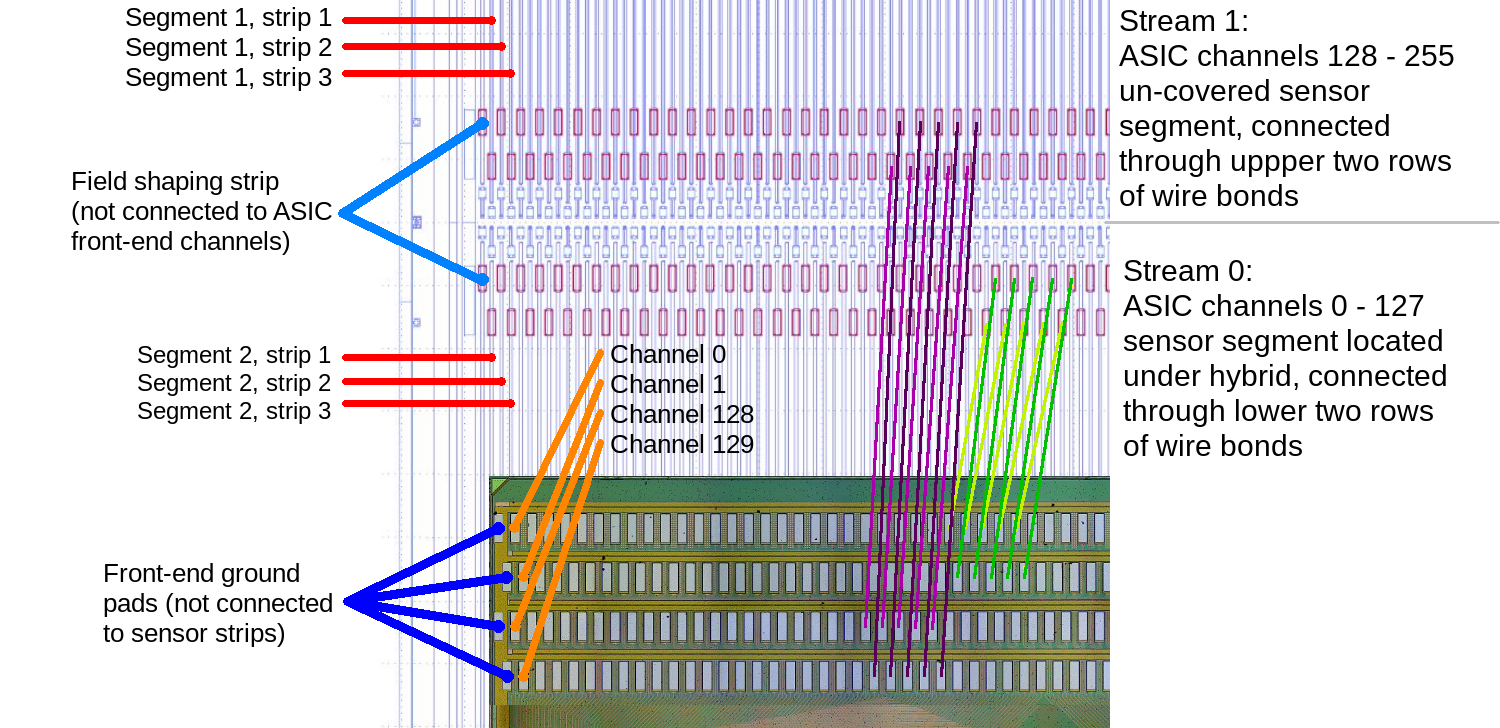}
\caption{Front-end wire bonding scheme mapping sensor strip numbers and front-end channel numbers (not to scale). The 256 ABC130 readout channels are split into two streams of 128 channels each, with each stream corresponding to strips located on one sensor segment.}
\label{fig:bond_map_frontend}
\end{figure}
Though the ABC130 ASICs have 256 channels, half of these are bonded to
strips running under the ASIC and half are bonded running away from
the ASIC (see figure~\ref{fig:bond_map_frontend}). Thus a pair of histograms is produced for occupancy plots for each of these sets of channels. Where plots show 128 channels per ASIC, an arbitrary choice between these sets has been made.

\subsubsection{Characterisation Tests} 
\label{chartests}

In order to characterise both hybrids and modules electrically, a
sequence of tests is performed. This starts with digital tests where
the response data is expected to be either all or nothing.

Following this are a series of analogue tests with a variable
response. As the hit decision is binary, analogue values are extracted
by sending and accumulating data from multiple triggers.

The current set of digital tests are as follows:

\begin{itemize}
 \item Capture HCC and ABC IDs\\
 This function tests whether communication with all ASICs is possible. If successful, the ID numbers assigned to each ABC130 and HCC130 ASIC are read out.
 \item NMask\\
 This diagnostic test changes the setting of the mask register and uses the
send mask feature to produce a deterministic pattern on the output. 
\end{itemize}

The analogue tests mostly involve using the charge self-injection function of
the ABC130 (see section~\ref{sec:ABC}). This involves sending a particular command to the ASIC, followed
by an L0 trigger. This simulates a strip hit using the discharge of an
internal capacitor via a timed pulse. The timing
of this pulse (aka strobe) within a clock period can be adjusted by passing
through some number of buffers. The pattern of strips into which charge is injected can
be changed arbitrarily, which allows different patterns of
neighbouring strips to be injected independently. In this case, an
additional loop is applied so that charge is injected into all strips when
integrated over the full scan. How many strips enabled at each step is
configurable, in addition to the number of triggers in each loop.

The current set of analogue tests are as follows:

\begin{itemize}
 \item Strobe Delay \\ 
   Before using the calibration injection for
   other tests an appropriate delay value is chosen. The correct
   setting varies between ASICs due to process variations and over
   time due to sensitivity to conditions such as temperature.  During
   a Strobe Delay scan, a charge of approximately \unit[4]{fC}
   is injected into each readout channel, which
   is subsequently read out repeatedly at a readout threshold of
   approximately \unit[2]{fC}.  The varying parameter is the delay in
   the injection strobe (between the clock edge and the pulse
   generation), over the full range of potential delays (6-bits
   representing approximately \unit[80]{ns}).
   The compression mode is set to detect the edge of the pulse, so 
   this finds a window of strobe delay units in which the injected
   charge is registered in a particular clock (see
   figure~\ref{fig:StrobeDelay}). The correct setting, for subsequent tests,
   is chosen based
   on the timing of the edges of this window.  The delay is set for
   each individual ASIC at \unit[57]{\%} of the distance from the
   rising edge. This value was selected based on a more detailed scan
   of the pulse shape and the noise and gain at different delay values.
 \begin{figure}
\centering
\includegraphics[width=0.8\linewidth]{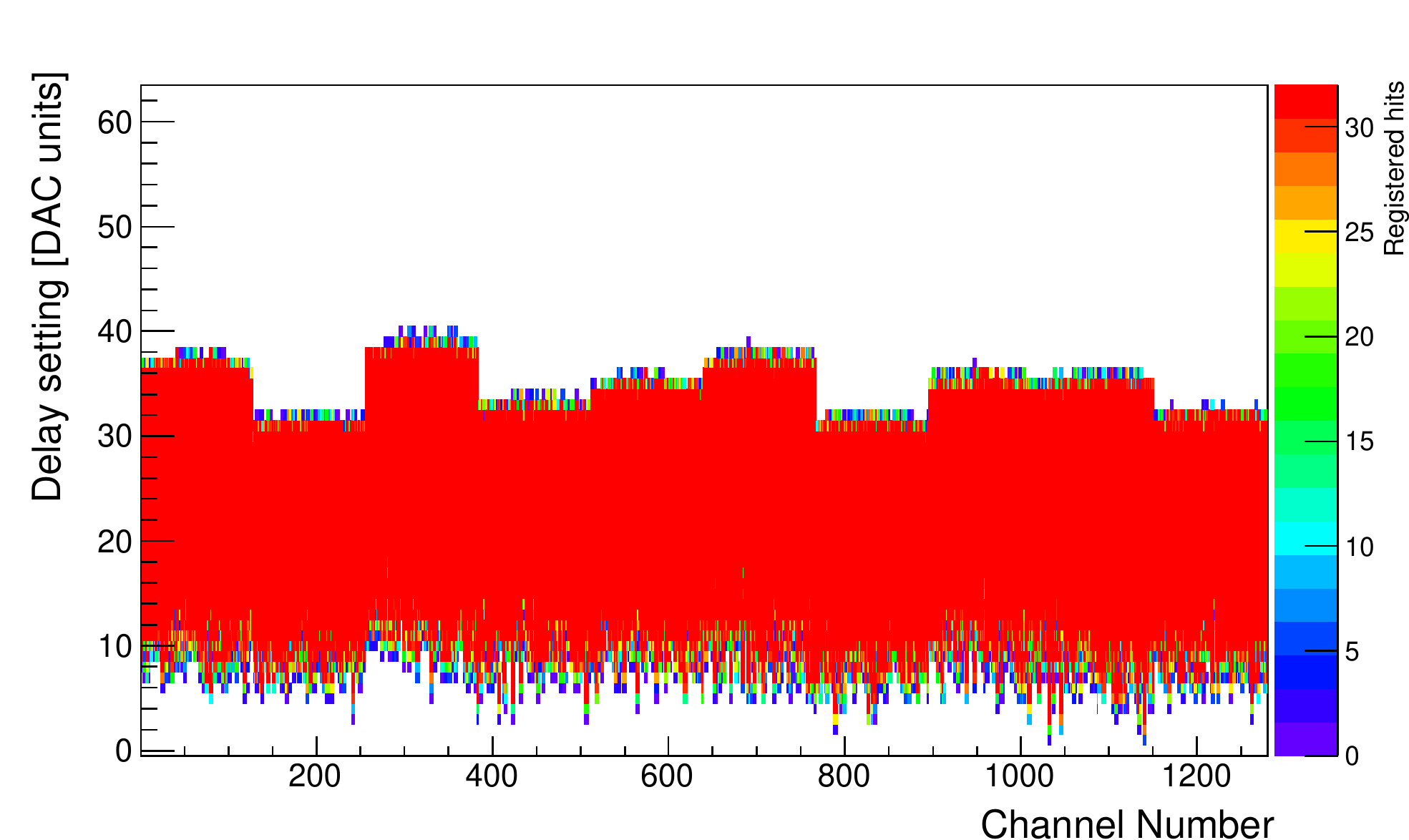}
\caption{Example of a Strobe Delay scan for a hybrid with ten readout
  chips. For a range of delay settings, signals are injected to each
  channel 32 times per setting. Increasing delay means moving
  from the falling edge (where channels are not aligned) to the rising
  edge (where channels are more sharply aligned) of the pulse. Each
  channel registers all injected signals over a range of about 26 DAC
  units. The delay for each ASIC is set at {\unit[57]{\%}} of the
  pulse width from its rising edge.}
\label{fig:StrobeDelay}
\end{figure}
\item Three Point Gain\\ 
  The response of the amplifier for each
  readout channel is measured using a sequence of threshold scans,
  where a different charge is injected for each. During the threshold
  scan, the discriminator threshold is varied. For each injected
  charge, the resulting distribution is expected to be a step function, which becomes an ``S-curve'' (a complimentary error function) due to smearing from noise effects (see
  figure~\ref{fig:TPG1}). The shape and slope of the S-curve can be
  used to determine the noise and Vt50 of each input channel (see figure~\ref{fig:TPG2}). Vt50 describes the mean amplifier response
  for an injected charge, i.e. the point of the curve where
  \unit[50]{\%} of readout triggers lead to a hit being registered. By performing
  threshold scans for different input charges (e.g. \unit[0.5]{fC},
  \unit[1.0]{fC} and \unit[1.5]{fC}), the relation between input
  charge and readout threshold can be mapped and each readout
  channel's gain be determined using a linear fit (see
  figure~\ref{fig:TPG3}). Additionally, the offset of the gain
  function of each channel is determined. Figure~\ref{fig:TPG4} shows
  the resulting noise distribution for one hybrid with ASICs. This
  gain can be used to convert the output noise (the measured width of
  the S-curve) into the input-referred noise (the derived noise at the input of
  the amplifier), which is then reported in electrons (of equivalent
  noise charge).

\begin{figure}
\centering
\begin{subfigure}{.48\textwidth}
 \centering
 \includegraphics[width=\linewidth]{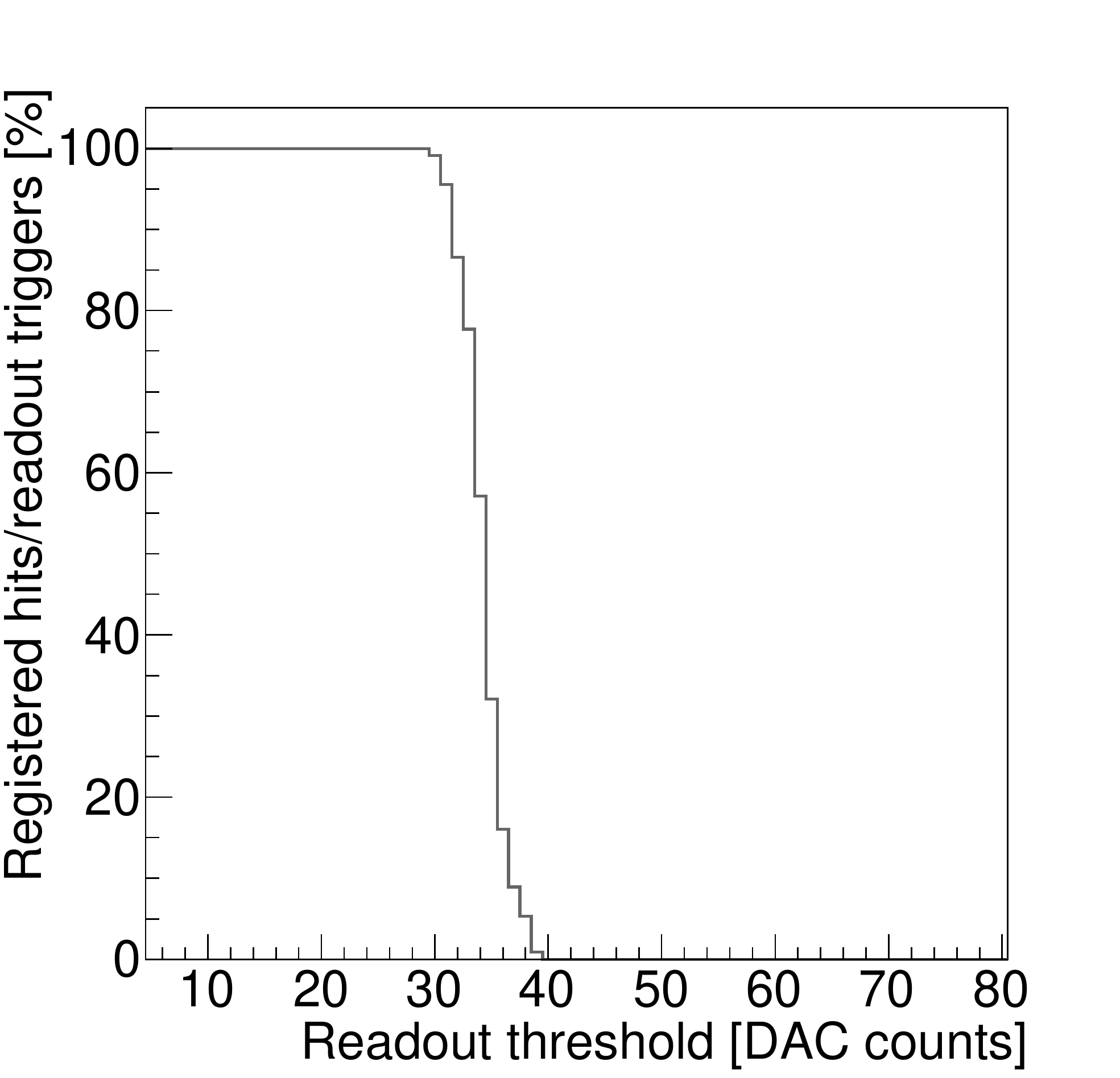}
 \caption{S-curve obtained from a threshold scan of a readout channel.}
 \label{fig:TPG1}
\end{subfigure}
\begin{subfigure}{.02\textwidth}
\hfill
\end{subfigure}
\begin{subfigure}{.48\textwidth}
 \centering
 \includegraphics[width=\linewidth]{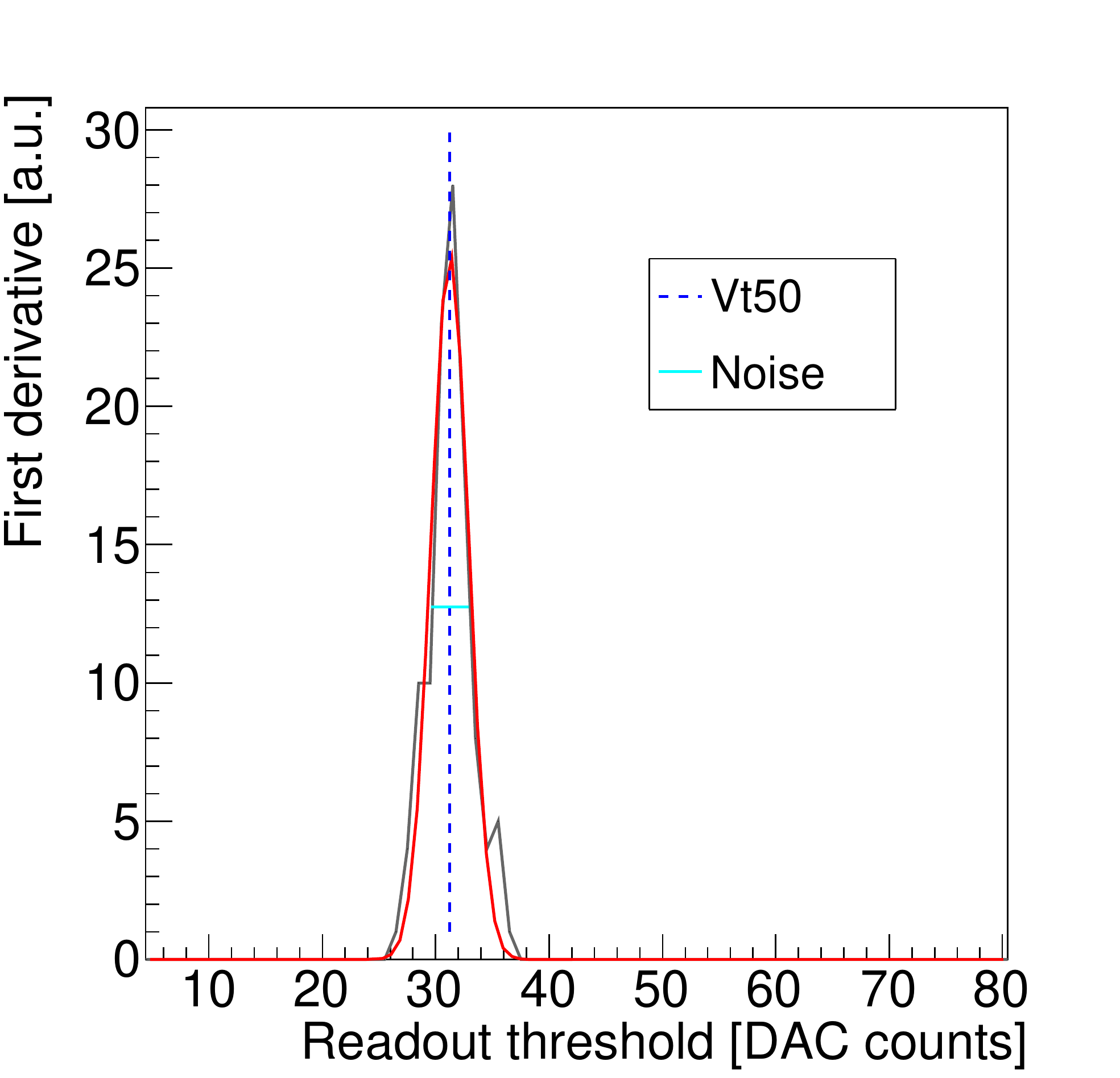}
 \caption{First derivative of an S-curve with Vt50 and noise.}
 \label{fig:TPG2}
\end{subfigure}
\begin{subfigure}{.48\textwidth}
 \centering
 \includegraphics[width=\linewidth]{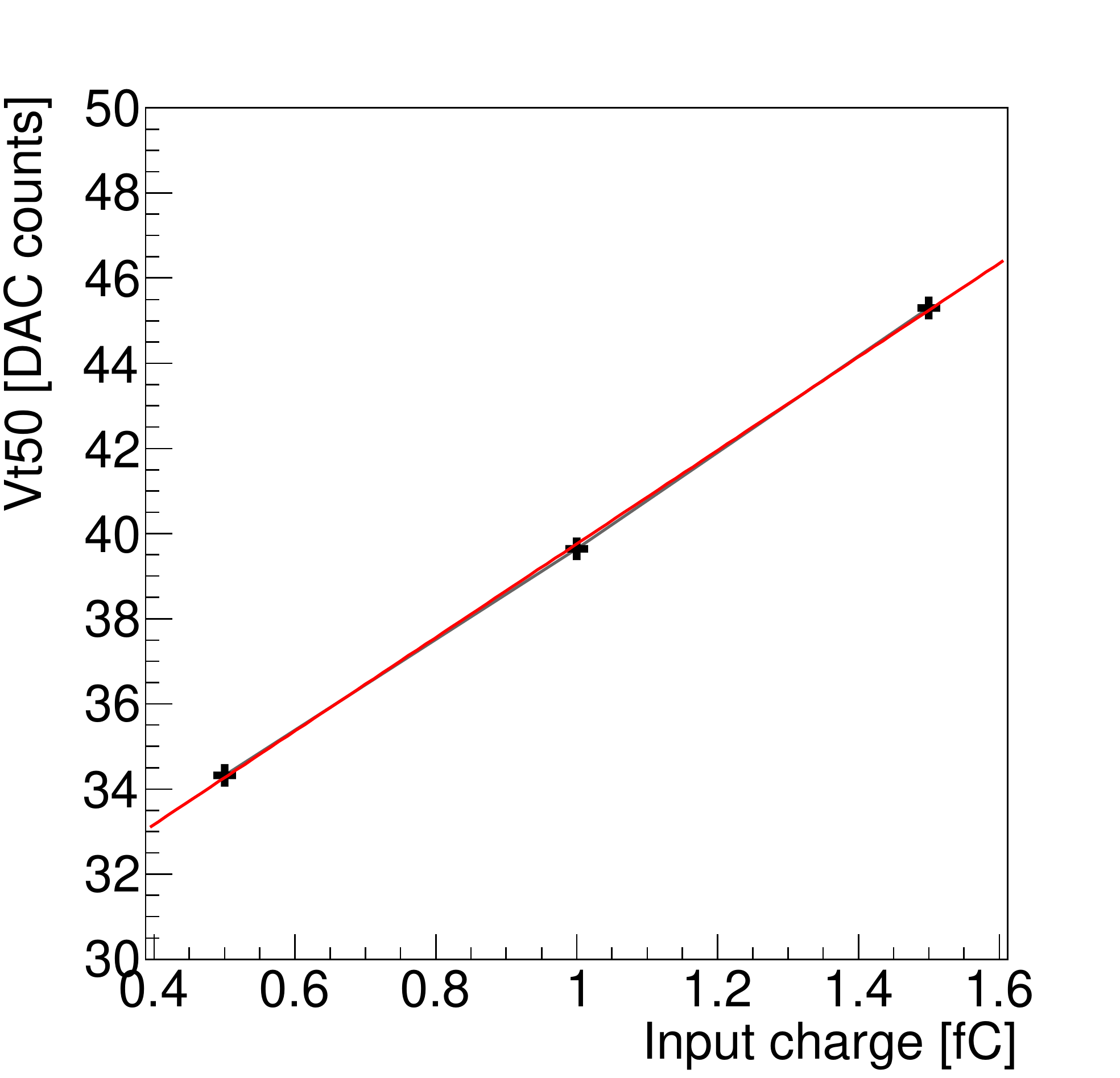}
 \caption{Gain of individual readout channel, from Vt50 measurements.}
 \label{fig:TPG3}
\end{subfigure}
\begin{subfigure}{.02\textwidth}
\hfill
\end{subfigure}
\begin{subfigure}{.48\textwidth}
 \centering
 \includegraphics[width=\linewidth]{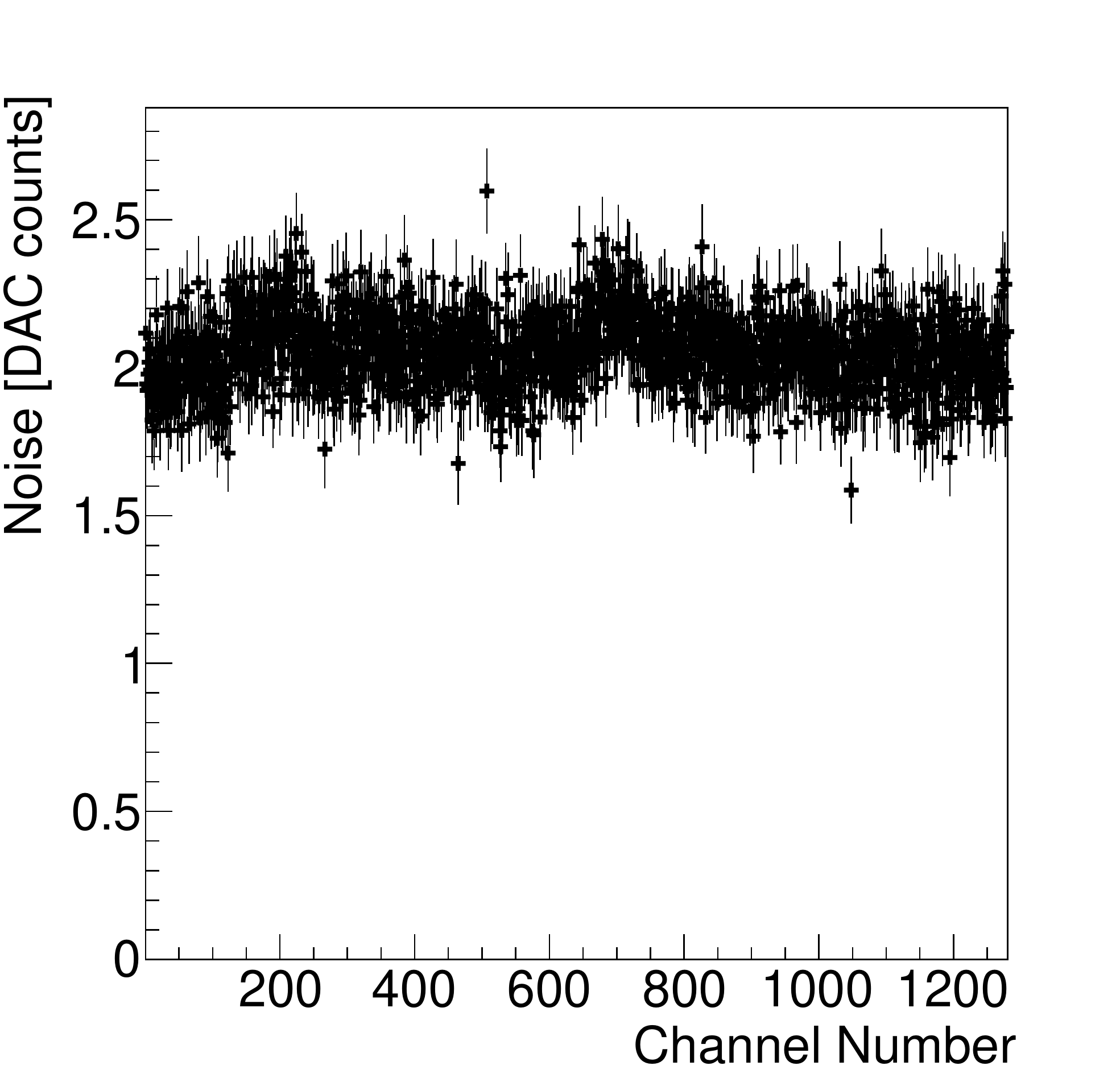}
 \caption{Noise of channels from all ASICS on one hybrid (about {\unit[400]{ENC}}).}
 \label{fig:TPG4}
\end{subfigure}
\caption{Examples from a Three Point Gain measurement of one hybrid:
  measurement of the S-curve of an individual channel, mean of
  noise and Vt50 for that channel, gain calculation from measurements
  at different input charges and noise distribution for one hybrid.}
\label{fig:TPG}
\end{figure}
It should be noted that while threshold tests and their analysis are
performed based on ASIC DAC counts (referring to bit register
settings), the corresponding threshold (measured in mV) does not
increase linearly with DAC counts over the full range of thresholds (see figure~\ref{fig:ABC130_thresh}) and is only converted into mV during
the last step of the analysis.
\item High statistics Three Point Gain\\ 
  While the standard Three
  Point Gain performed on hybrids (see section~\ref{subsec:test_hyb})
  is sufficient to identify dead channels, the uncertainties of
  parameters derived from a fit of the obtained S-curve (see
  figure~\ref{fig:TPG1}) depend on its statistics. Increasing the
  number of triggers used for the measurement of an S-curve leads to
  more reliable channel characteristics (see
  figure~\ref{fig:modtest_highstats}).
 \begin{figure}
\centering
\includegraphics[width=\linewidth]{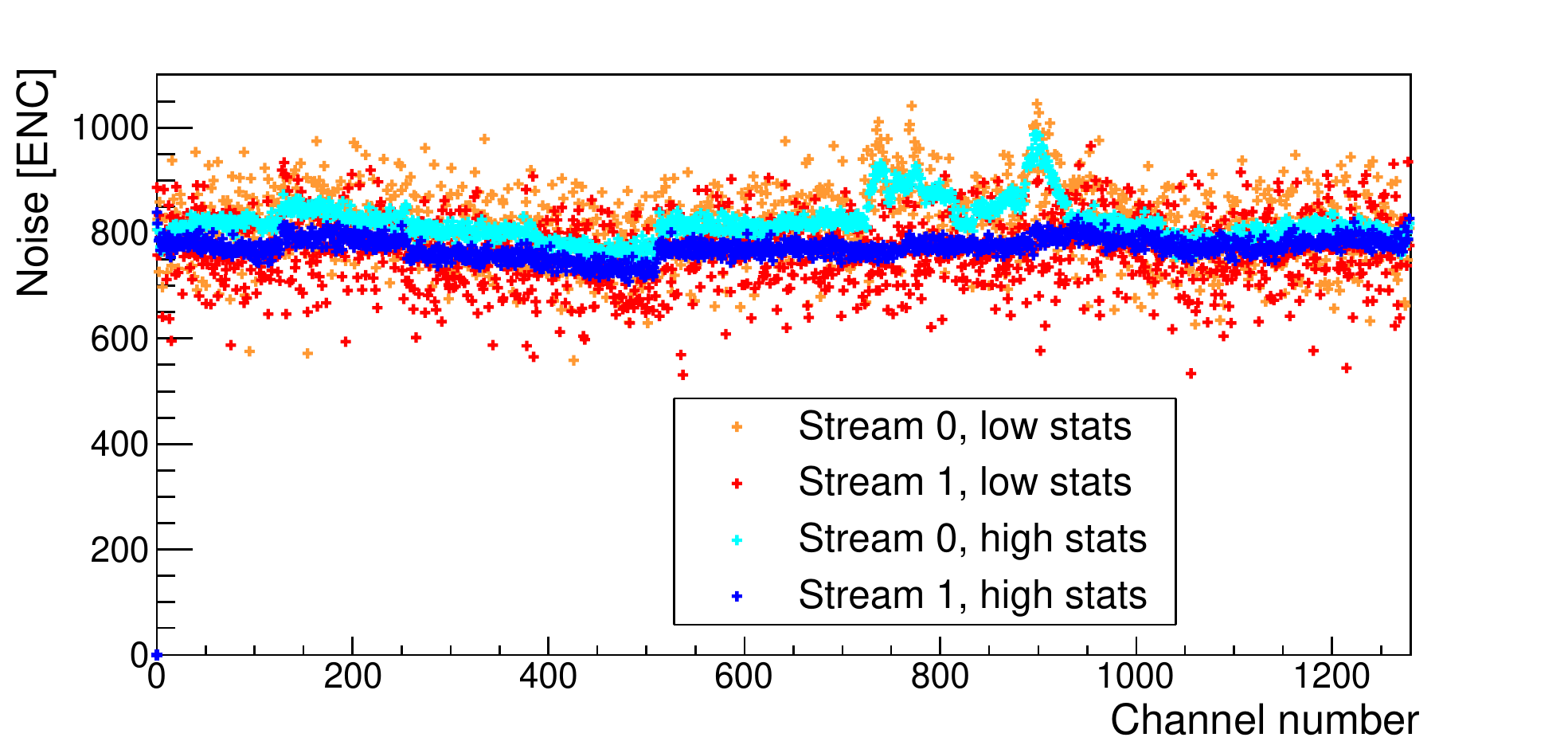}
\caption{Comparison of channel noise measured for one hybrid on a
  barrel module from Three Point Gains with high (1000 triggers per
  threshold) and low (50 triggers per threshold) statistics. Noise
  fluctuations were found to be reduced by the number of applied
  triggers, which shows increased noise in certain module regions (caused by insufficient powerboard shielding, see
  section~\ref{sec:selectModule:PBEffect}) that are hidden in overall
  noise fluctuations for a measurement with low statistics.}
\label{fig:modtest_highstats}
\end{figure}
Due to the time consumption of full threshold scans with high
statistics, modules were tested with low statistics thresholds scans
first to check their overall functionality before performing an
additional high statistics Three Point Gain.
\item Trim Range \\ 
  During the operation of a module, readout
  thresholds are not set for individual channels, but for full readout
  chips. An operating threshold is chosen to be as low as possible
  while also minimising noise occupancy ($\unit[<1]{\%}$). S-curves
  from different channels from the same module show a large spread
  over the threshold range (see figure~\ref{fig:modtest_notrim}),
  which makes the selection of an operating threshold less efficient,
  as a threshold with less than \unit[1]{\%} noise occupancy for all
  channels leads to a wide range of distances between operating
  threshold and Vt50 point.
 \begin{figure}
\centering
\includegraphics[width=\linewidth]{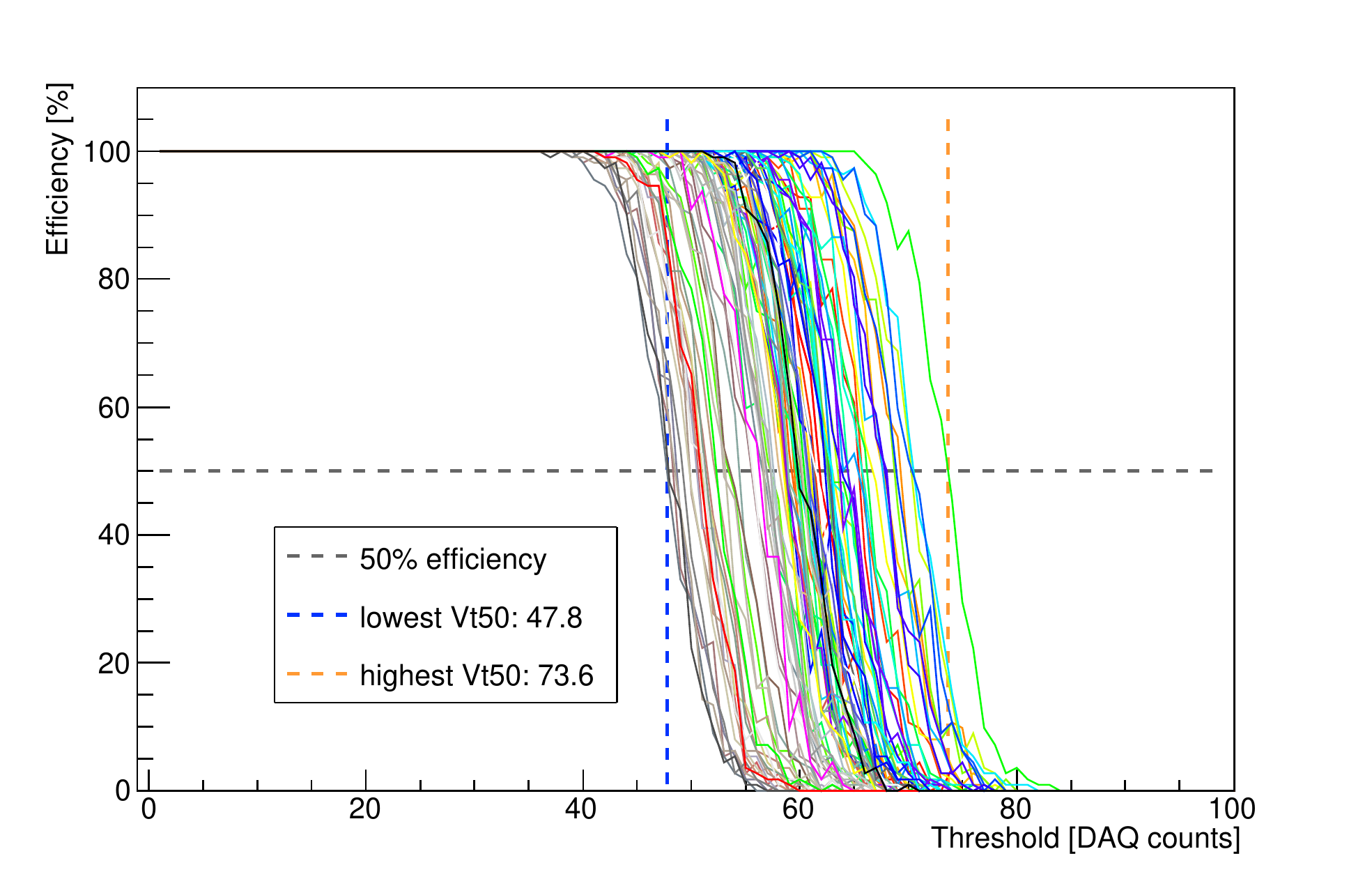}
\caption{S-curves for 100 sequential channels on an ABC130 module
  without trimming: the positions of their Vt50 points are distributed
  over a range of {\unit[26]{DAC counts}}.}
\label{fig:modtest_notrim}
\end{figure}
In order to ensure a uniform response of all module channels to which
the same readout threshold is applied, S-curve positions can be
shifted in the threshold range. While an efficiency curve can not be
moved towards lower thresholds, it can be moved towards higher
thresholds by adding an offset, which has to be
determined by channel, to the pedestal. In order to find a threshold to which a
majority of channels can be trimmed, a scan over the TrimDAC values and 
the Trim Range is performed. A chosen charge is injected into the front-ends
and the trims adjusted so that the thresholds align for a particular
chip-level threshold. The Trim Range (the scale of the trim changes) is
chosen to be allow as fine tuning as possible
while including as many channels as possible. Thus, all channels
on a chip are trimmed by adding the tuned offset value to their
threshold, leading to a uniform
distribution of Vt50 on all channels of the same chip (see
figure~\ref{fig:modtest_trim}).
 \begin{figure}
\centering
\includegraphics[width=\linewidth]{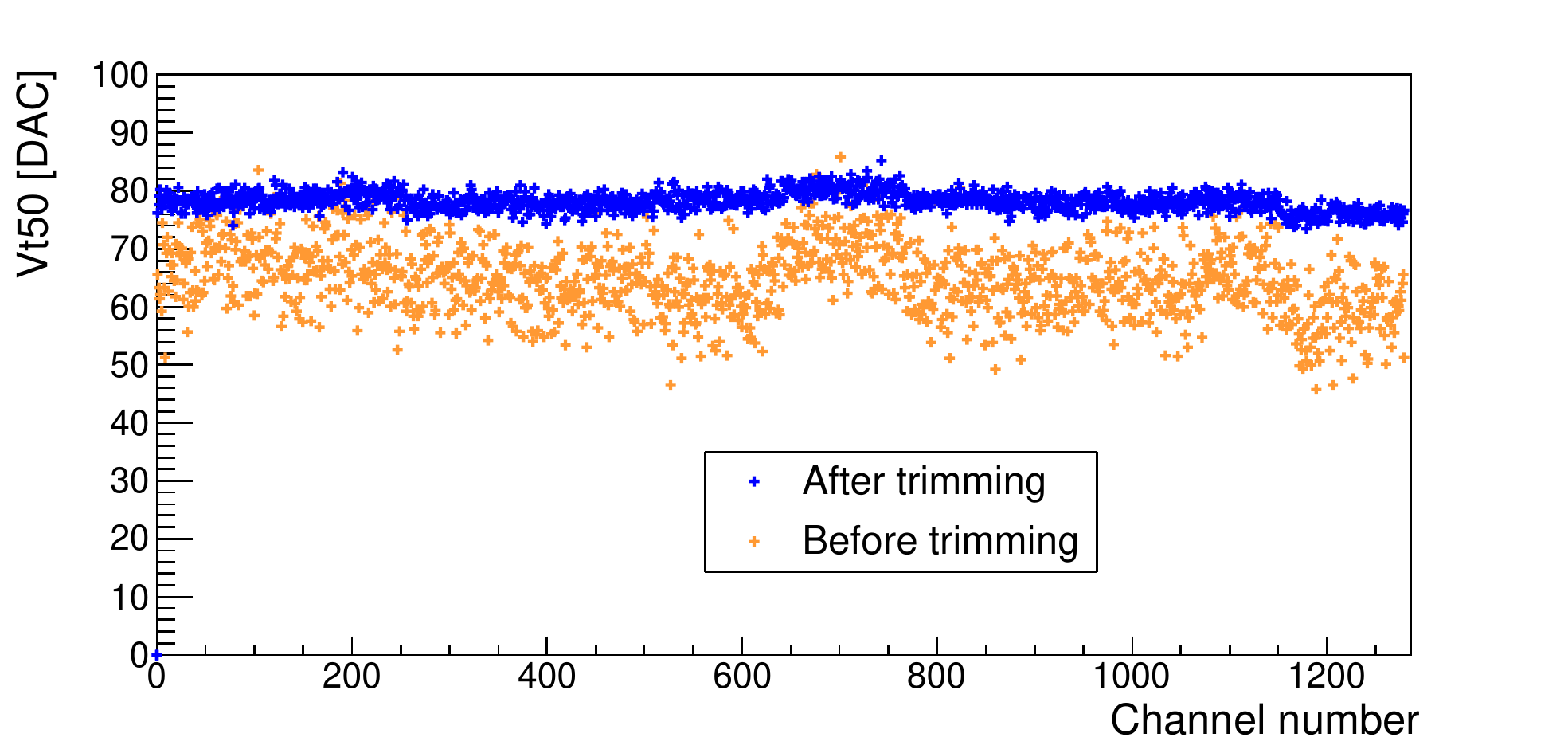}
\caption{{\unit[50]{\%}}-efficiency point (Vt50) of channel S-curves
  before and after performing a Trim Range: before trimming, Vt50s of
  channels show a large spread. After trimming, channels on the same
  readout chip show a flat distribution of Vt50 at a higher value,
  which is achieved by adding to the pedestal of S-curves with low
  Vt50.}
\label{fig:modtest_trim}
\end{figure}
\item Noise Occupancy\\ 
  The noise occupancy test records channel
  occupancy with no injected charge. This is carried out for a series
  of thresholds in order to extract the noise curve of the pedestal.
  A variety of options are provided for the timing and number of these
  triggers.

\item Response Curve \\ 
  In the response curve test, the
  correspondence between input charge and threshold is further characterised, beyond the linear
  regime of the Three Point Gain test (see
  figure~\ref{fig:ABC130_RC}), for higher input charges and corresponding
  thresholds and
  where the relationship becomes non-linear. This uses ten threshold
  scans over a range of input charges up to \unit[6]{fC}.
  The correspondence between DAC count
  and threshold voltage is approximately linear over a large range of
  these settings, an option is provided to make a correction base on
  based on a simulation of the relationship
  (see figure~\ref{fig:ABC130_thresh}), for greater accuracy.
  
 \begin{figure}
\centering
\includegraphics[width=0.7\linewidth]{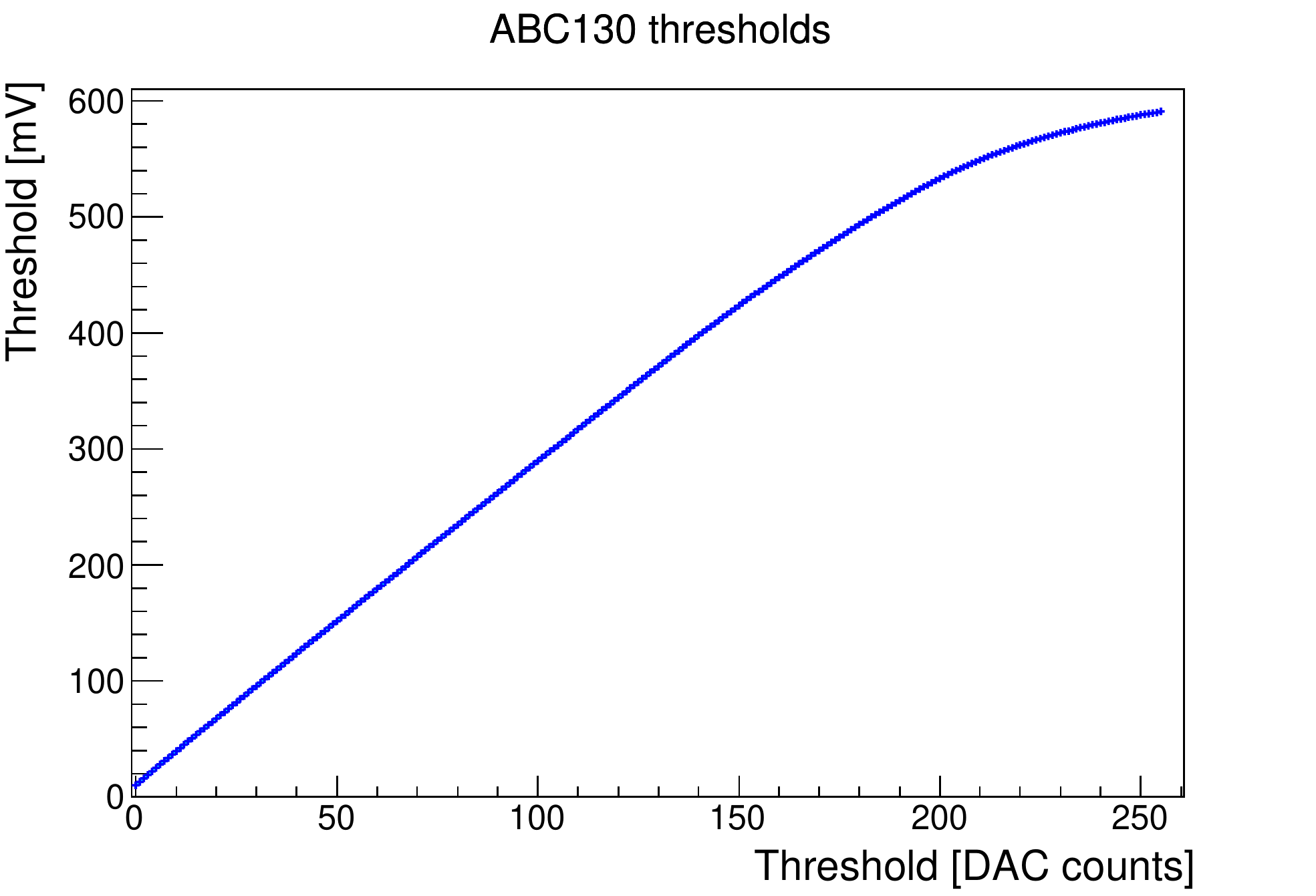}
\caption{Calibrated relation between ABC130 threshold setting and
  corresponding physical threshold: the corresponding threshold
  increase is linear for low thresholds, but becomes non-linear for
  high thresholds.}
\label{fig:ABC130_thresh}
\end{figure}
 \begin{figure}
\centering
\includegraphics[width=0.7\linewidth]{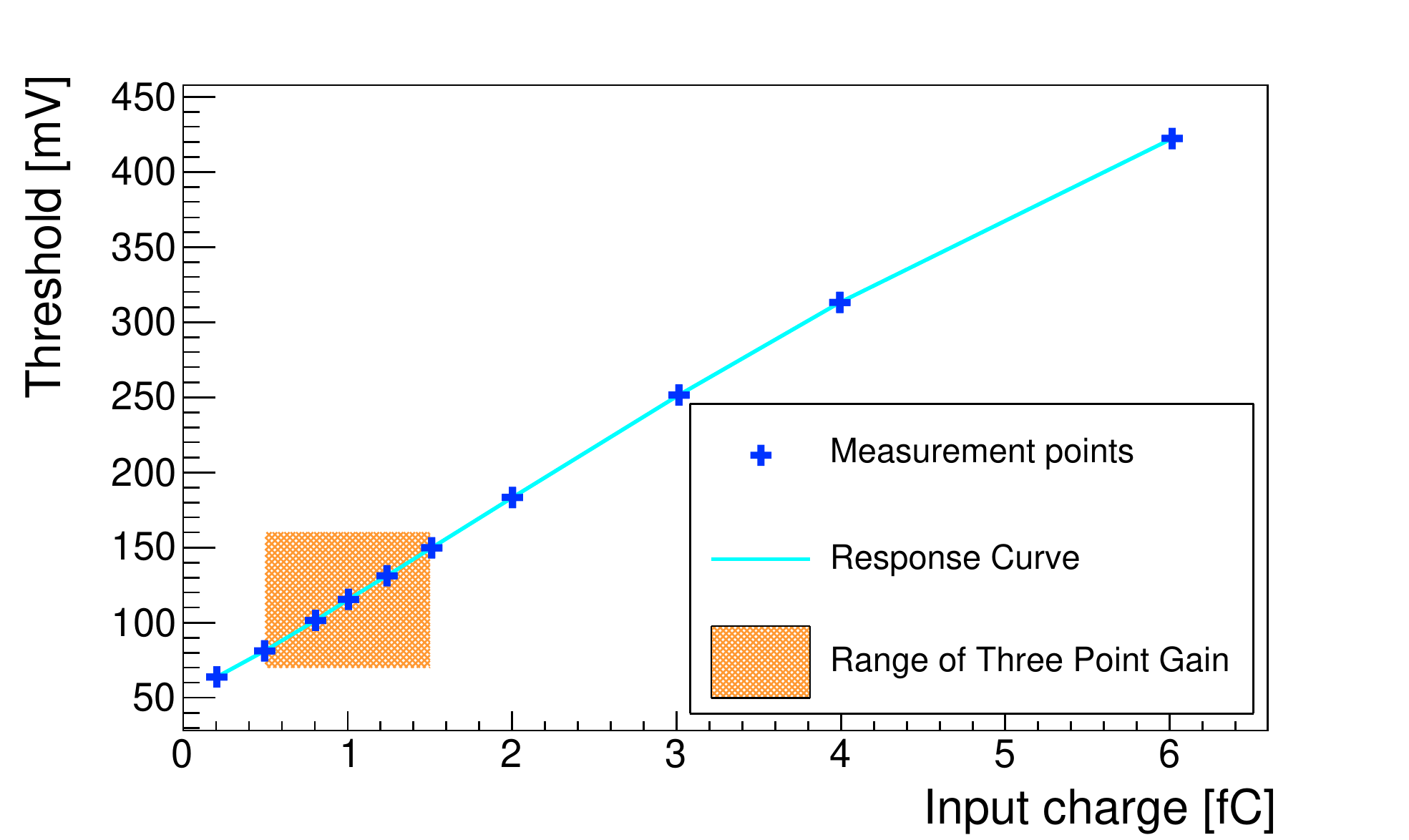}
\caption{Response Curve for an average of all readout channels on an
  ABC130 readout chip: the relation between input charge and readout
  threshold is linear for low threshold and becomes non-linear for
  high thresholds. The input charge range covered in a Three Point
  Gain scan (performed at {\unit[1]{fC}}) is highlighted in orange.}
\label{fig:ABC130_RC}
\end{figure}
Performing a Response Curve allows to relate the ABC130 threshold
setting, physical readout threshold and input charge.
\end{itemize}

\subsection{Hybrid tests}
\label{subsec:test_hyb}

Flexes with ABC130 ASICs and an HCC130 chip can be tested for electrical
functionality, which is used to identify nonfunctional hybrids prior
to the assembly of hybrids, powerboards and sensor into a module. In
order to test assembled hybrids, hybrids are electrically connected to
a hybrid panel, which serves as a test structure using wire bonds to
supply power and read out data.

Each panel provides test positions for eight hybrids, which can be
tested in parallel, provided that the HCC130 on each hybrid has been
assigned an HCC130 ID different from HCC130 IDs of the other seven hybrids.

A test sequence for hybrids includes the following steps (described in
detail in section~\ref{subsec:test_elec}):
\begin{enumerate}
 \item Capture HCC and ABC IDs
 \item Strobe Delay
 \item Three Point Gain
\end{enumerate}

A hybrid is only mounted on a sensor if it has passed all stages of
electrical testing.

\subsection{Powerboard assembly}
\label{subsec:assem_pb}

The powerboard v2 is produced in a thin FR4 based stack-up and loaded
with SMDs in a typical reflow process. The DCDC inductor due to its
shape is loaded manually, as is the shield box enclosing the DCDC
circuit. Special attention is given to fully seal the shield box with
a continuous solder seam to avoid leakage of radiated noise. Once
proper SMD loading has been verified in a visual inspection, the bare
die ASICs, the AMAC and HVmux, can be glued to the powerboard v2 and
wire bonded.

\begin{figure}
\centering
\includegraphics[width=0.9\linewidth]{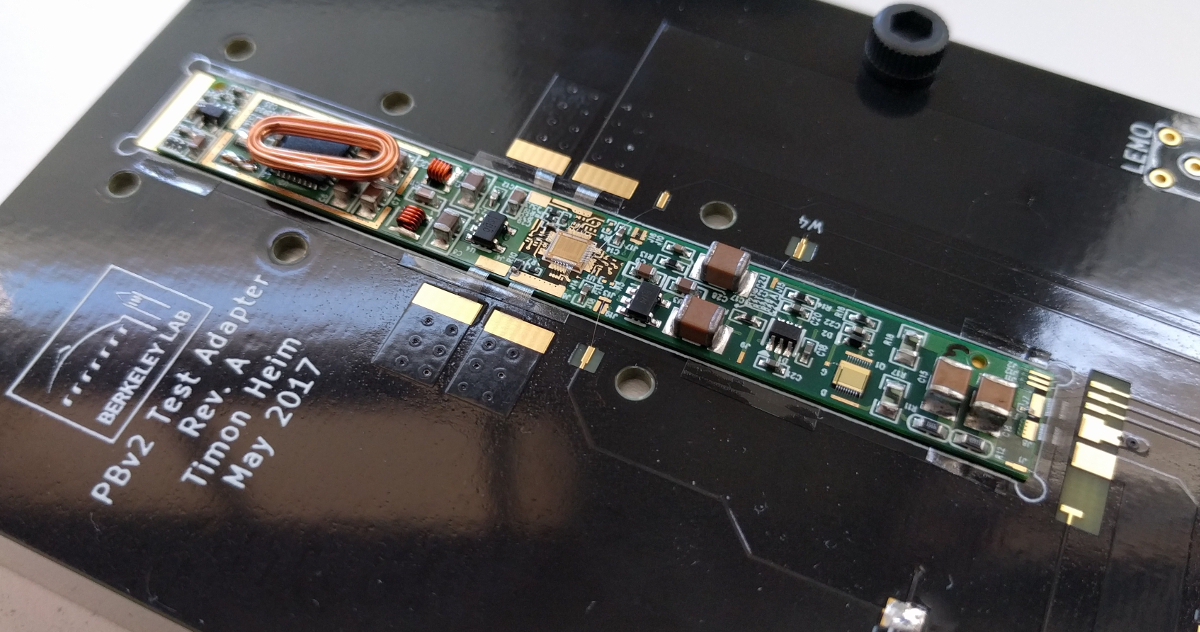}
\caption{Photograph of a powerboard v2 mounted to the test carrier PCB.}
\label{fig:pbv2_on_test_carrier}
\end{figure}

For testing the powerboard is temporarily wire bonded to a test
carrier PCB, which can be seen in
fig.~\ref{fig:pbv2_on_test_carrier}. This carrier can now be connected
to test equipment to test the functionality of the powerboard before
loading it onto a module. The test carrier also provides passive
cooling, which is necessary to run the DCDC circuit at high load
current without overheating the FEAST chip.

\subsection{Powerboard tests}
\label{subsec:test_pb}

To test the functionality of the powerboard before loading it onto the
module and to calibrate the AMAC the following tests are performed:
\begin{itemize}
\item AMAC communication test: ensures reliable communication with the AMAC
\item LV turn on/off: ensures functionality of switching \unit[1.5]{V}
  of DCDC to hybrid on/off
\item HV turn on/off: ensures functionality of switching HVmux to
  supply high voltage to a sensor
\item DCDC efficiency: measures DCDC efficiency for load currents from
  0 to \unit[4]{A}
\item $V_\text{in}$ calibration: varies input voltage, which is
  measured by the AMAC
\item $I_\text{out}$ calibration: varies output load, which is measured
  by the AMAC via an amplifier measuring the voltage drop over the
  inductor in the output $\pi$-filter
\item HV sense calibration: varies the current sources by the HV power supply, which is measured by a current-to-voltage converter in the
  AMAC
\item Temperature calibration: vary the output load to change the
  temperature of the powerboard and measure this via the thermistor
  inside the shield volume and temperature sensor within the FEAST chip
\end{itemize}

All of the test results are saved and a calibration is derived for
each monitored value. During testing of 100 powerboard v2 boards only
hard failures have been observed, typically caused by errors during
SMD reflow or dead ASICs, as chips were not tested before loading.

\subsection{Sensor tests}
\label{subsec:test_se}

\begin{figure}
\begin{subfigure}{.47\textwidth}
 \centering
 \includegraphics[width=\linewidth]{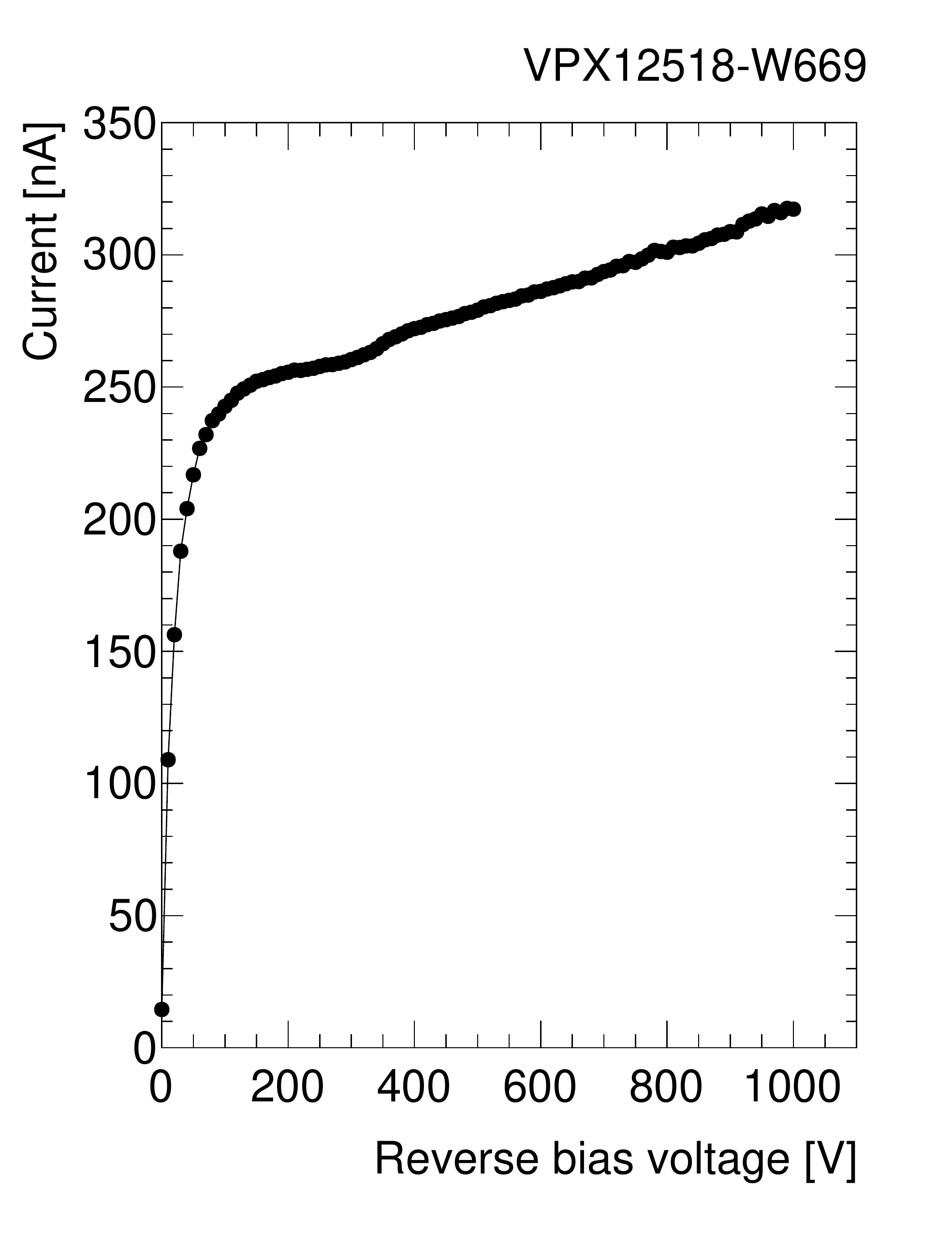}
 \caption{Sensor leakage current measurement (IV)}
 \label{fig:MA:IV}
\end{subfigure}
\begin{subfigure}{.04\textwidth}
\hfill
\end{subfigure}
\begin{subfigure}{.47\textwidth}
 \centering
 \includegraphics[width=\linewidth]{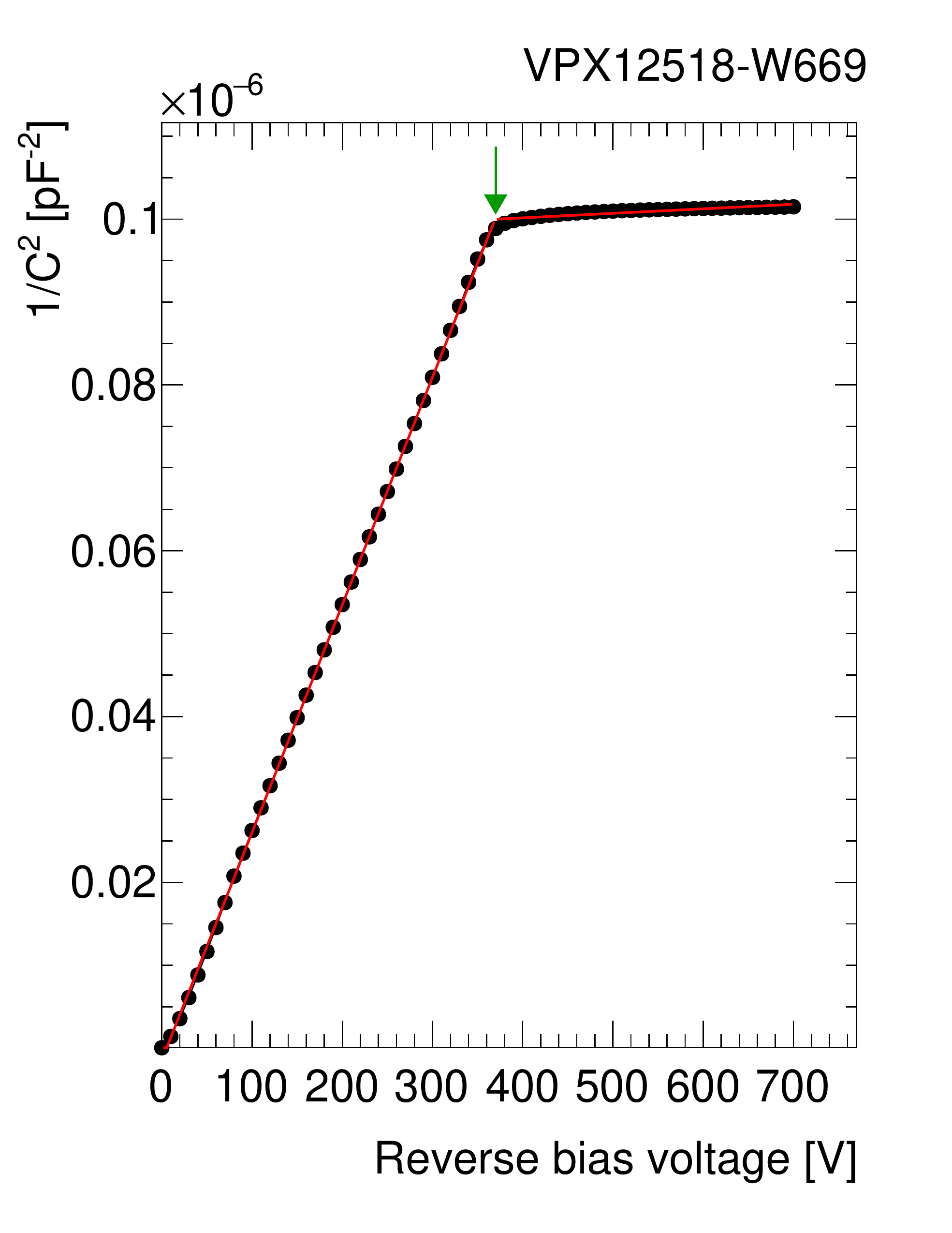}
 \caption{Sensor bulk capacitance measurement (CV)}
 \label{fig:MA:CV}
\end{subfigure}
\caption{IV and CV curves for \mbox{ATLAS12SS} sensor. The sensor depletion voltage is derived from the CV measurement (indicated by red arrow).}
\end{figure}

Sensors are electrically tested upon reception from the vendor, and additional tests are carried out after shipment to module assembly sites.
Reverse bias leakage current (IV) characteristics are determined by raising the bias voltage in \unit[10]{V} steps, observing a \unit[10]{s} delay before reading the current. An example IV curve is plotted in figure~\ref{fig:MA:IV}: the leakage current is well behaved up to~\unit[-1000]{V}.

Reverse bias capacitance curves are measured by an LCR meter using a frequency between \unit[1 and 5]{kHz} and \unit[100]{mV} amplitude. The full depletion voltage ($V_{D}$) is extracted from the intersection of two straight line fits to the
curve of $1/C^2$ versus voltage: one line is fitted to the linear
slope below the $V_{D}$, the other is fitted to
the flat section above $V_{D}$.
The depletion voltage is indicated by the red arrow shown in
figure~\ref{fig:MA:CV}.

\begin{figure}
 \centering
 \includegraphics[width=0.6\linewidth]{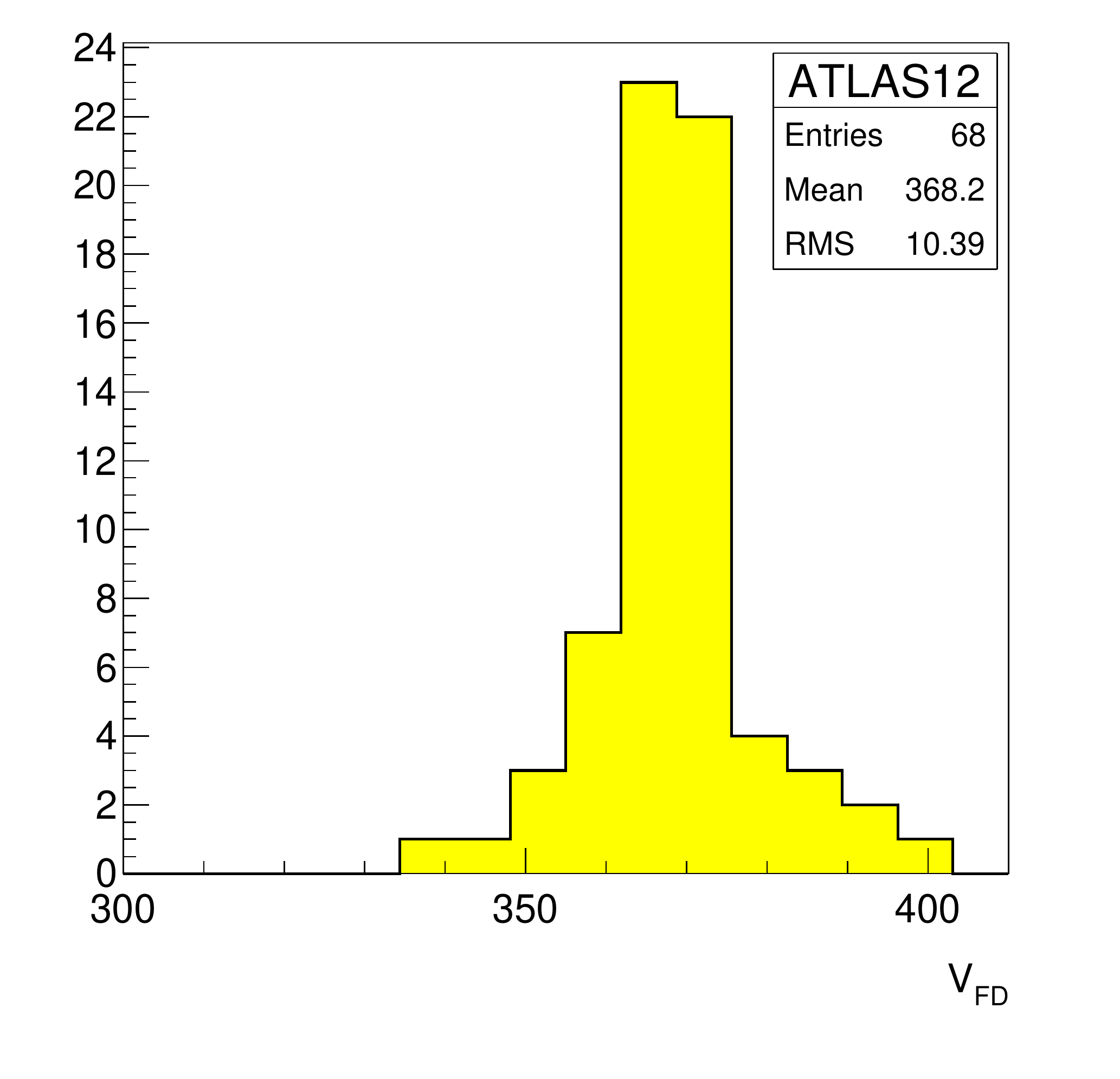}
 \caption{Depletion voltages determined from bulk capacitance measurements performed on 68 \mbox{ATLAS12} sensors. An average depletion voltage of {\unit[$368\pm10$]{V}} was determined.}
 \label{fig:depletions}
\end{figure}

The depletion voltage was extracted for all sensors and found to be \unit[-370]{V} on average for the investigated sensors. An overview for a subset of 68 \mbox{ATLAS12} sensors is shown in figure~\ref{fig:depletions}.

Full strip test measurements were carried out on the barrel sensors, whereby
successively \unit[-10]{V}, \unit[-100]{V} is applied across the strip metal and
bias rail to check for short circuits or oxide pinholes, respectively. An LCR
meter (\unit[1]{kHz}, \unit[100]{mV}) is used to determine the
R$_{\text{bias}}$ and C$_{\text{coupling}}$ values of the AC circuit formed by
the bias resistor and capacitance between the strip implant and strip metal.

\begin{figure}
 \centering
 \includegraphics[width=0.8\linewidth]{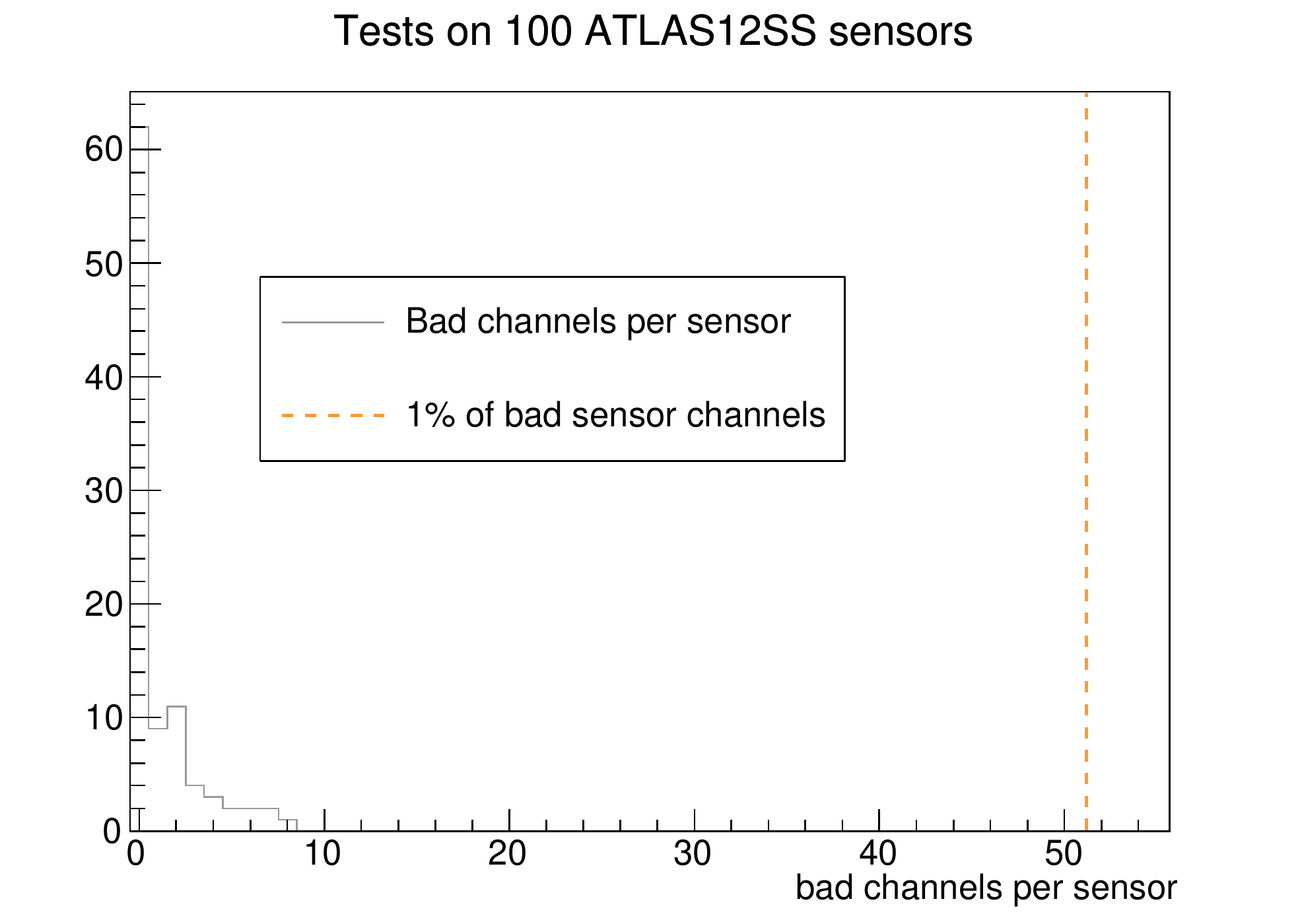}
 \caption{Number of bad channels per sensor found during indidividual strip tests of 100 \mbox{ATLAS12SS} sensors.}
 \label{fig:badchannels}
\end{figure}

The specification for the sensor bias resistance is \unit[1.5$\pm$0.5]{M$\Omega$} and the strip AC coupling capacitance is required to be \unit[$>$20]{pF/cm}. The measurements discussed above were compared against these values, and the number of channels outside these specification was found to be on average 5 out of 1280 channels (with the specification requiring a minimum of \unit[98]{\%} good channels).
Figure~\ref{fig:badchannels} shows an overview of the number of bad channels per sensor found on 100 sensors.

In addition, the maximum number of pinholes was required to be at most seven per 1280 sensor strips, which were required to not form a cluster of eight or more for an individual sensor segment.

\subsection{Module assembly \label{subsec:MA}}

\begin{figure}
\centering
\includegraphics[width=0.6\linewidth]{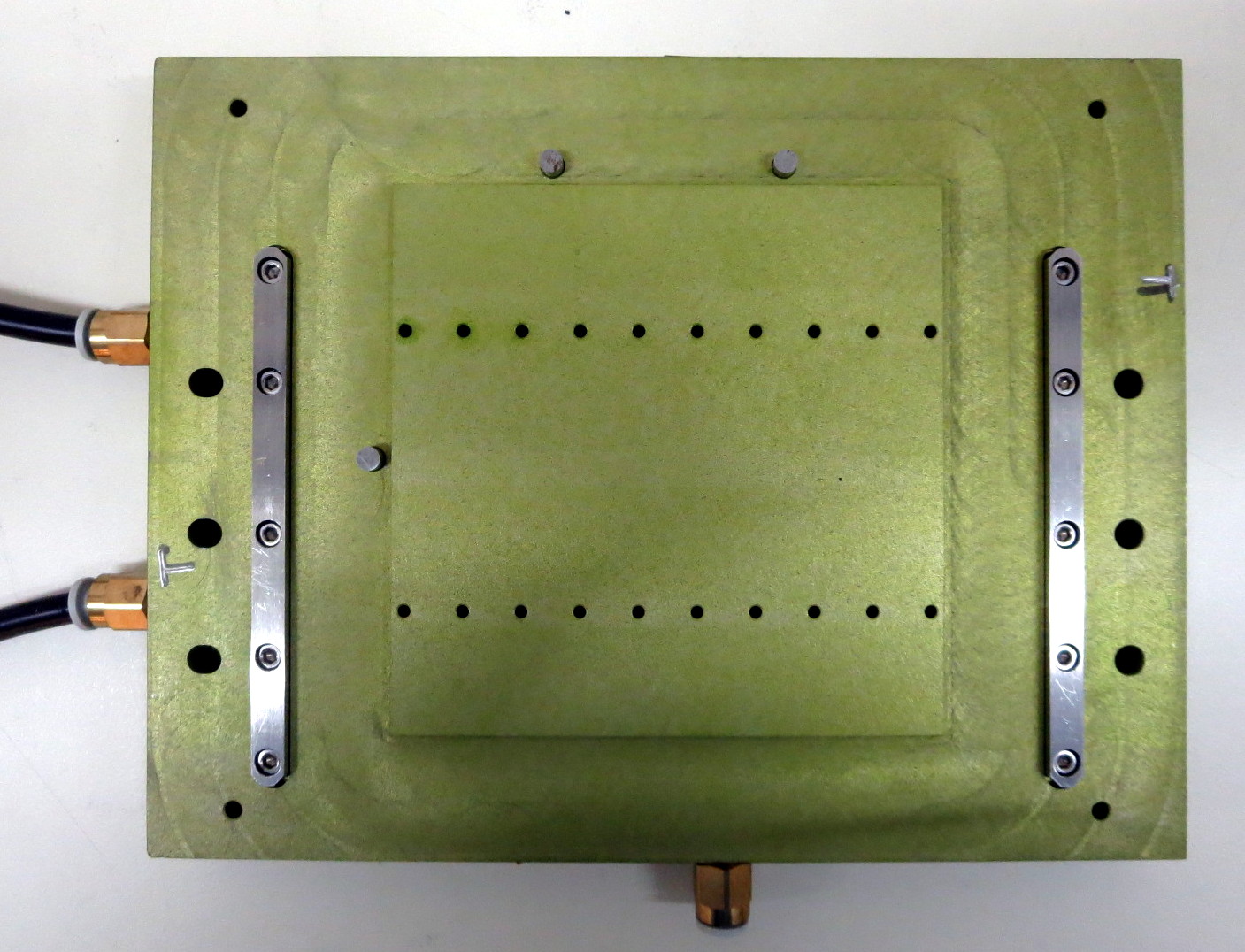}
\caption{Vacuum jig for module assembly: a sensor is positioned with respect to alignment holes by using it against three alignment pins located around the central vacuum area. Hybrids are placed on top of the sensor using the same hybrid pick-up tool also used for hybrid assembly, which is positioned using precision alignment holes in the vacuum jig.}
\label{fig:modulejig}
\end{figure}

Successfully tested hybrids, powerboards and sensors were assembled into modules in a gluing process similar to the assembly of hybrids:
First, sensors are aligned on a precision vacuum jig using three alignment pins (see figure~\ref{fig:modulejig}). After positioning the sensor, vacuum is applied to the sensor backside to hold the sensor in position and keep it flat during the assembly process.

Hybrids had to be mounted on the sensor surface before the powerboard, as the powerboard shield box on the powerboard prevents the positioning of the hybrid pick-up tool at the correct height. At this stage, hybrids were still located on individual positions on hybrid panels, where they were populated with ASICs and tested. Prior to assembly into modules, the wire bonds connecting hybrids to the power and data lines of a panel need to be removed.

Hybrids are then lifted from the hybrid panel using a hybrid pick-up tool (see figure~\ref{fig:pickuptool}), which is located using the hybrid panel alignment holes (see figure~\ref{fig:hyb_panel}) and thereby ensures the correct hybrid position with respect to the pick-up tool alignment pins.

A two-component epoxy (Epolite FH-5313) was used to attach hybrids and
powerboards to sensors. Over the course of the barrel module programme, an
extensive study was conducted to investigate potential epoxy glues for module
assembly~\cite{Cole}, which yielded two additional candidates: Eccobond F-112 and Polaris
PL-5313. Polaris PL-5313, the successor of Epolite FH-5313, was
chosen as the baseline for module assembly during production and Eccobond
F-112 as the alternative.

During the ABC130 barrel module programme, eight modules were assembled with F-112 and PL-5313 adhesives each, while the majority of modules was constructed using Epolite FH-5313.
Since no difference was observed for modules assembled
with these three adhesives, they will not be distinguished further in the text.

\begin{figure}
\centering
\includegraphics[width=0.9\linewidth]{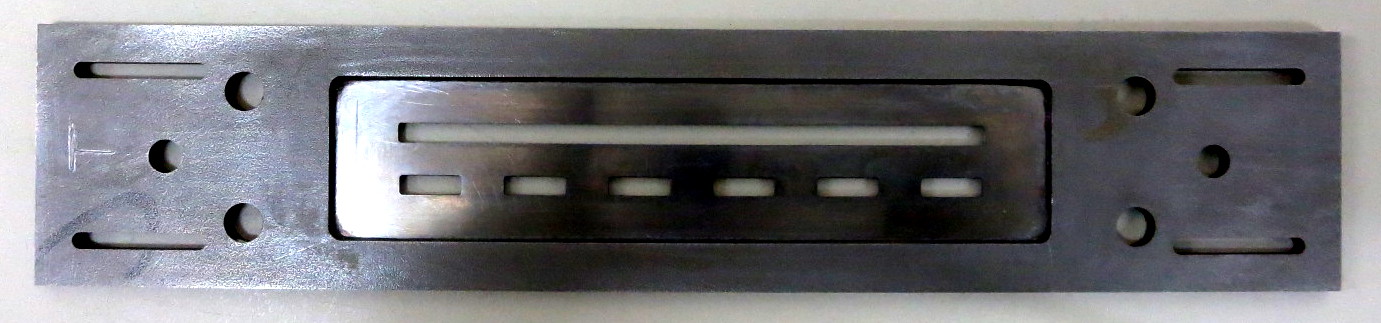}
\caption{Glue stencil used to apply glue to the backside of a hybrid before mounting it on a sensor. A continuous glue line under all ABC130 ASICs was used to ensure good mechanical support for wirebonding and thermal connection. Alignment holes in the stencil frame allow to position the stencil on the hybrid using the pick-up tool alignment pins.}
\label{fig:stencil}
\end{figure}

The mixed adhesive was applied to the hybrid backside using a glue stencil mounted on the hybrid pick-up tool (see figure~\ref{fig:stencil}). In order to ensure a sufficient glue viscosity for the use of a stencil, a waiting time of up to \unit[20]{min} was used between mixing both epoxy components and applying the glue.
After dispensing the adhesive, the stencil was removed and the hybrid positioned on the sensor and held in place using a brass weight. A stencil thickness of \unit[250]{$\upmu$m} was chosen to produce corresponding glue layers. By positioning the hybrid pick-up tool on dedicated landing pads in the module assembly jig, the glue layer between hybrids and sensor was compressed to a thickness that was set using adjustment screws on the pick-up tool. The target glue thickness between hybrid and sensor was chosen to be \unit[$120 \pm 40$]{$\upmu$m}, which corresponded to a \unit[50]{\%} thickness compression (assuming a \unit[100]{\%} stencil fill factor) and therefore a doubling of the glue area for good mechanical support below the hybrid.
A minimum curing time of \unit[6]{hours} was used before turning off the vacuum holding hybrid and sensor at a defined distance.
Since glue spreading over the sensor bias ring was found to cause early sensor breakdowns in several cases~\cite{Cole}, the sensor current was monitored after individual gluing steps.

\begin{figure}
\centering
\includegraphics[width=0.8\linewidth]{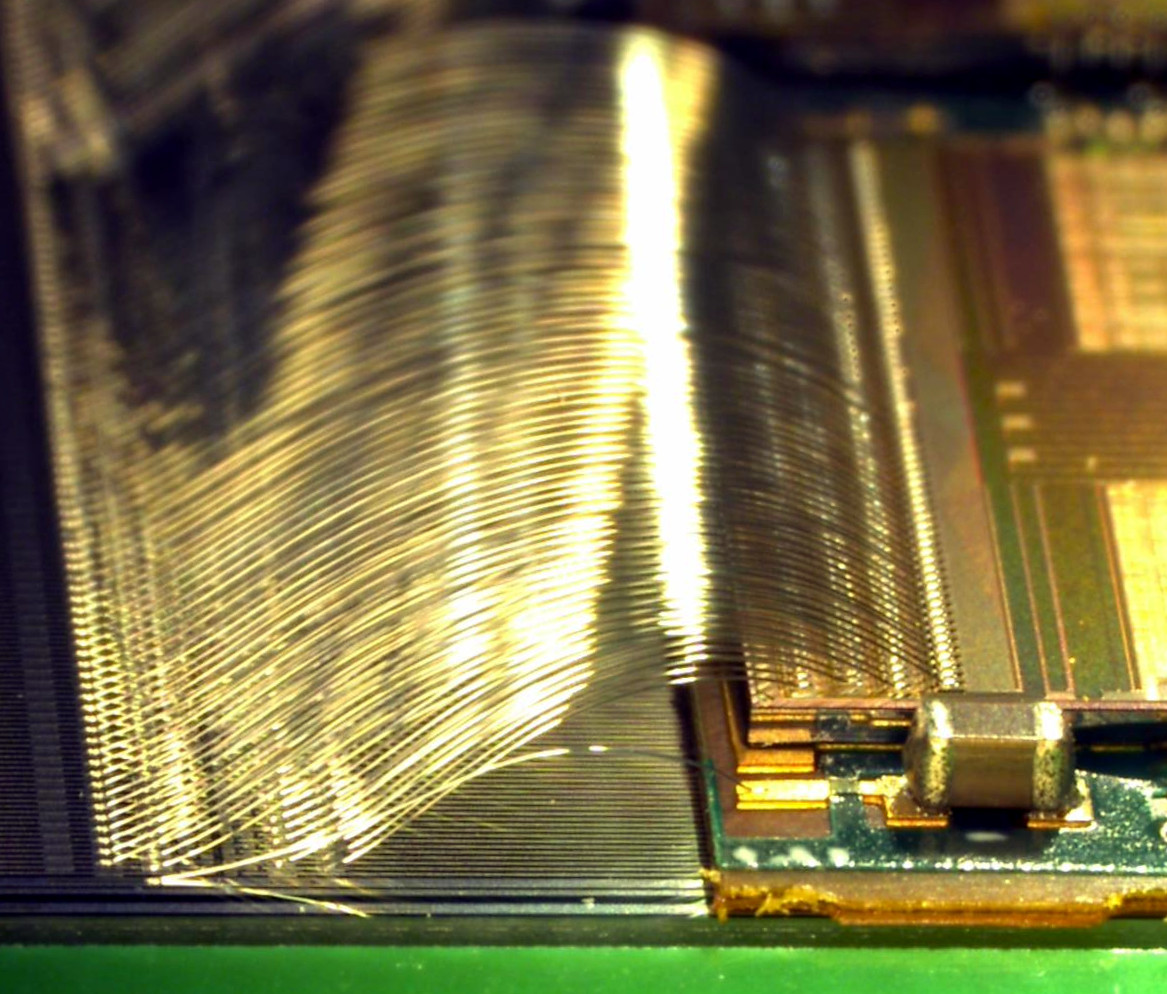}
\caption{Front-end wire bonds connecting the readout channels of ABC130 readout chips and sensor strips: wire bonds are arranged in four layers, with the lower two rows attached to the sensor segment located below the readout chips and the upper two rows attached to the neighbour sensor segment.}
\label{fig:sensorwires}
\end{figure}

ABC130 barrel modules can be tested (see section~\ref{subsec:test_mod}) without a powerboard attached to the module. An additional module test was therefore performed between hybrid and powerboard attachment, so that the impact of mounting a powerboard on a module could be studied directly (see section~\ref{sec:selectModule:PBEffect}).
In order to test the module performance, each ABC130 readout channel was connected to a silicon sensor strip using aluminium wire wedge bonding. Front-end bonds (i.e. wire bonds connecting the analogue ABC130 readout channels to the sensor) were drawn in four rows arranged in layers (see figure~\ref{fig:sensorwires}). Out of the four rows of staggered sensor bond pads, the lower two rows (64 wires each) were attached to the inner strip segment of an SS module (or the sensor segment located beneath hybrid and powerboard of LS modules), the upper two rows were connected to the outer segment of an SS sensor (or sensor segment without hybrid and powerboard on an LS module), see figures~\ref{fig:module_SS} and~\ref{fig:module_LS}.

\begin{figure}
\centering
\includegraphics[width=0.9\linewidth]{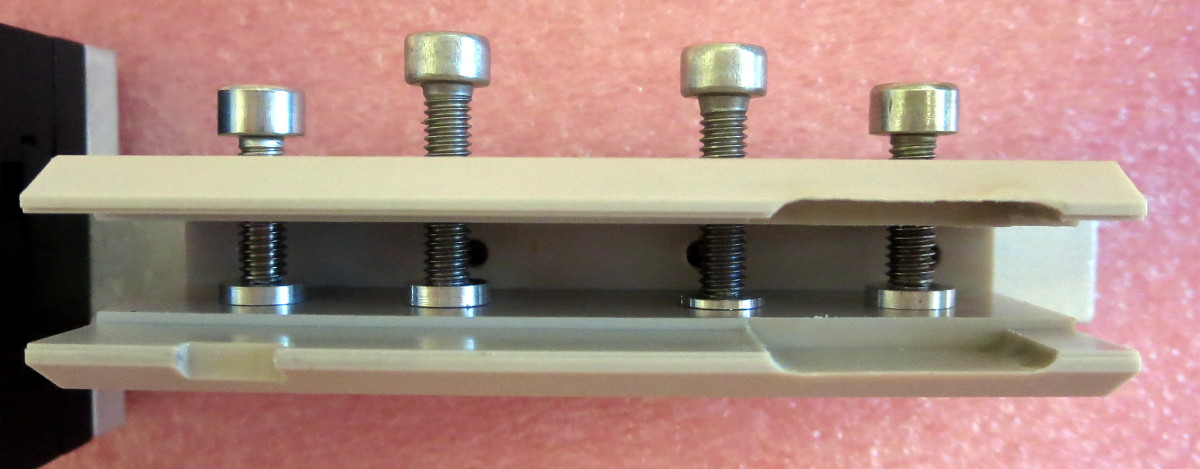}
\caption{3D printed holder to be fastened around the edges of a powerboard for positioning on a module. Four screws along the powerboard were used to close or open the tool and account for variations in the PCB width. In order to avoid squeezing the shield box mounted on the powerboard, the corresponding shape was carved out of the holder (right side).}
\label{fig:pbtool}
\end{figure}

\begin{figure}
\centering
\begin{subfigure}{.47\textwidth}
 \centering
 \includegraphics[width=\linewidth]{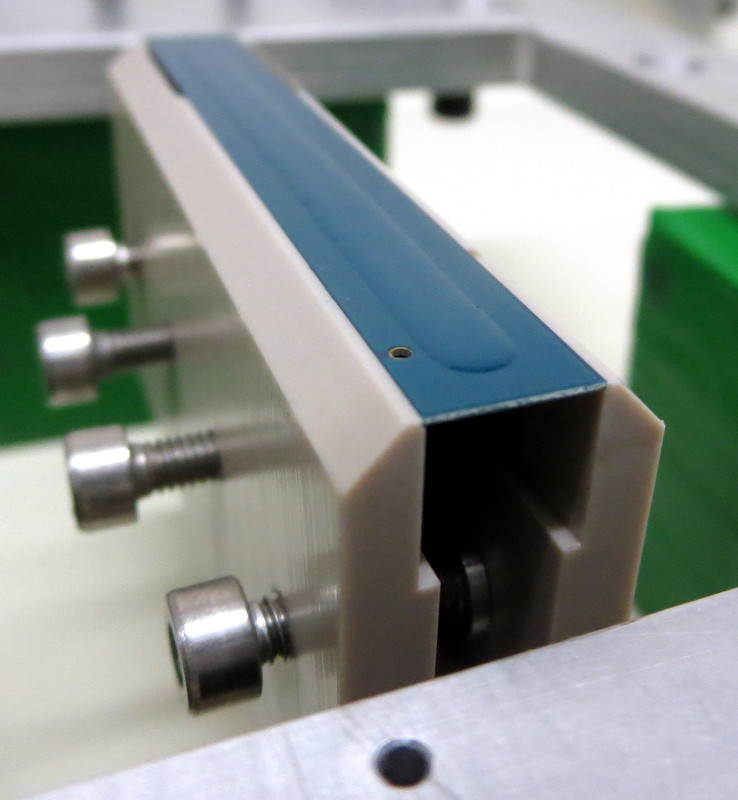}
 \caption{Glue layer on backside of an upside down powerboard mounted in a powerboard holder.}
 \label{fig:pbwithglue}
\end{subfigure}
\begin{subfigure}{.04\textwidth}
\hfill
\end{subfigure}
\begin{subfigure}{.47\textwidth}
 \centering
 \includegraphics[width=\linewidth]{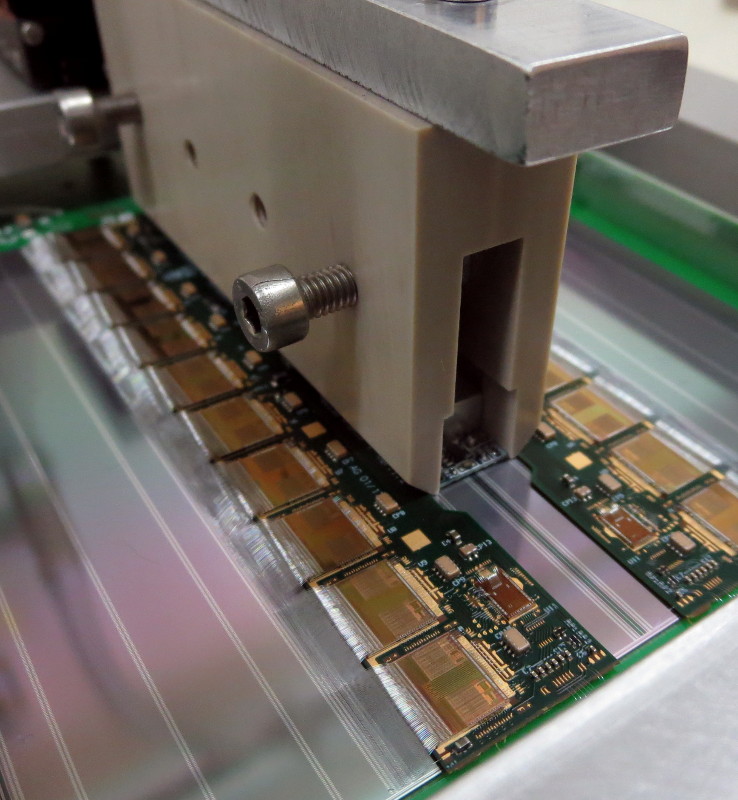}
 \caption{Powerboard held in powerboard holder positioned between hybrids on a sensor.}
 \label{fig:pbmounting}
\end{subfigure}
\caption{Assembly of a powerboard between two hybrids on a sensor.}
\end{figure}

\begin{figure}
\centering
\includegraphics[width=0.7\linewidth]{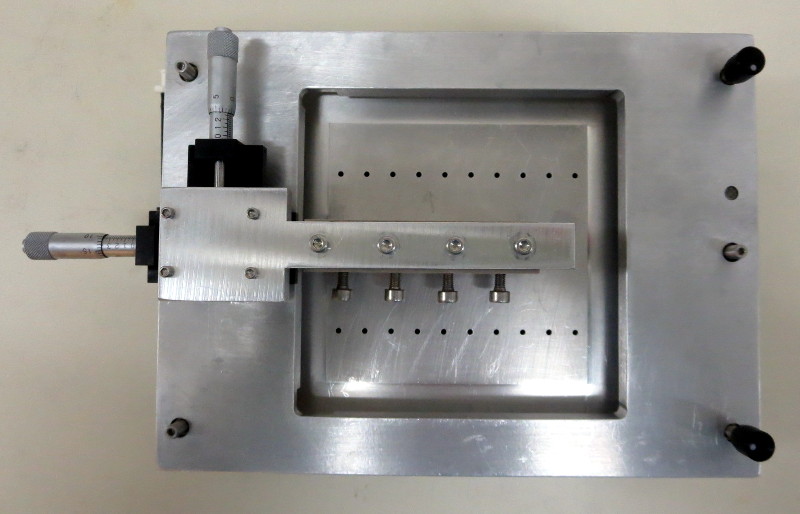}
\caption{Full powerboard gluing tool assembly: while the module with hybrids is held in position using a vacuum jig, a powerboard with glue is positioned above it. The powerboard is held in a clamp, which is aligned with respect to hybrid and sensor edges using micrometer screws on the holding frame. Screws on the frame allow to set the glue height between powerboards and hybrid to the target height.}
\label{fig:pbjig}
\end{figure}

Different from hybrids, the layout of powerboards did not allow them to be picked up using vacuum pick-up tools, as the high density of components mounted on the powerboard did not leave enough space for reliable vacuum connections. Powerboards were therefore held along the PCB edges using a width adjustable tool that could be tightened around the powerboard edges using screws (see figure~\ref{fig:pbtool}).
After picking up the powerboard and fixing it in position, glue was applied to the powerboard's backside (see figure~\ref{fig:pbwithglue}). Powerboards were attached to sensors using the same two-component epoxy used between hybrids and sensors: Epolite FH-5313. 
Afterwards, the powerboard was placed between two hybrids on an SS module (see figure~\ref{fig:pbmounting}) or next to one hybrid on an LS module, with a gap of about \unit[1]{mm} between hybrid and powerboard edge to facilitate wire bonding.
Similar to the tools used to mount hybrids on sensors, powerboard gluing tools can be adjusted to set the target glue height of \unit[$120 \pm 40$]{$\upmu$m} prior to assembly (see figure~\ref{fig:pbjig}).
After positioning a powerboard on a module, vacuum pressure was maintained throughout a curing time of at least six hours.

Since powerboard assembly tools were still under development during the ABC130 barrel module programme, part of the modules were assembled without the tools being available. In these instances, modules were assembled by placing powerboards on the module by hand.

During the ABC130 barrel module programme, the method of holding a powerboard along its edges for assembly was found to lead to uneven glue thicknesses in case of warped powerboards or irregularities along the powerboard edges. Subsequent versions of powerboard pick-up tool were therefore designed to use vacuum pickup pins (similar to hybrid pick-up tools). Additionally, the powerboard design for the next generation of readout ASICs was modified to increase the size of areas suitable for vacuum pick-up.

\subsubsection{Sensor metrology after module assembly}

\begin{figure}
\centering
\begin{subfigure}{.47\textwidth}
 \centering
 \includegraphics[width=0.9\linewidth]{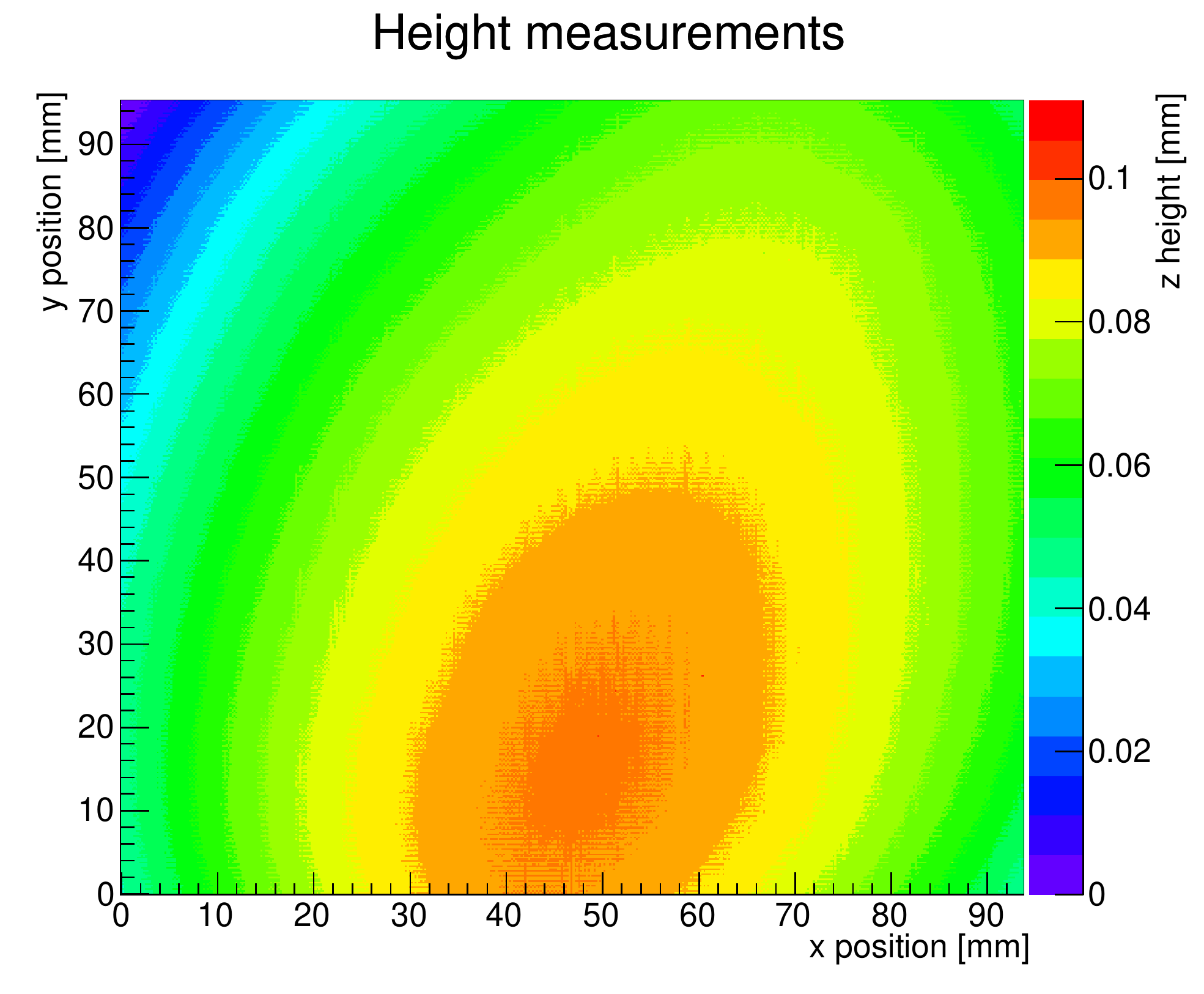}
 \caption{Absolute height measurement grid on sensor}
 \label{fig:bow1}
\end{subfigure}
\begin{subfigure}{.04\textwidth}
\hfill
\end{subfigure}
\begin{subfigure}{.47\textwidth}
 \centering
 \includegraphics[width=\linewidth]{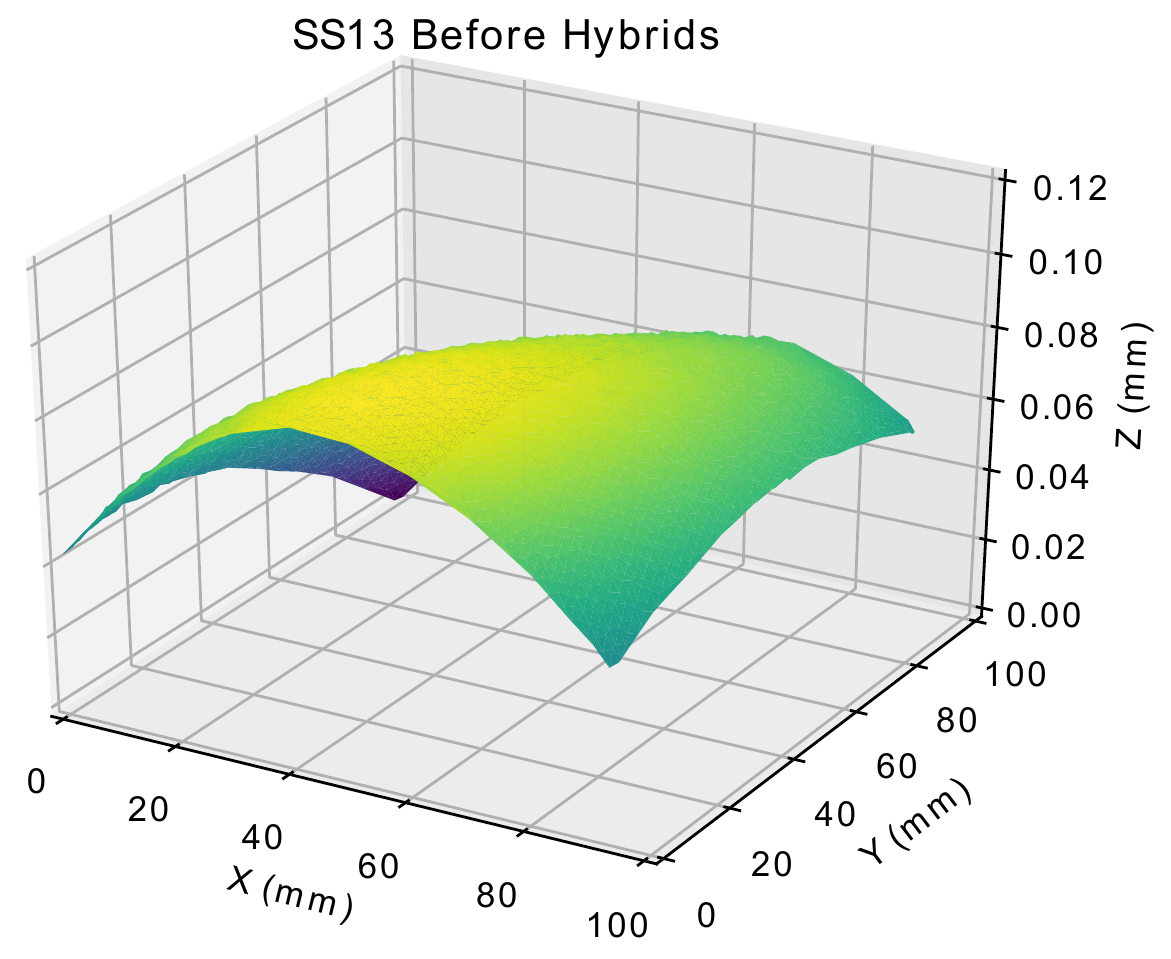}
 \caption{Overall sensor shape after tilt correction}
 \label{fig:bow2}
\end{subfigure}
\begin{subfigure}{.47\textwidth}
 \centering
 \includegraphics[width=\linewidth]{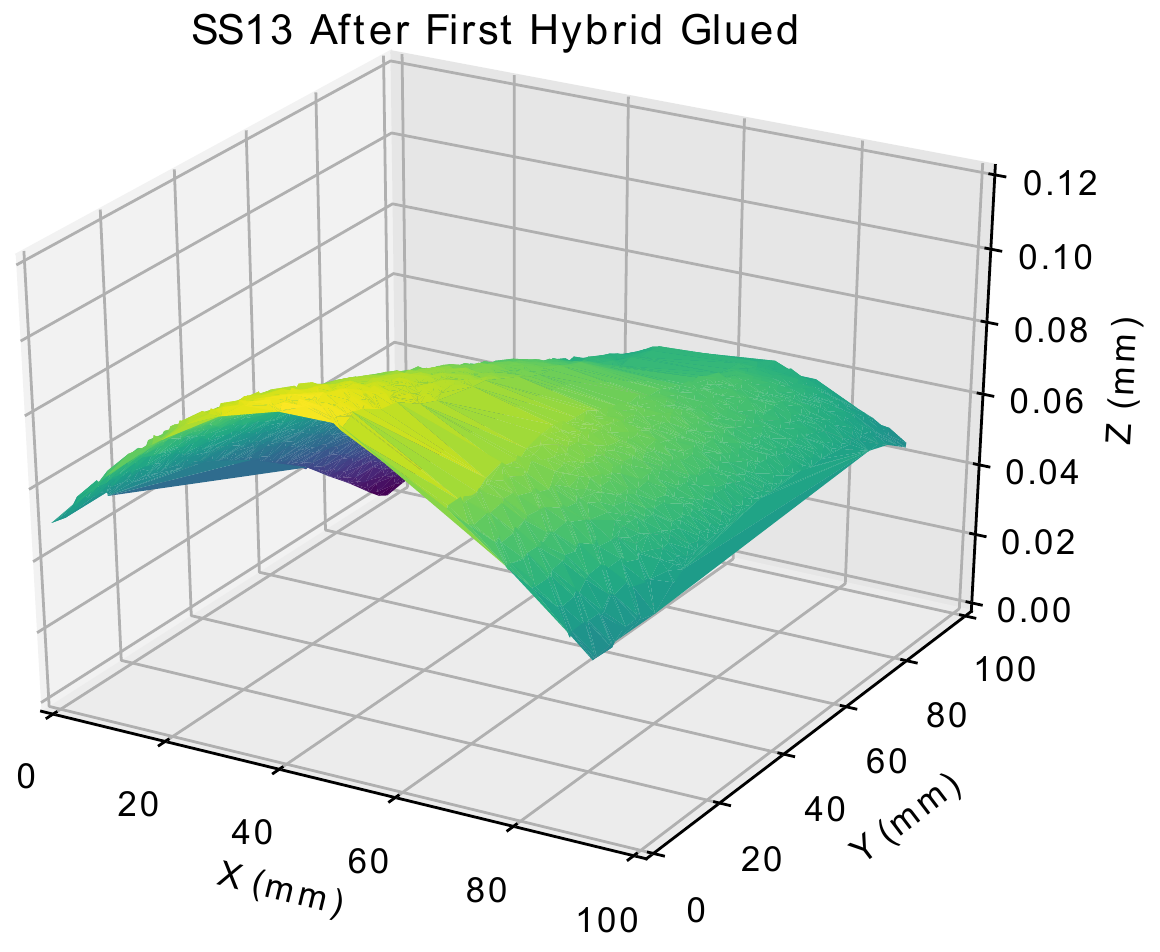}
 \caption{Sensor shape after attaching first hybrid}
 \label{fig:bow3}
\end{subfigure}
\begin{subfigure}{.04\textwidth}
\hfill
\end{subfigure}
\begin{subfigure}{.47\textwidth}
 \centering
 \includegraphics[width=\linewidth]{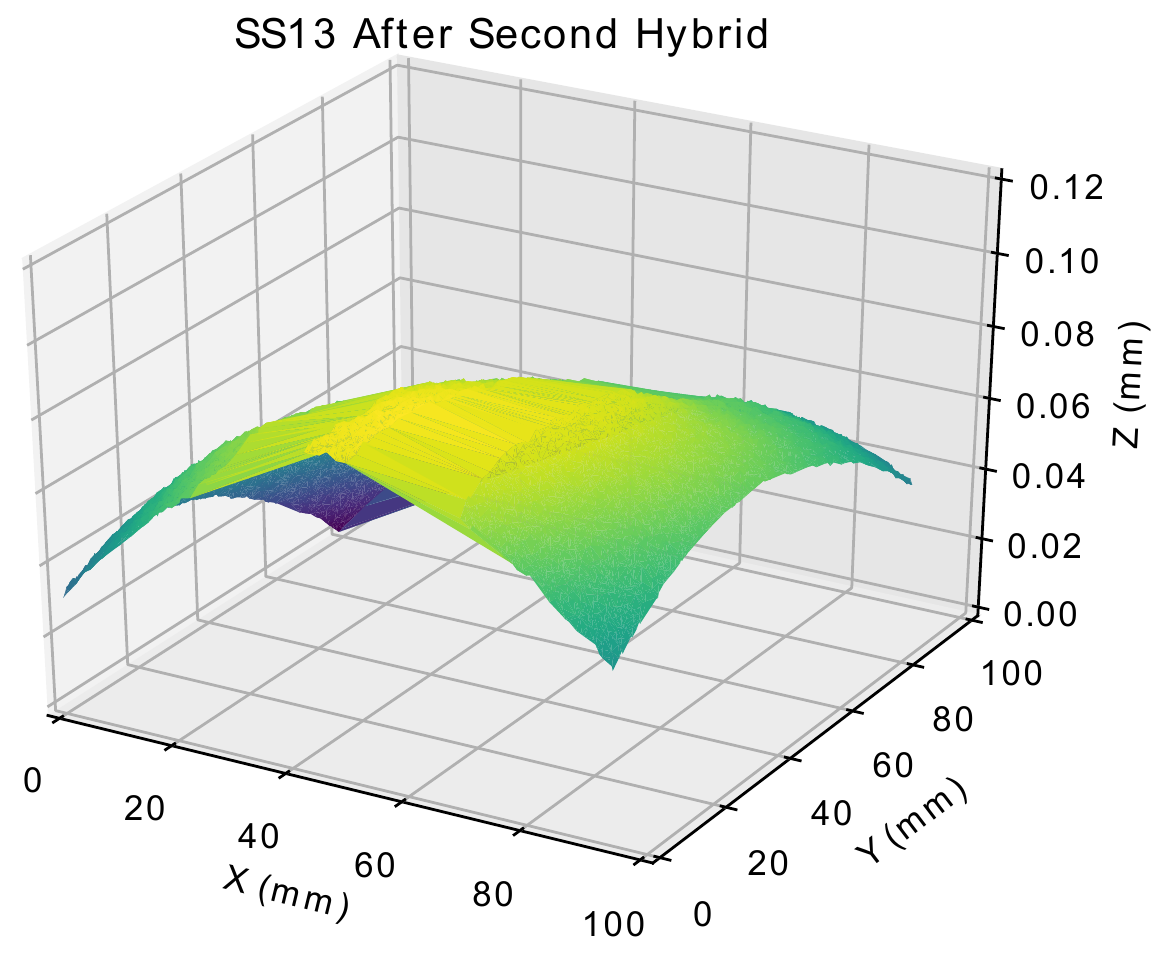}
 \caption{Sensor shape after attaching second hybrid}
 \label{fig:bow4}
\end{subfigure}
\caption{Height measurements performed on sensor at different stages of module assembly.}
\end{figure}

For the attachment of modules onto support structures, modules are required to have a bow of no more than $\unit[\pm^{150}_{50}]{\upmu\text{m}}$, where positive numbers refer to the module centre being below the edges. 
While sensors themselves typically conform with this envelope upon delivery, the assembly process of modules, during which sensors and PCBs are held flat using vacuum tools, can affect the overall sensor bow. In order to monitor the impact of gluing on the sensor shape, dedicated metrology measurements were performed at different stages of module assembly (see figures~\ref{fig:bow1} to~\ref{fig:bow4}): using a white light interferometer system, the absolute sensor height was measured for a fine grid of inspection points (see figure~\ref{fig:bow1}). The sensor shape was mapped based on the measured heights and the sensor bow was calculated based on a fit through the sensor plane.

After each measurement, the overall sensor shape and maximum height deviations were calculated. Examples for four modules are shown in table~\ref{tab:bows}.
\begin{table}
\centering
\begin{tabular}{c|c|c|c}
 & \multicolumn{3}{c}{Maximum difference, $[\upmu\text{m}]$} \\
Module & Sensor & First hybrid & Second hybrid \\
\hline
SS12 & 76 & 83 & 103 \\
SS13 & 103 & 101 & 93 \\
SS14 & 84 & 85 & 95 \\
SS15 & 96 & 119 & 98 \\
\end{tabular}
\caption{Distance between maximum and minimum height deviation from the sensor plane determined in optical measurements for four short strip modules.}
\label{tab:bows}
\end{table}

Measurements did confirm that the attachment of hybrids to a sensor affected the sensor shape and changed the distance between the maximum and minimum deviation from the sensor plane by up to \unit[27]{$\upmu$m}. All modules were found to be within the specifications at all stages of assembly.

\subsection{Module tests}
\label{subsec:test_mod}

\begin{figure}
\centering
\includegraphics[width=0.8\linewidth]{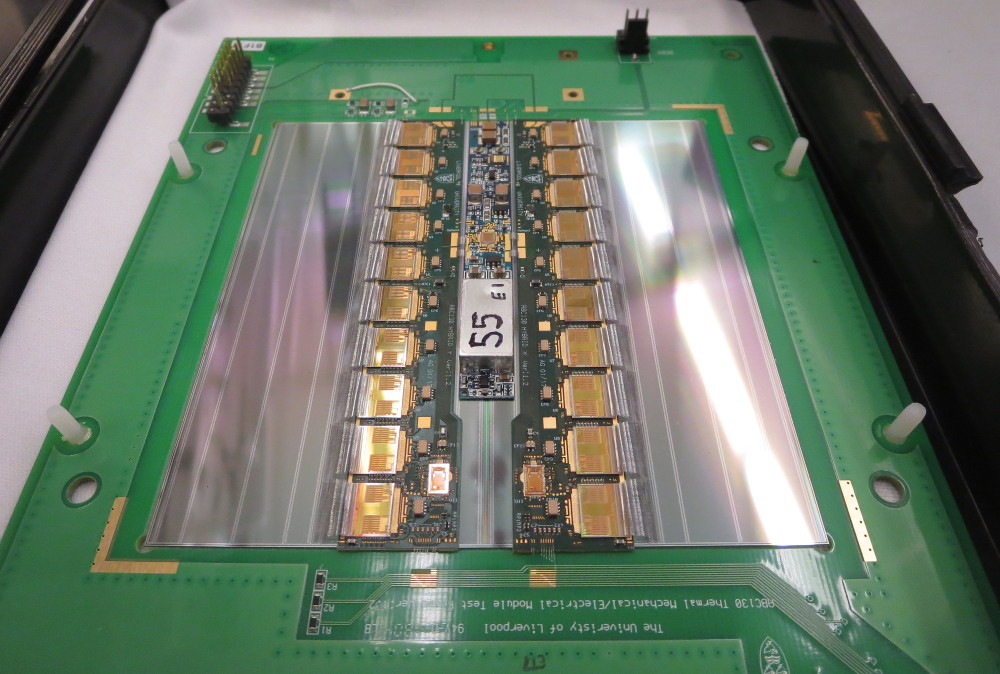}
\caption{ABC130 short strip barrel module on a test frame PCB: wire bonds connect hybrids and powerboards to the test frame for data readout (bottom side) and power supply (top side). The sensor rests on a ledge surrounding a cutout in the test frame, through which high voltage and cooling are applied to the sensor backplane.}
\label{fig:testframe}
\end{figure}

For electrical tests, modules are connected to dedicated test frames (see figure~\ref{fig:testframe}), through which power is supplied and data is read out.
In order to apply high voltage to the sensor backside, test frames have a cutout that allows a direct contact between the sensor backside and a testing jig. During testing, vacuum holes hold the sensor on the high voltage testing jig. Additionally, a cooling loop embedded in the testing jig is used to maintain a constant sensor temperature during testing. Condensation on the cooled module is prevented by flushing the test setup with dry air or nitrogen.

\begin{figure}
\centering
\includegraphics[width=\linewidth]{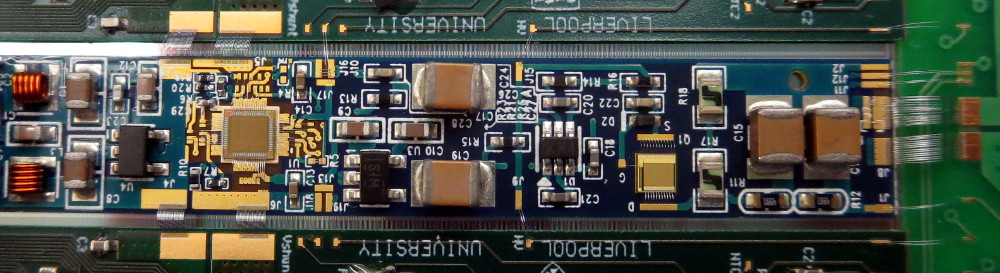}
\caption{Power supplying bonds on a short strip module with a powerboard: current is supplied from the test frame to the powerboard through bond wires (right side). Hybrids are powered through wire bonds from the powerboard (left side).}
\label{fig:powerbonds}
\end{figure}

The powering concept of ABC130 modules allows the testing of directly powered hybrids as well as hybrids powered through a powerboard (see figure~\ref{fig:powerbonds}). 
Since the attachment of hybrids and a powerboard are separate steps in the construction of modules, which can lead to sensor damage, causing e.g. an early sensor breakdown, modules were tested both after hybrid attachment and after powerboard attachment, permitting a comparison of the module performance at both stages (see section~\ref{sec:selectModule:PBEffect}).

For the full electrical test of a module, the tests performed on electrical hybrids (Capture HCC and ABC IDs, Strobe Delay and Three Point Gain - see section~\ref{subsec:test_elec}) are repeated with a fully depleted sensor. Results from the Three Point Gain are used to grade each individual module channel (see table~\ref{tab:3pg_grades}) and determine the number of good channels on a module.
\begin{table}
\centering
\begin{tabular}{c|c|c}
 & Gain & Noise \\
\hline
\multirow{2}{*}{High} & \textbf{high gain} & \textbf{high noise} \\
 & $\unit[>125]{\%}$ average chip gain & $\unit[>115]{\%}$ average chip noise \\
 \hline
\multirow{2}{*}{Low} & \textbf{low gain} & \textbf{dead} \\
 & $\unit[<75]{\%}$ average chip gain & \unit[0]{ENC}
 \end{tabular}
\caption{Criteria for grading of module channels based on results from Three Point Gain.}
\label{tab:3pg_grades}
\end{table}

In addition to the test sequence above, more tests can be performed for a full module characterisation:
\begin{itemize}
\item High statistics Three Point Gain
\item Trim Range
\item Noise Occupancy
\item Response Curve
\end{itemize}
These tests provide a complete characterisation of a module such that its quality can be
graded.

\section{Selected module test results}

The aim of the ABC130 barrel module programme was to allow tests of the proposed procedures for assembly, readout and cooling concepts of both modules and integrated structures. Due to the extent of the programme, it was possible to gather statistics on assembly yields and test results and to validate component designs for subsequent component generations.

This section summarises some of the module test results obtained in electrical tests of modules as well as individual components.

In addition to the investigations of individual electrical characteristics presented in the following, performance evaluations of full electrical modules were conducted in particle beams at the DESY-II and CERN SPS facilities~\cite{testbeam}.

\subsection{Dependence on number of strobed channels and triggers}
\label{nstrobe_ntrig}

During an internal charge injection test (either three point gain or response curve), two important parameters can be changed in order to optimise the accuracy of the extracted results and the time that the measurement takes. These are the number of triggers (readout requests after charge injections) per threshold value and the number of channels in the ABC130 chip strobed simultaneously.

\begin{figure}
\centering
\includegraphics[width=0.9\linewidth]{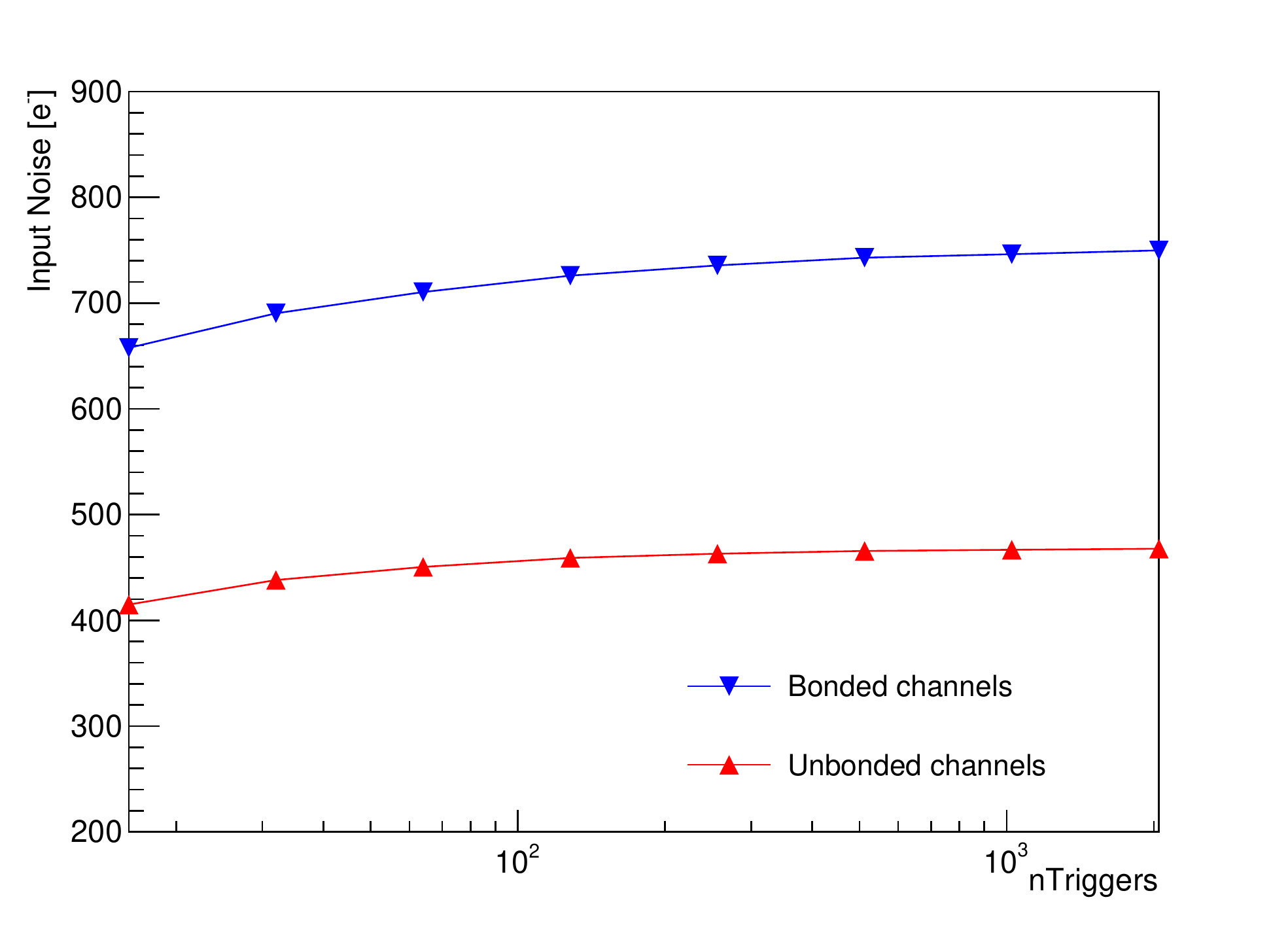}
\caption{Calculated input noise as a function of the number of triggers per scan used in a response curve measurement for both bonded and unbonded channels on a short strip module.}
\label{fig:nTrig}
\end{figure}

Figure~\ref{fig:nTrig} shows the noise extracted at \unit[1.5]{fC} from a response curve test as a function of the number of triggers included in the scans. As can be seen, the noise is underestimated when a small number of triggers is used but plateaus as the number is increased. This is because with a small number of triggers the low occupancy tail in the S-curve is underpopulated, which results in S-curve fits returning a narrower noise profile. As the number of triggers is increased, the tail is better populated, resulting in increased, and more correct, noise measurements.  This effect has been verified using Monte Carlo simulation.

\begin{figure}
\centering
\begin{subfigure}{.6\textwidth}
 \centering
 \includegraphics[width=\linewidth]{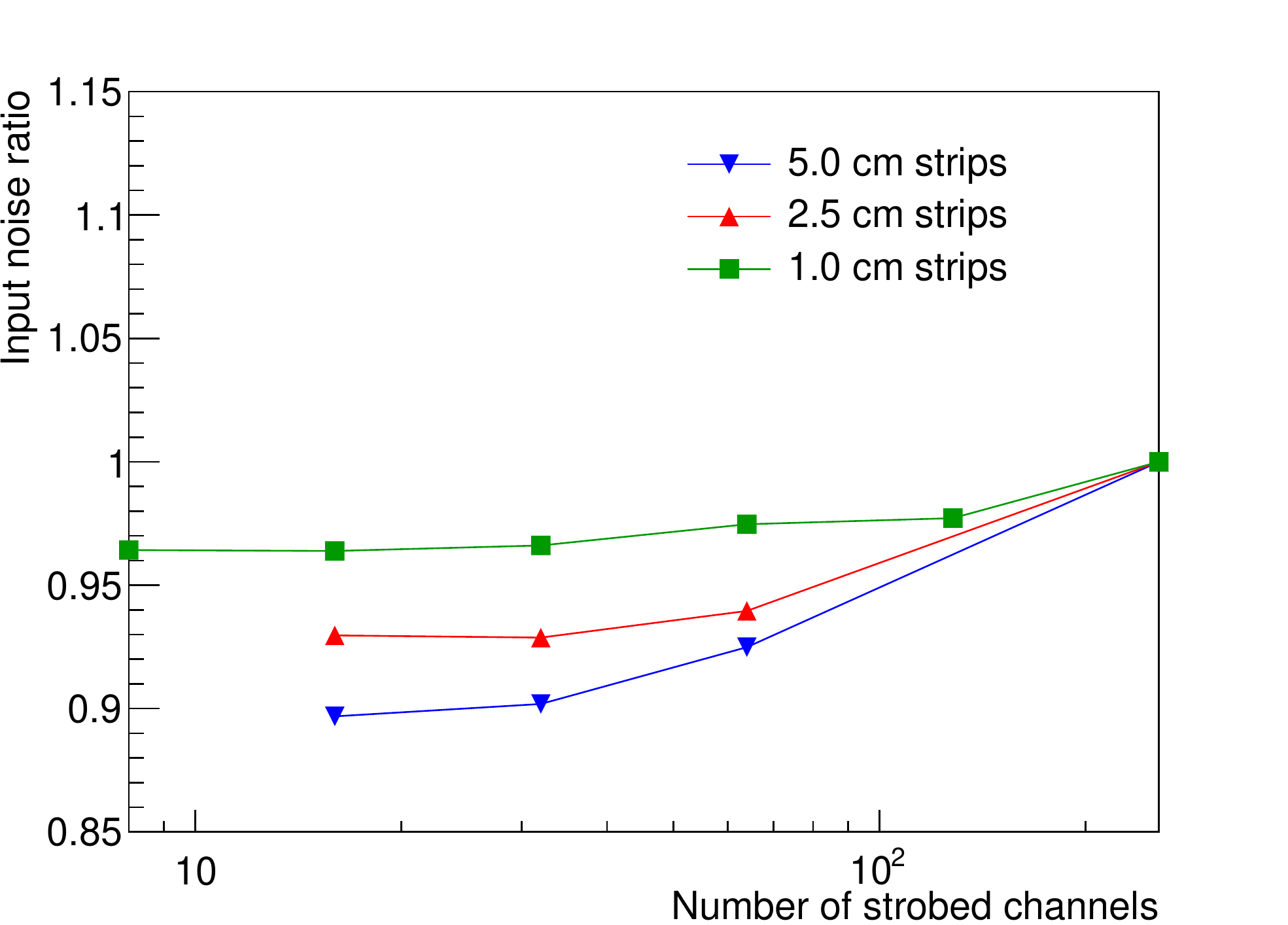}
 \caption{Input noise}
 \label{fig:nStrobed_innse}
\end{subfigure}
\begin{subfigure}{.6\textwidth}
 \centering
 \includegraphics[width=\linewidth]{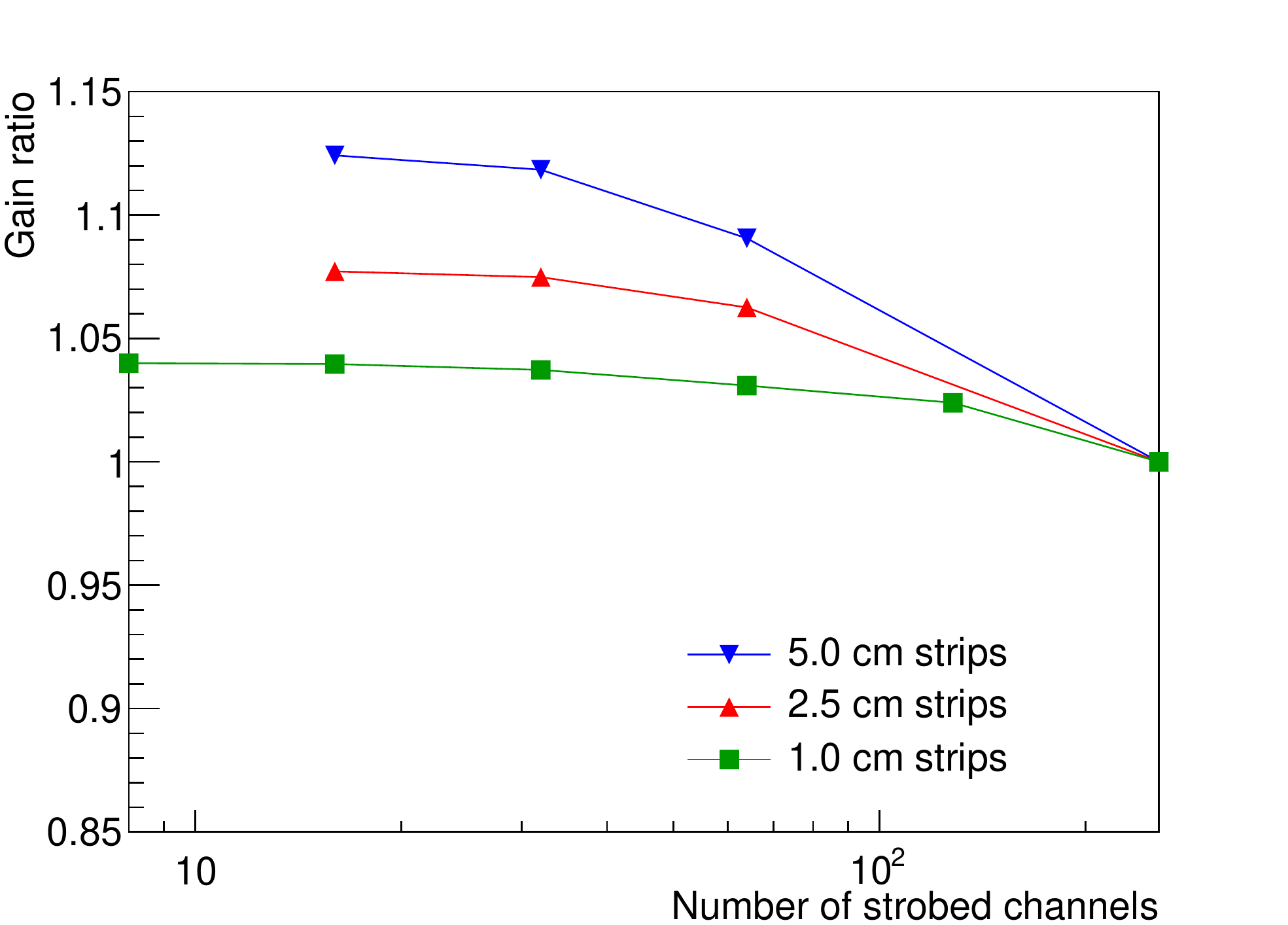}
 \caption{Gain}
 \label{fig:nStrobed_gain}
\end{subfigure}
\begin{subfigure}{.6\textwidth}
 \centering
 \includegraphics[width=\linewidth]{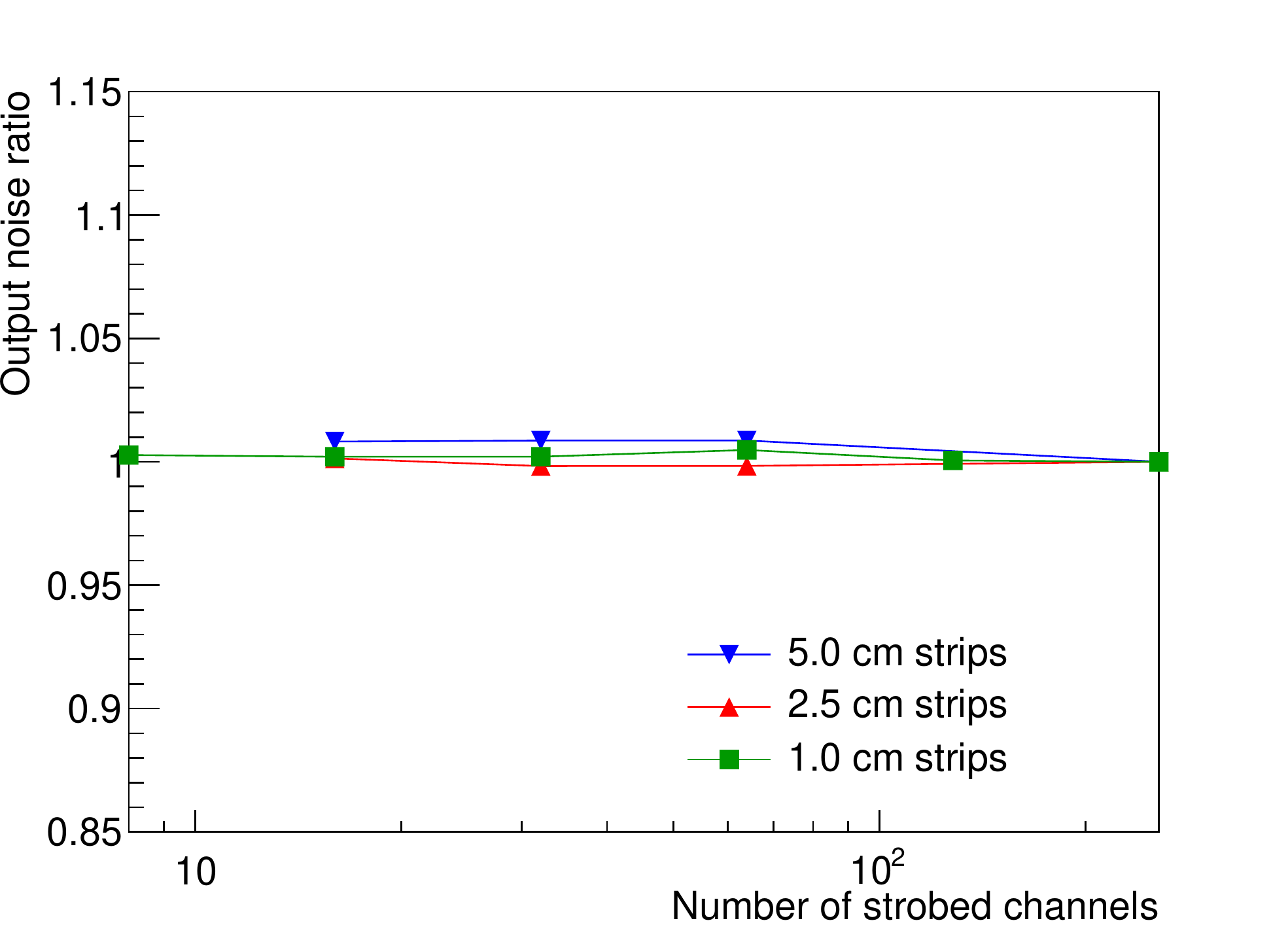}
 \caption{Output noise}
 \label{fig:nStrobed_outnse}
\end{subfigure}
\caption{Input noise, gain and output noise as a function of the number of strobed (charge injected) channels. Results are shown as a ratio compared to the case when all 256 channels are strobed simultaneously. Results for 2.5 and \unit[5.0]{cm} strips are taken on an ABC130 module whilst the \unit[1.0]{cm} strip result uses a mini-sensor connected to an ABC130 hybrid with a single chip.}
\label{fig:nStrobed}
\end{figure}

Figure~\ref{fig:nStrobed} shows the input noise, gain and output noise extracted at \unit[1.5]{fC} from a response curve test as a function of the number of channels strobed (charge injected) at once compared to that extracted when all 256 channels are strobed simultaneously. The extracted input noise can be seen to reduce as the number of strobe channels decreases. In the response curve analysis, the input noise is calculated as the ratio of the measured output noise (in mV) and gain (in mV/fC) extracted from the response curve. As such, it is useful to look also at the output noise and gain in addition to the input noise. It is seen that the measured gain increases with reducing number of strobed channels whilst the output noise remains constant.

This measurement shows that the charge injection circuitry within the chip does not manage to inject as much charge as expected when strobing many channels at once. This results in a decreased measured gain. In turn, this decreased gain results in increased calculated input noise and thereby explains the decreased noise observed with reduced number of strobed channels. It can also be seen that increasing the capacitive load on the channels by increasing the strip length increases the size of the observed effect.

As shown above, the extracted noise depends on both the number of triggers per scan and the number of channels probed simultaneously. During module testing a choice must therefore be made between speed of test, preferring low numbers of triggers and high numbers of strobed channels, and accuracy of the extracted input noise and gain, preferring high numbers of triggers and low number of strobed channels. As a result the decision was made that for quality control module testing, 192 triggers and 256 strobed channels would be the default whilst for the detailed module or chip characterisation tests shown below, 1024 triggers and 16 strobed channels is preferred to ensure that measurements are taken in the plateau of both distributions. Unless otherwise stated, all results below are done with the high number of triggers and low number of strobed channels required to get an accurate measure of input noise and gain.

\subsection{Module noise and strip capacitance}

The ABC130's amplifier input capacitance is the leading cause for an increased module noise (before irradiation) compared to the hybrid noise value. Therefore, the ABC130 chip set and modules have been fully characterised by measuring noise and gain as a function of load capacitance using the internal charge injection circuitry. This has been done on 32 channel prototype chips~\cite{Kaplon2012FrontEE}, on single chip test boards with either capacitor or mini sensor loads and on short strip and long strip barrel modules. In the case of the prototype, the front end was initially designed for positive signals rather than the negative signals coming from n-on-p sensors and so, results for both positive and negative signals are shown. In addition, results from an irradiated module are included which will be discussed further in Section~\ref{subsec:irradModules}. All non-prototype measurements shown are results are taken using the standard measurement configuration described in section~\ref{nstrobe_ntrig}.

 \begin{figure}
\centering
\includegraphics[width=0.8\linewidth]{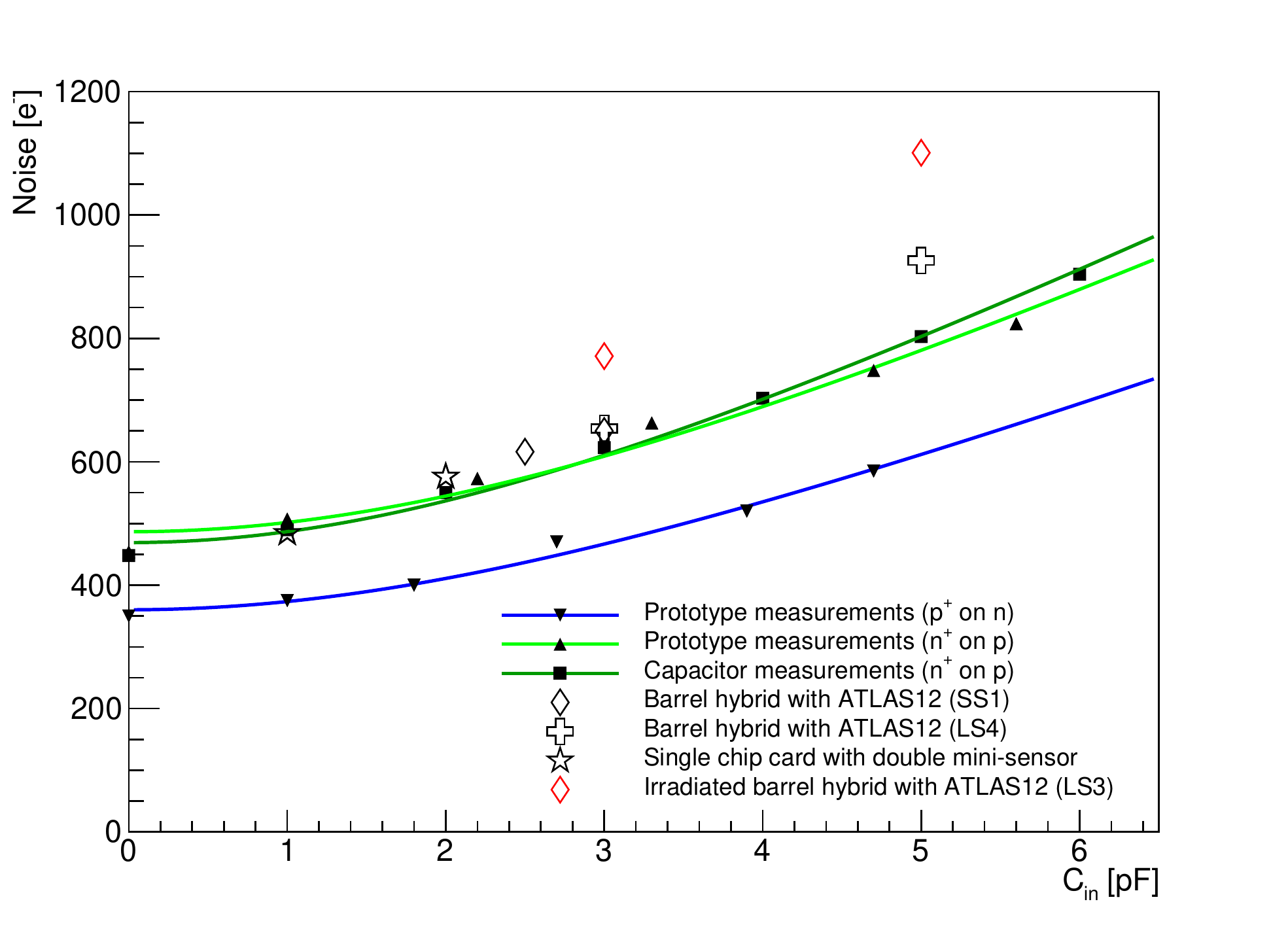}
\caption{Input noise as a function of capacitance for FE prototype measurements with positive polarity (inverted triangles and blue fit) and negative polarity (triangles and light green fit), single chips loaded with capacitors (squares and dark green fit), single chips loaded with mini sensors (open stars), hybrids on short strip modules (open diamonds), hybrids on long strip modules (open crosses) and hybrids on mixed strip length irradiated modules (red open diamonds).  The capacitance values for sensors is taken as the sum of the strip-backplane capacitance and interstrip capacitance (not including next-to-nearest neighbours) as taken from probing measurements of the sensors.}
\label{fig:NoisevsCint}
\end{figure}
Figure~\ref{fig:NoisevsCint} shows a summary of the noise measurements made as part of the ASIC characterisation. Capacitance values for tests performed including sensors are taken as the sum of the inter-strip capacitance and strip-backplane capacitance as measured during sensor probing. In addition, an extra \unit[0.5]{pF} is included for those strips running under the hybrid, taken from calculation of the expected capacitance increase due to the ground plane running above the strips. Good agreement is seen between prototype measurements and capacitive load measurements performed on full chips. An increase in noise is seen between the capacitor measurements and the short strip and long strip modules. A small increase, at the level of a few percent, is expected for such factors as the bias resistance, metal strip resistance and inclusion of next-to-nearest neighbour inter-strip capacitance. The leakage current contribution is negligible before irradiation, but it increases the noise by \unit[5-10]{\%} at full fluence. Finally, a significant increase in noise is seen during irradiation, which is corroborated by single chip measurements and is a TID dependent effect, which has led to a redesign of the front-end for the next ASIC generation.

\begin{figure}
\centering
\begin{subfigure}{.48\textwidth}
 \includegraphics[width=\linewidth]{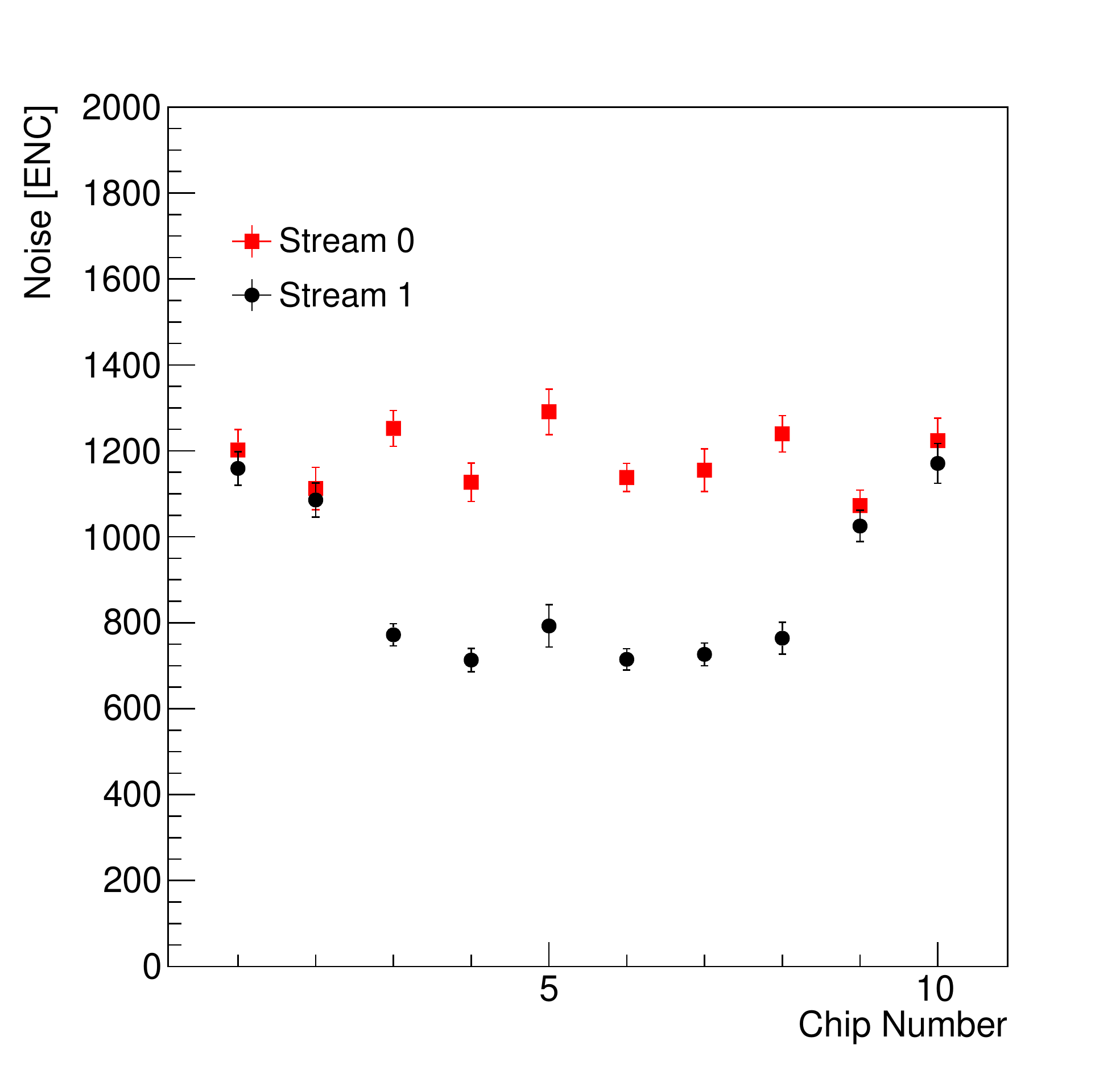}
 \caption{Input noise}
\end{subfigure}
\begin{subfigure}{.48\textwidth}
 \includegraphics[width=\linewidth]{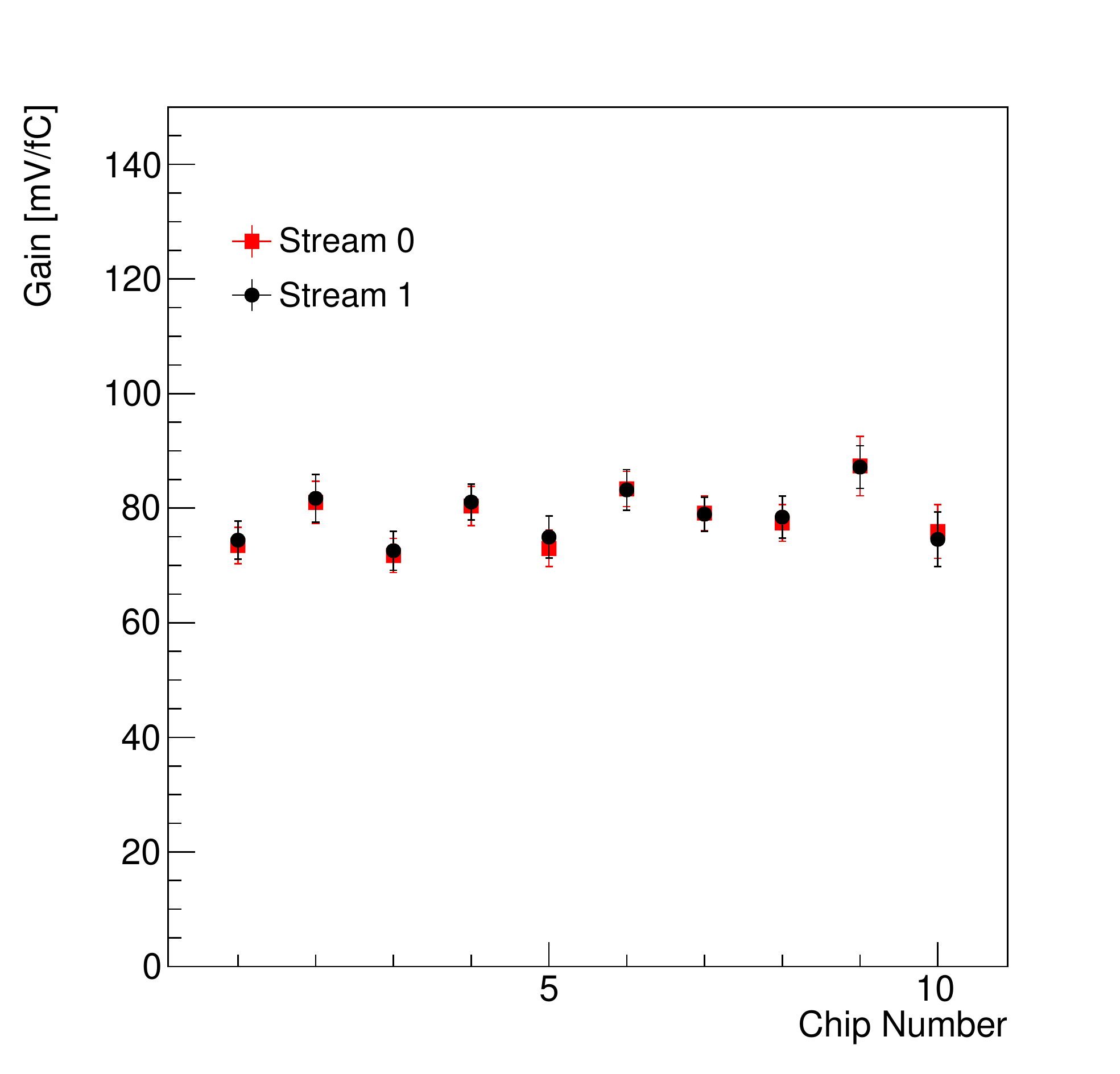}
 \caption{Gain}
\end{subfigure}
\caption{Noise and gain results for a mixed strip module taken from a response curve test at {\unit[1.5]{fC}} injected charge using 192 triggers and strobing all channels simultaneously. Stream 0 channels run under the hybrid and are ganged long strips whilst stream 1 channels run away from the hybrid. Stream 1 channels on the two chips at either end of the hybrid are connected to ganged long strips whilst the six central chips are connected to short strips. Results shown are a chip-by-chip average whilst the error bars show the RMS spread of noise or gain across each chip.}
\label{fig:LS2}
\end{figure}

Figure~\ref{fig:LS2} shows the noise and gain of an early barrel module. This module was built as a long strip module but using a short strip \mbox{ATLAS12} sensor. Short strips were then ganged together in such a way so that all channels running under the hybrid (stream 0) are long strips. In addition, strips running away from the hybrid (stream 1) and connected to the two chips at either end of the hybrid were ganged as long strips with the remaining six chips of channels running away left as short strips. This mixed strip module allows simultaneous characterisation of short and long strip channels at once. Results in this plot were taken at \unit[1.5]{fC} injection charge during a response curve test of 192 triggers and strobing all channels at once.

Firstly, the increased noise associates to channels connected to long strips can immediately be observed comparing the two streams on the six central chips. In addition, the extra \unit[30-50]{ENC} due to the presence of the hybrid above strips can be seen looking at the two chips at either end, which have long strips connected to both streams but show a slightly increased noise on those channels running under the hybrid. Finally, the gain results demonstrate that the measured gain is independent of the connected load.

\subsection{Noise occupancy results}

\begin{figure}
\centering
\begin{subfigure}{.75\textwidth}
 \centering
 \includegraphics[width=\linewidth,trim={0 0 0 0.75cm},clip]{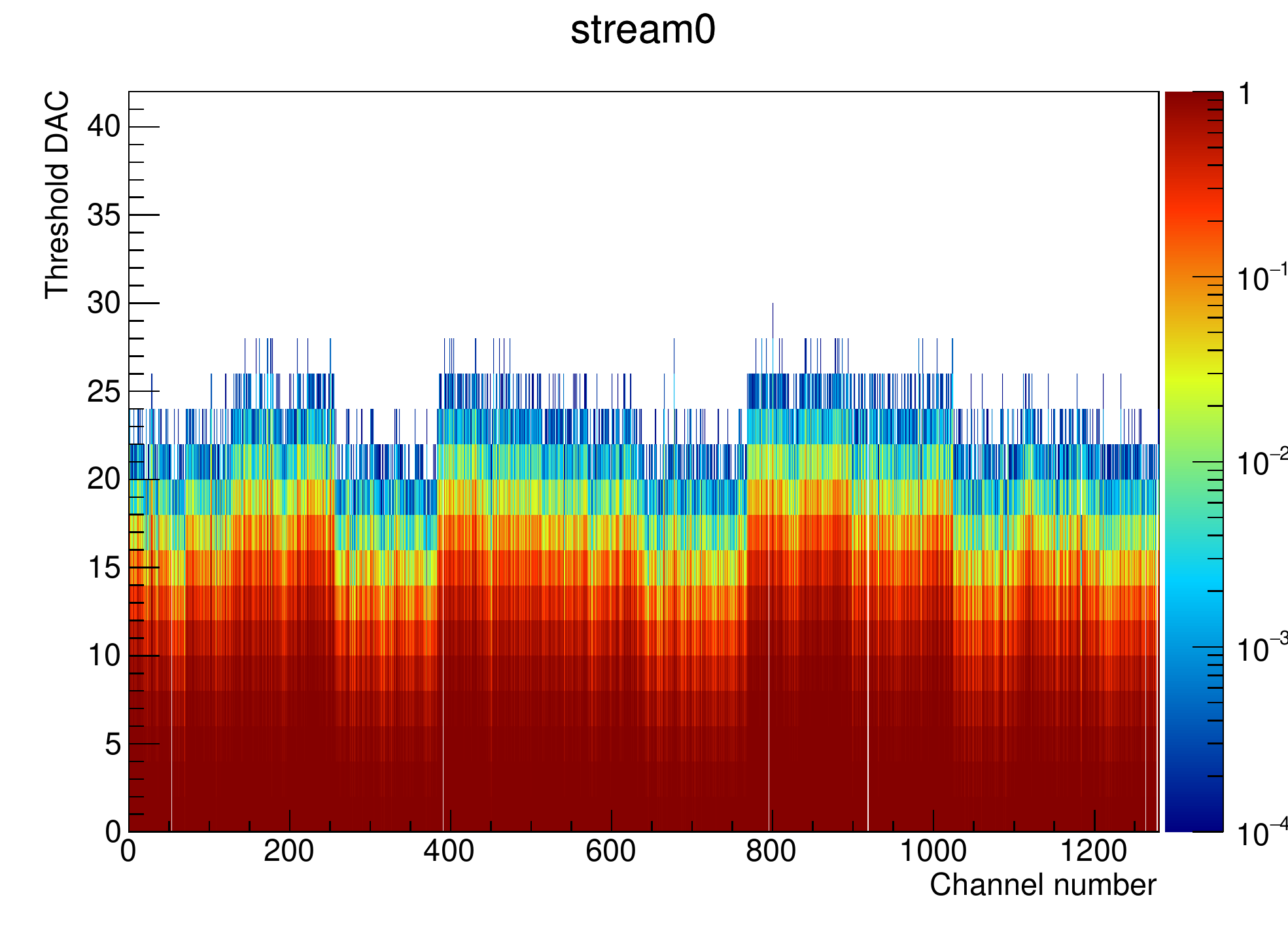}
 \caption{{\unit[2.5]{cm}} strips under hybrid}
 \label{fig:NOresult1}
\end{subfigure}
\begin{subfigure}{.75\textwidth}
 \centering
 \includegraphics[width=\linewidth,trim={0 0 0 0.75cm},clip]{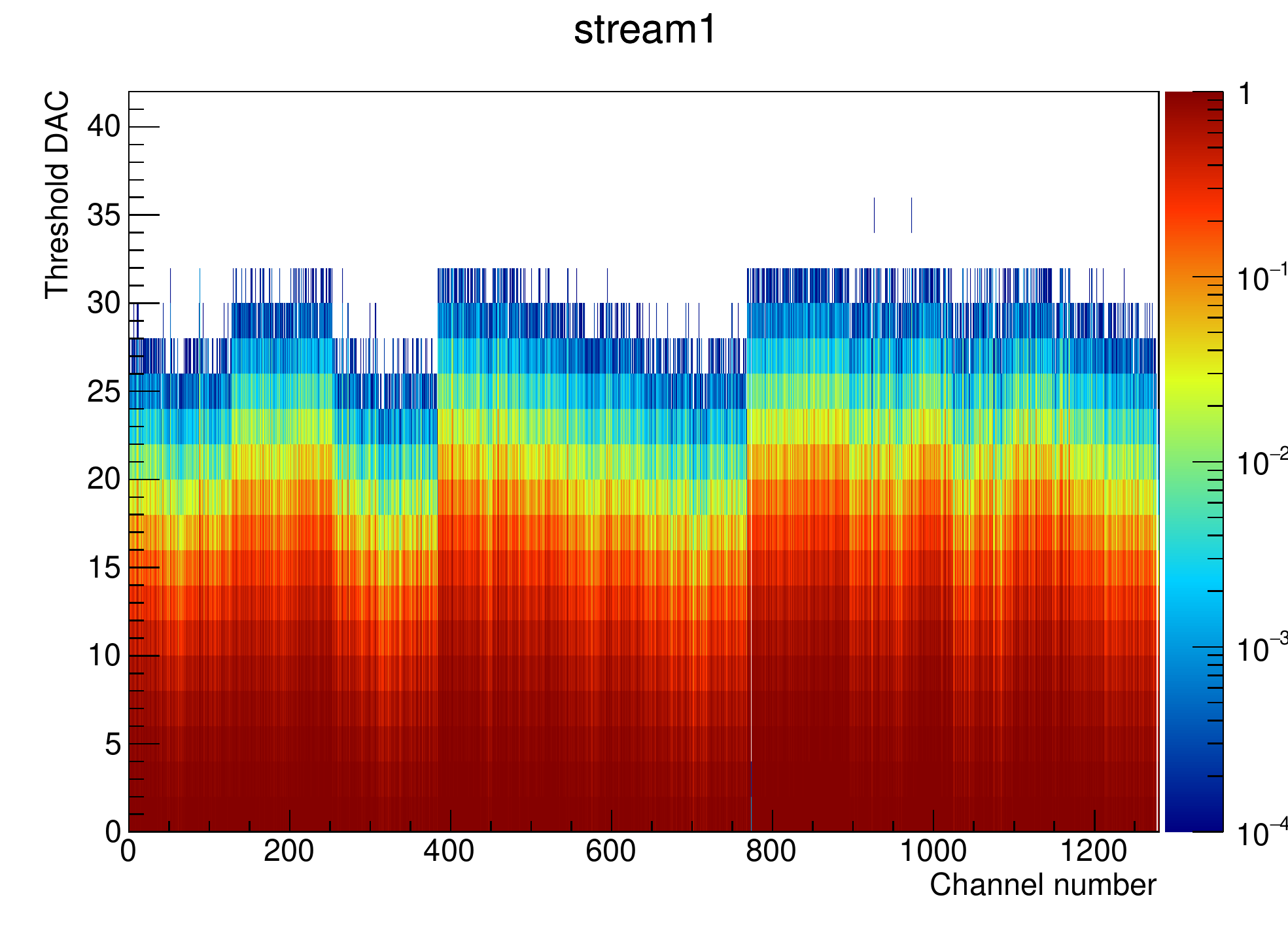}
 \caption{{\unit[5.0]{cm}} strips away from hybrid}
 \label{fig:NOresult2}
\end{subfigure}
\caption{Noise occupancy as a function of threshold setting and channel number on a mixed strip length module. The module has been trimmed such that the response of each channel is equalised for the case where {\unit[1]{fC}} of charge is injected, corresponding to a threshold DAC value of approximately 40.}
\label{fig:NOresults}
\end{figure}

Noise occupancy scans were run on a module in test beam with mixed strip lengths. Strips running under the hybrid were short strips whilst strips running away from the hybrid were ganged together as long strips. Noise occupancy scans without external charge injection are shown in figures~\ref{fig:NOresult1} and~\ref{fig:NOresult2}. These results were taken with the module trimmed to 1fC injected charge. Results are shown as a function of the 8-bit register setting of the threshold.

\begin{figure}
\centering
\includegraphics[width=0.8\linewidth]{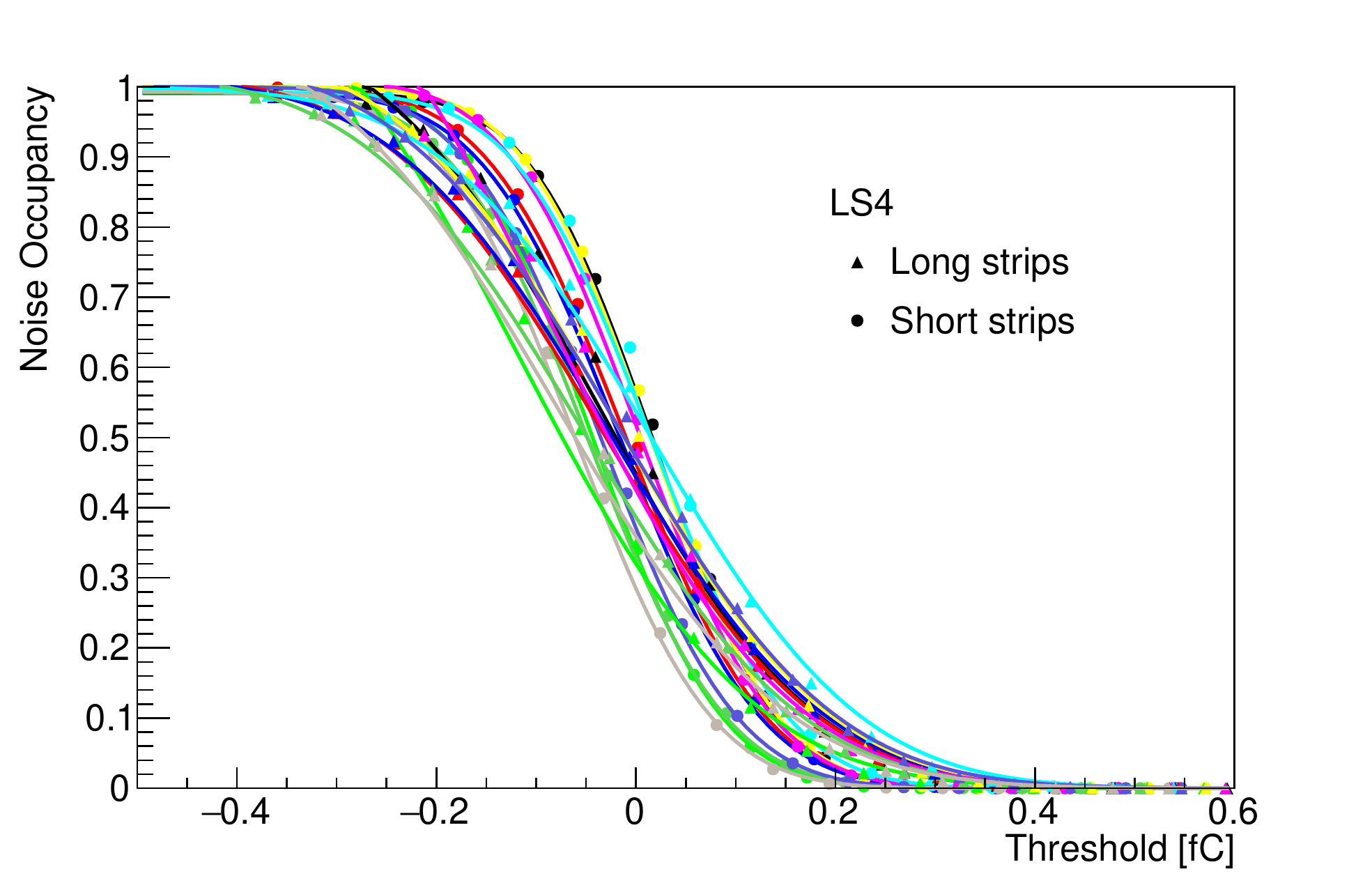}
\caption{Noise occupancy against threshold on a mixed strip length module. Results are shown chip-by-chip as an average over the connected channels and separated into long strips and short strips (128 channels of each per chip). The results are fitted with a complementary error function. The {\unit[50]{\%}} noise occupancy point is centred around {\unit[0]{fC}} by construction as it is the {\unit[50]{\%}} point, which is calibrated in the response curve analysis from which the charge calibration is derived.}
\label{fig:NO_ERFC}
\end{figure}

Chip-by-chip averages across 128 channels per chip are shown in figure~\ref{fig:NO_ERFC}. Threshold settings in DAC have been converted to mV using calibrations from simulation whilst conversion from threshold voltage to charge have been taken from the chip-by-chip response curve results obtained using the internal charge injection circuitry using 1024 triggers and strobing 16 channels at a time. The chip-by-chip results are then fitted with the complementary error function, the width of which can be used to extract the input noise of the system.

\begin{figure}
\centering
\includegraphics[width=0.8\linewidth]{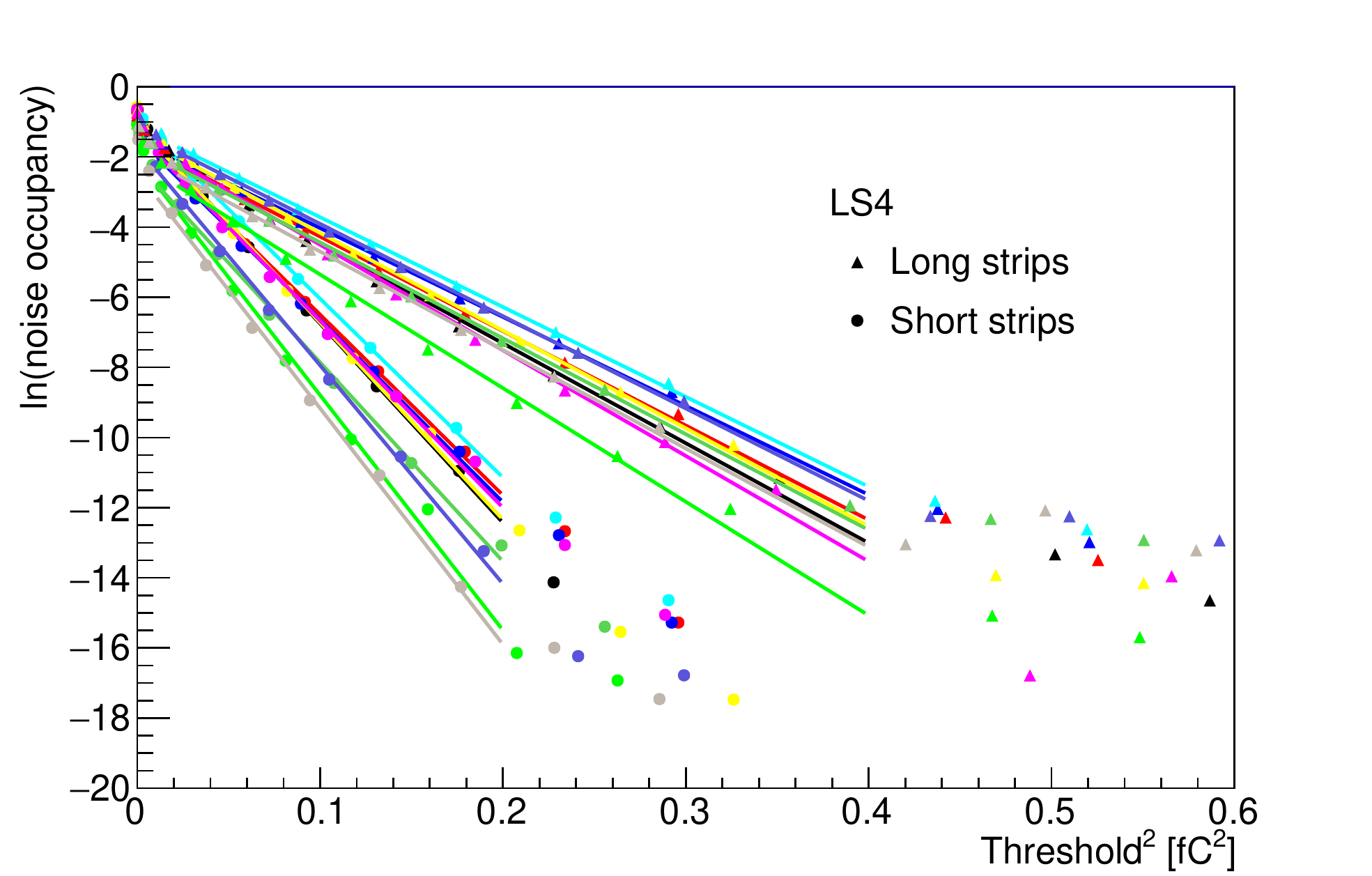}
\caption{The natural logarithm of noise occupancy against threshold squared on a mixed strip length module. The results are fitted with linear functions. Two populations of results can be clearly seen, arising from those channels connected to short strips and those to long strips. The linear agreement shown demonstrates the consistency of the tails of the noise occupancy with a Gaussian shape.}
\label{fig:NO_lnOcc}
\end{figure}

Assuming that the shape of the noise occupancy is Gaussian, the tail of the noise occupancy can also be used to extract noise from this data by plotting the natural logarithm of the noise occupancy versus the square of the threshold. The result of this is shown in figure~\ref{fig:NO_lnOcc}. As expected, the chip-by-chip averages clearly separate into two populations dependent on long-strip or short-strip. It can also be seen to a very good approximation that the tails of the noise occupancy are indeed Gaussian as demonstrated by the linear trends shown down to very low noise occupancy. In addition, the gradient of linear fits to this data can be used to extract the measured noise of the system.

\subsection{Comparison between Noise Occupancy and Three Point Gain}

Three methods by which noise can be extracted from module testing results have been shown above:
\begin{itemize}
 \item that extracted from internal charge injection measurements (extracted at \unit[1.5]{fC} from a response curve)
 \item that extracted from a complementary error function fit of the noise occupancy data
 \item that extracted from a linear fit to the natural logarithm of occupancy versus the square of threshold
\end{itemize}
The latter two, although based on the same data are driven by different parts of the distributions as the complementary error function extraction is dominated by the core of the noise occupancy S-curve whilst the linear fit is driven by the shape of the tail of the S-curves.

\begin{figure} 
\centering
\includegraphics[width=0.8\linewidth]{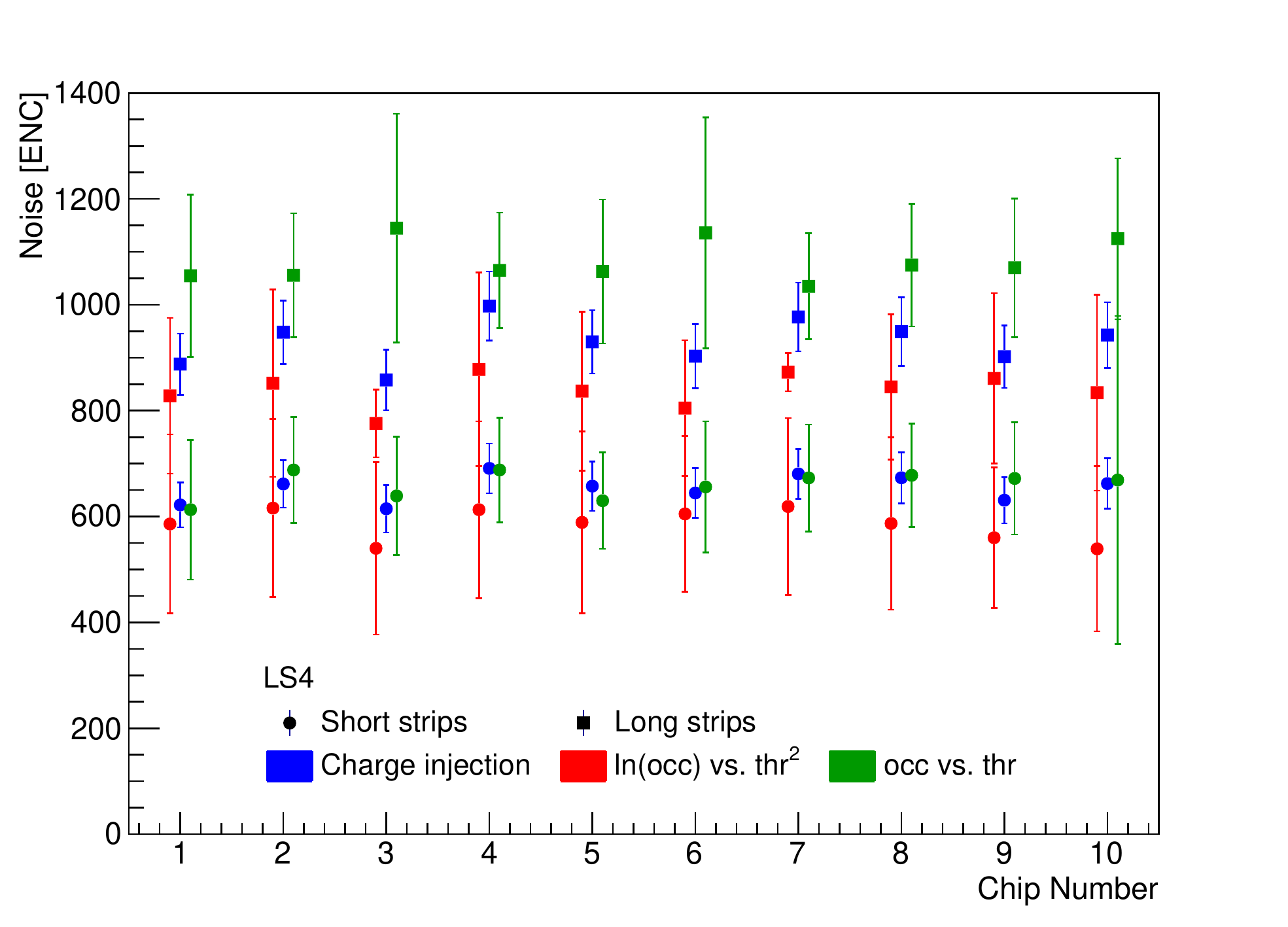}
\caption{A comparison of different noise extraction techniques used on a mixed strip length module. Results are shown for short and long strips. The error bars shown arise from the statistical errors on the fits used to extract the results.}
\label{fig:CompareNoise}
\end{figure}

Figure~\ref{fig:CompareNoise} shows a comparison of the noise extracted from the three methods. For the short strips on the module, all three methods agree very well whilst in the long strip case the complementary error function analysis tends to return a larger noise than the other two methods, although within the statistical error bars shown the three methods generally agree everywhere.

\subsection{Sensor hysteresis}

\begin{figure}
\centering
\begin{subfigure}{.8\textwidth}
 \centering
 \includegraphics[width=\linewidth]{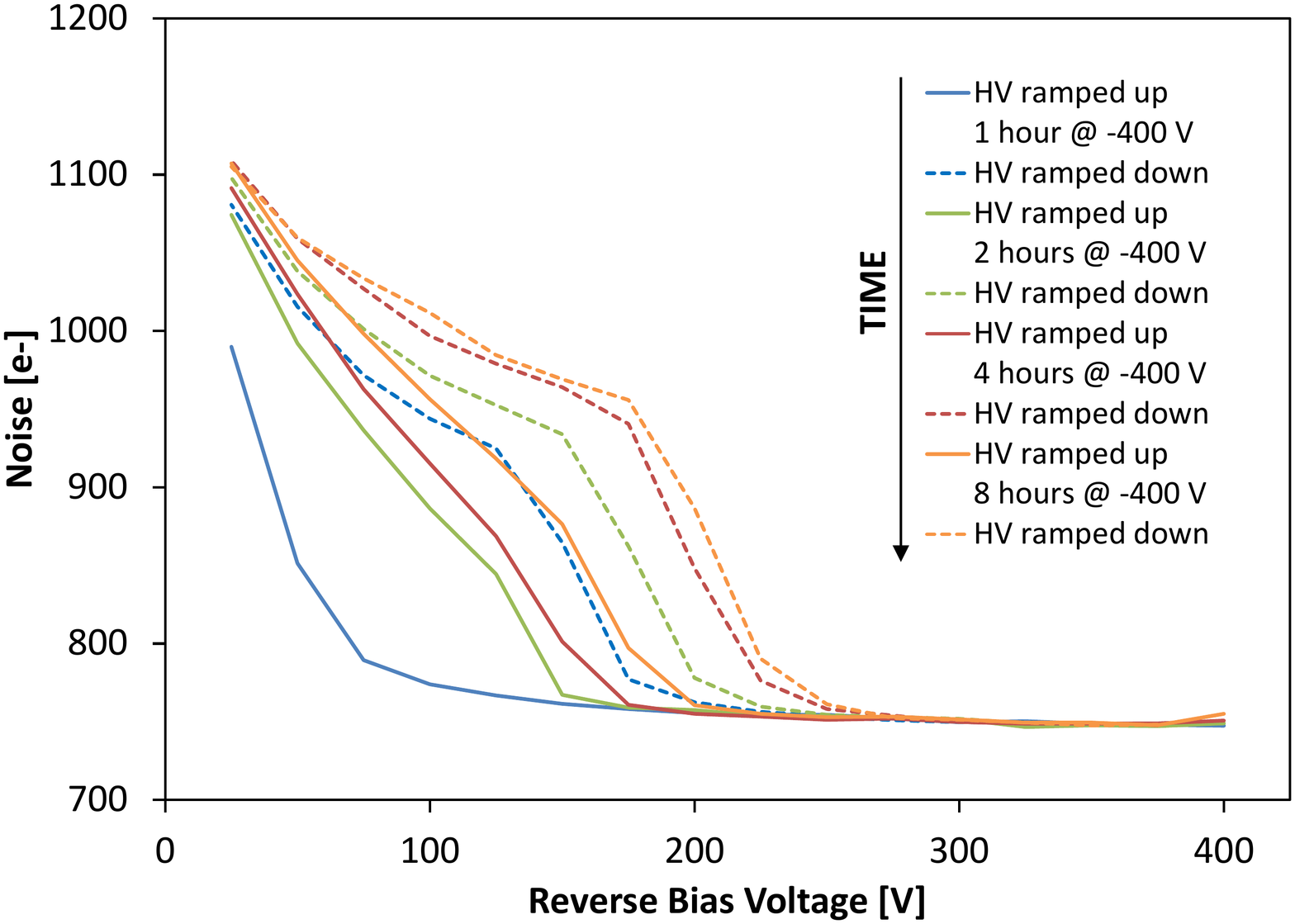}
 \caption{Hysteresis effect of ramping the sensor up and down with varying lengths of time holding the sensor reverse biased at \unit[-400]{V}.}
 \label{fig:hysteresis_1}
\end{subfigure}
\begin{subfigure}{.8\textwidth}
 \centering
 \includegraphics[width=\linewidth]{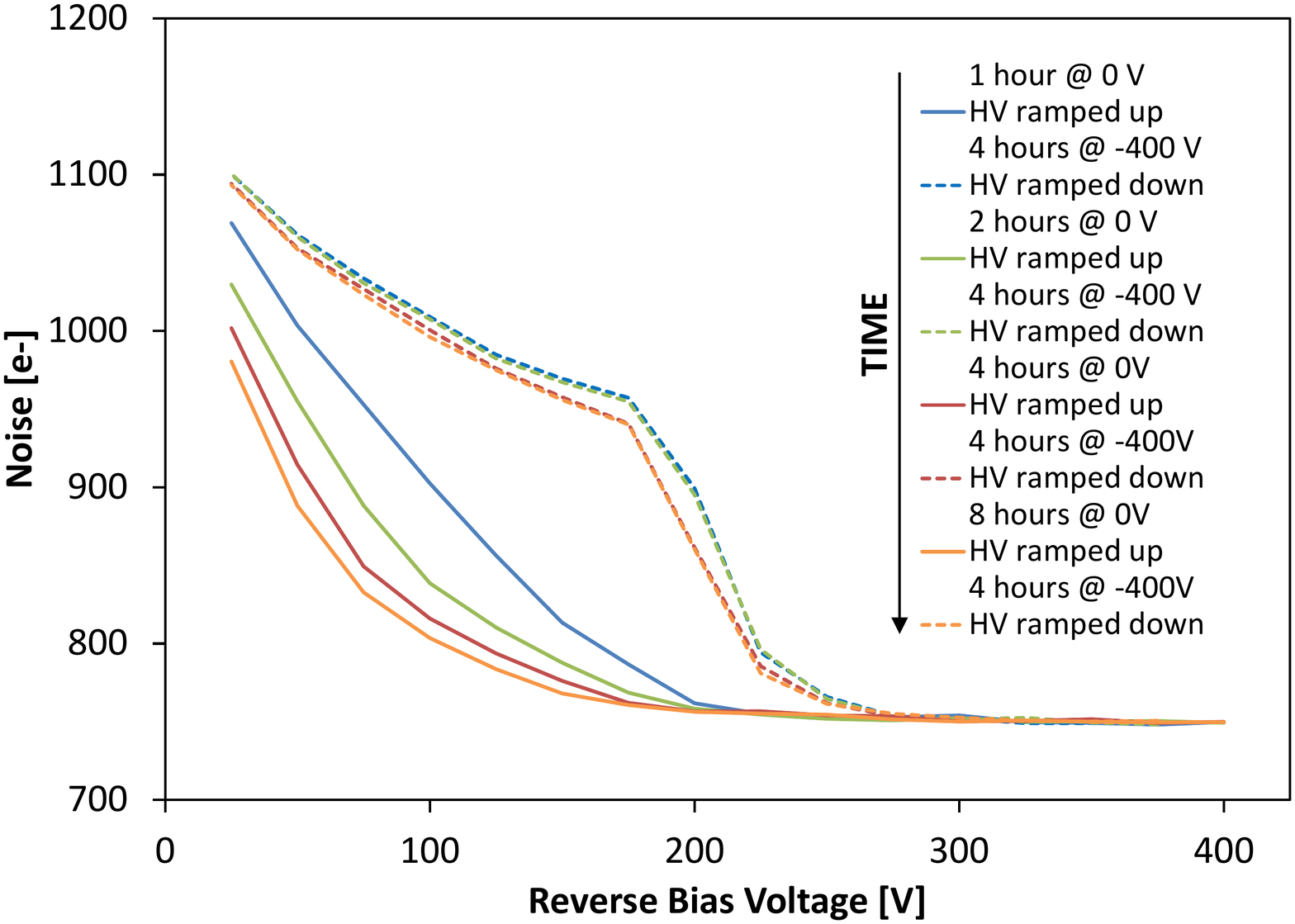}
 \caption{Hysteresis effect of changing the length of time for which the sensor was held unbiased between tests.}
 \label{fig:hysteresis_2}
\end{subfigure}
\caption{Input noise versus reverse bias voltage as a function of the bias voltage history of the module. Before these tests, the sensor was held unbiased for at least \unit[24]{hours}. In both cases, the sensor was held at \unit[-400]{V} for four hours. The bias voltage history is shown in the legend on each plot.}
\label{fig:hysteresis}
\end{figure}

Figure~\ref{fig:hysteresis} shows the result of running internal charge injection tests when changing the reverse bias voltage of a short strip module. The applied reverse bias voltage was changed in \unit[25]{V} steps from \unit[-25 to -400]{V}, where the full depletion voltage for this sensor is \unit[-370]{V}. The bias ramp speed was \unit[2]{V/s} and tests were run immediately after the bias voltage was reached. In the first test, the bias voltage was ramped down immediately after ramping up. The time the sensor was held at \unit[-400]{V} was changed from one to eight hours. As can be seen from the plots, the shapes of both the upward and downward ramp change depended on the biasing history of the sensor. In the second test, the bias voltage was held at \unit[-400]{V} for four hours, and the length of time it was kept unbiased was varied from zero to eight hours. As can be seen, the shape of the downward ramp is independent but the upward ramp shape strongly depends on the time the sensor was left unbiased. Note that prior to the first test, the sensor had been unbiased for more than \unit[24]{hours}.

These variations are attributed to the increase in inter-strip capacitance (C$_{\text{int}}$), which dominates the preamplifier input noise, for reverse bias voltages below the set bias voltage. The effect is described in detail in~\cite{thesis_CK}, and is probably caused by localised charge build-up in the surface layers of the sensor. Prolonged biasing of the sensor exacerbates the effect, and hence when ramping down, the noise levels increase compared to those measured when ramping up the bias voltage.

\subsection{EMI pick up studies \label{sec:selectModule:Pickup}}

\begin{figure}[t]
\centering
\includegraphics[width=0.6\linewidth]{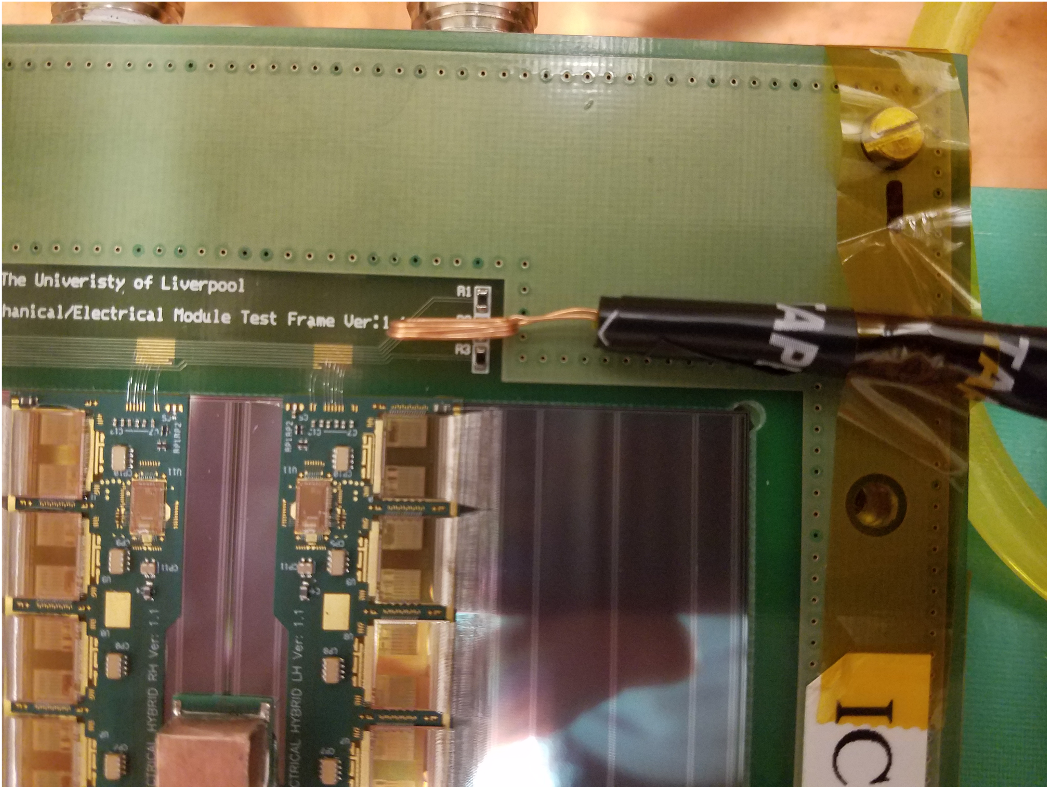}
\caption{The external coil and its orientation relative to the signal wire bonds that was used for the EMI pick up studies.}
\label{fig:EMI:CoilSetup}
\end{figure}

The effect of induced noise from electromagnetic fields generated by the powerboard components
was studied in more detail. In a first investigation, the module susceptibility to an EMI agressor was tested as a function of both distance and frequency.
The module was powered by a powerboard glued onto the sensor surface in its nominal position between the two hybrids.
Noise was intentionally induced using an external coil of the same type and dimension as the one used on powerboards.
The coil was placed with its flux aimed perpendicular to the directions of the signal wire bonds of one of the ABC130 chips, as shown in Figure~\ref{fig:EMI:CoilSetup}.
A frequency generator provided the input with an input power of \unit[-15]{dBm}.

In the first configuration, the coil was placed $\sim$~\unit[1]{cm} from the sides of the ABC130 wire bonds, and the frequency was scanned between 5 and \unit[50]{MHz}.
A three point gain test was run for each frequency step, and the mean chip noise was calculated for the exposed chip.
In addition, prior to turning on the frequency generator, a three-point gain test was run to establish a baseline noise.
The induced noise from the coil was assumed to be uncorrelated from existing noise sources without the coil, and therefore the noise from the coil was calculated by subtracting in quadrature with the baseline noise values:

\begin{equation}
\sigma_{\text{EMI}} = \pm\sqrt{\sigma_{\text{coil}}^{2} - \sigma_{\text{baseline}}^2}
\end{equation}

\noindent
where the sign is positive in cases where the baseline noise was less than the induced noise from the coil and negative otherwise;
the latter being possible in cases where very little noise was induced into the module, and due to statistical fluctuations the noise measured with the coil was less than the baseline.

\begin{figure}[ht]
\begin{subfigure}{.48\linewidth}
\centering
\includegraphics[width=\linewidth]{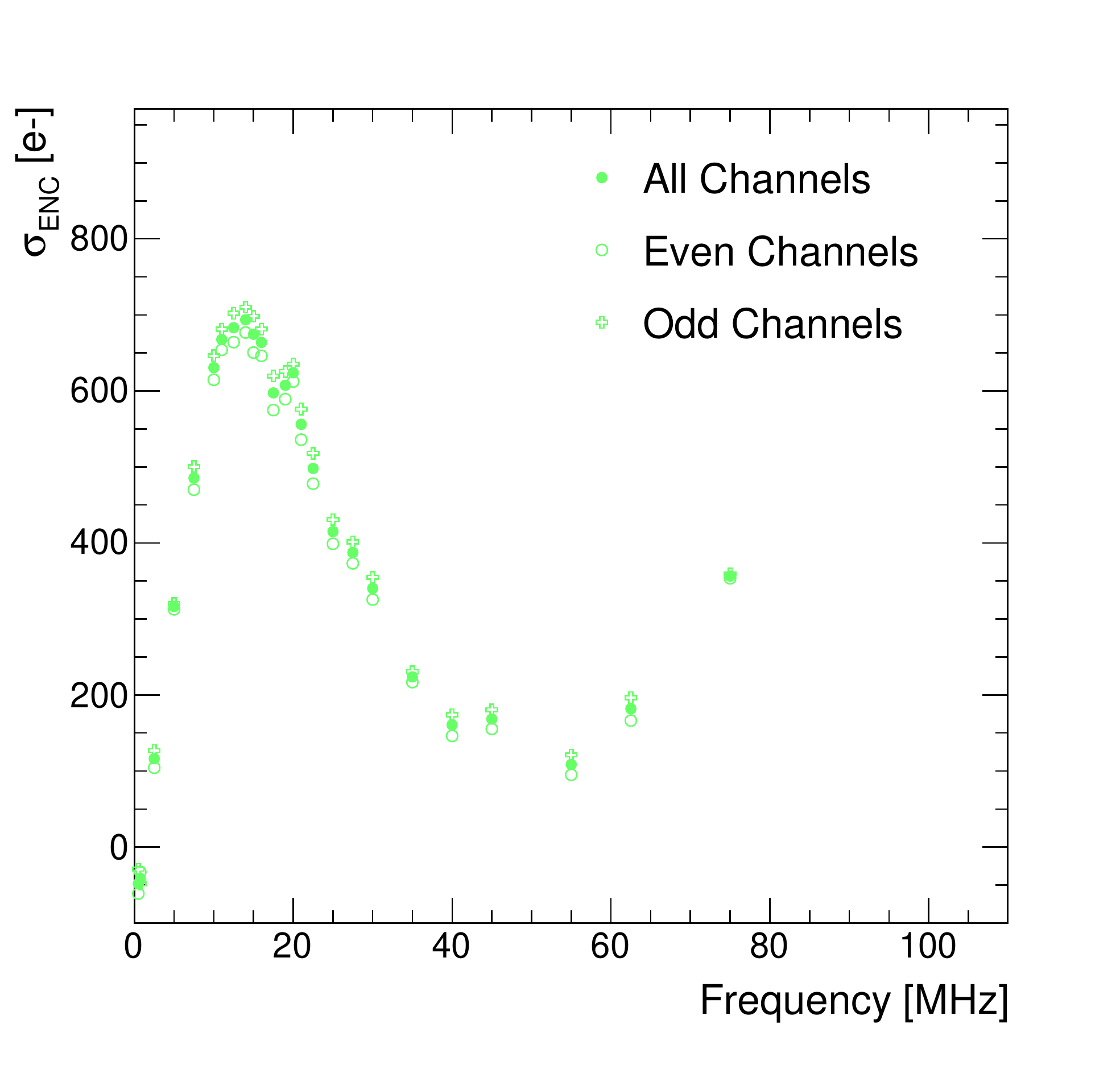}
\end{subfigure}
\begin{subfigure}{.48\linewidth}
\centering
\includegraphics[width=\linewidth]{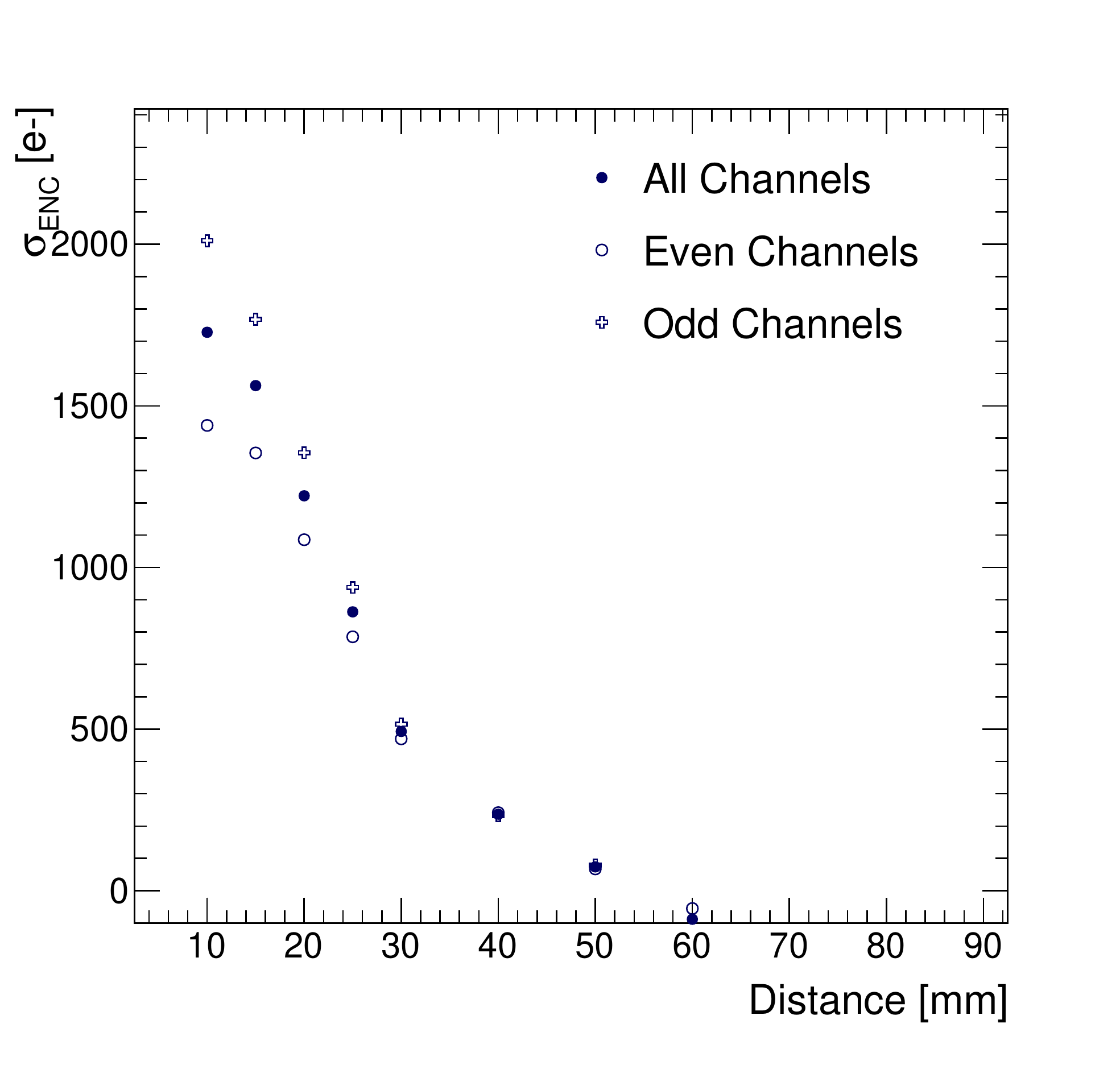}
\end{subfigure}
\caption{Measured noise values as a function of the frequency of applied EMI fields, and distance from the external powerboard coil.}
\label{fig:selectModule:pickup_bandpass}
\end{figure}

As shown in figure~\ref{fig:selectModule:pickup_bandpass}, the increase in noise is most pronounced for frequencies in the range of 10 to \unit[25]{MHz},
and has a strong dependence on proximity to the module. 
Noise values are shown separately for even and odd channels, corresponding to inner and outer signal bonds. 
For distances beyond~\unit[50]{mm}, the induced noise from the coil falls to zero. 
The general shape of the frequency dependence reflects the ABC130 front-end amplifier bandpass. 
The top wirebond rows have higher pickup than the bottom ones, indicating either an incomplete screening effect, or dependence on the wirebond loop area.

\begin{figure}
 \centering
\begin{subfigure}{.85\textwidth}
 \centering
 \includegraphics[width=\linewidth]{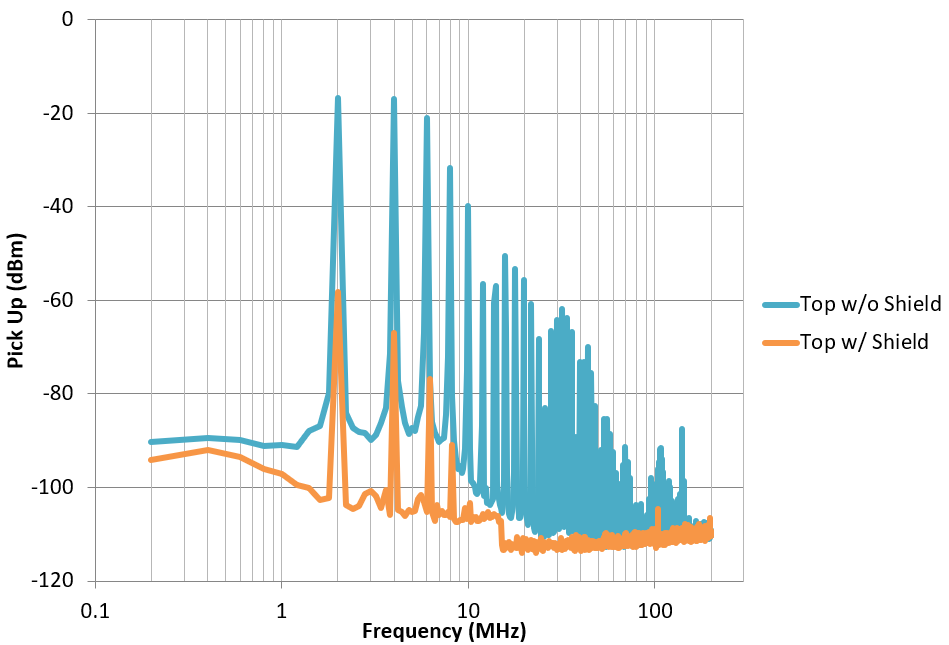}
 \caption{Measurements of powerboard EMI power spectrum with an open loop probe with and without the shield box.}
 \label{fig:EMIattenuation}
\end{subfigure}
\begin{subfigure}{.90\textwidth}
 \centering
 \includegraphics[width=\linewidth]{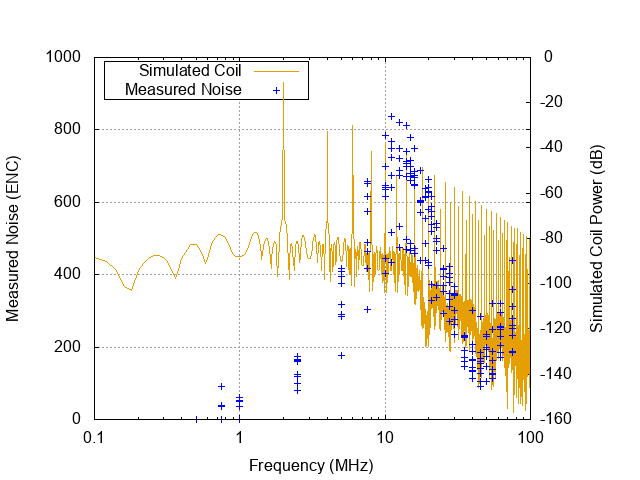}
  \caption{Simulated power spectrum from the coil in the powerboard circuit and the module noise pickup.}
 \label{fig:SPICE_EMI}
\end{subfigure}
\caption{Powerboard-induced EMI measurements and simulations as a function of frequency, illustrating the effect of the powerboard shield (a) and the presence of EMI harmonics in the bandpass of the ABC130 amplifiers (b).}
\end{figure}

The EMI spectrum from the powerboard was measured with an open loop probe and a network analyzer, both for the board with and without the shieldbox (figure~\ref{fig:EMIattenuation}). The primary emission spectrum has peaks descending with frequency, which are harmonics of the \unit[2]{MHz} driving frequency. The shield box provides an attenuation that grows with frequency, consistent with the skin effect properties. The attenuation is direction-dependent. It is about \unit[60]{dB} upwards and downwards of the powerboards, and about \unit[25]{dB} on the side. Although the \unit[2]{MHz} emission is significant, there is a strong reduction of the emission at the peak of the amplifier's acceptance. The general features of the spectrum have been simulated in SPICE using a model with a buck converter, as shown in figure~\ref{fig:SPICE_EMI}, where the measured EMI noise on a module is super-imposed above the simulated power spectrum of the coil. 
The measurements confirm the existence of many EMI emission harmonics in the front-end amplifier's bandpass, which can contribute to the readout noise. The strong attenuation provided by the shield box is essential for the unusual placement of the DC-DC converter directly on the module in close proximity to both sensor strips and the front-end circuitry.

\subsection{Effect of powerboard on module noise \label{sec:selectModule:PBEffect} }

\begin{figure}
\centering
\includegraphics[width=\linewidth]{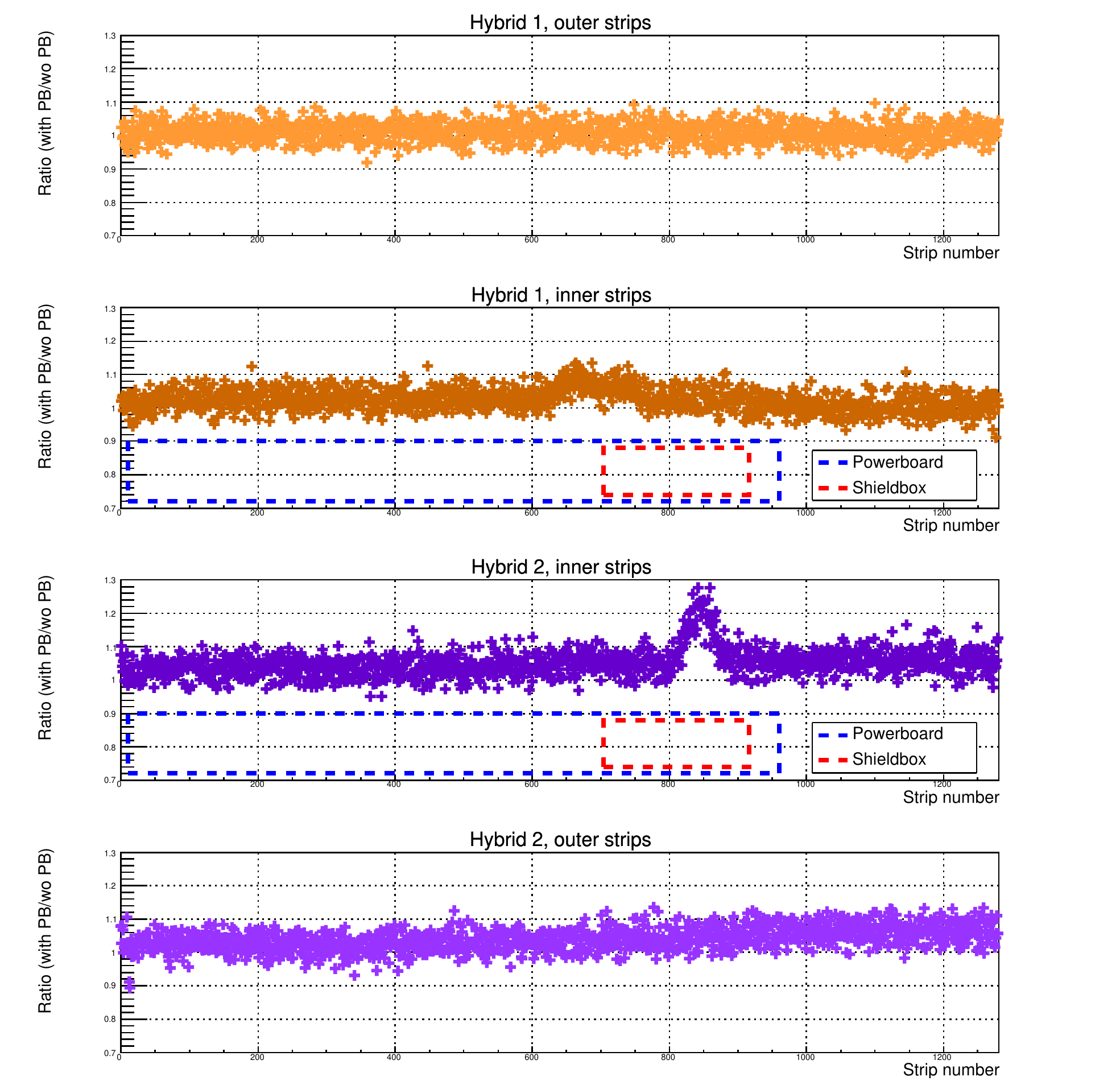}
\caption{Strip-by-strip ratios of channel noise with and without attached powerboard, per sensor segment. While the outer sensor strips show a mostly flat distribution of noise ratios, the inner two segments, which are partially covered by the powerboard, show an increased noise after powerboard attachment in the vicinity of the shield box. Hybrid 2 shows a distinct peak next to the shield box, hybrid 1 shows a flat increase next to it.}
\label{fig:PBnoise1}
\end{figure}

During the ABC130 barrel module program, each module was tested twice: once after hybrids were mounted, by powering hybrids directly, and again after its powerboard had been mounted, by powering hybrids through the powerboard as intended in the detector. While the additional test was performed to attribute causes of electrical failures during testing to either powerboard or hybrids, the repeated measurements allowed for a comparison of the module performance before and after its powerboard was mounted.
Figure~\ref{fig:PBnoise1} shows the ratio of noise measured before and after powerboard attachment for each sensor strip on all of the four strip segments.

A similar pattern was found for most of the produced barrel modules: while the outer sensor segments, which are not covered by the powerboard, showed mostly flat noise distributions, inner module segments, which are partially covered by the powerboard, consistently showed an increase in the vicinity of the shield box.

\begin{figure}
\centering
\includegraphics[width=\linewidth]{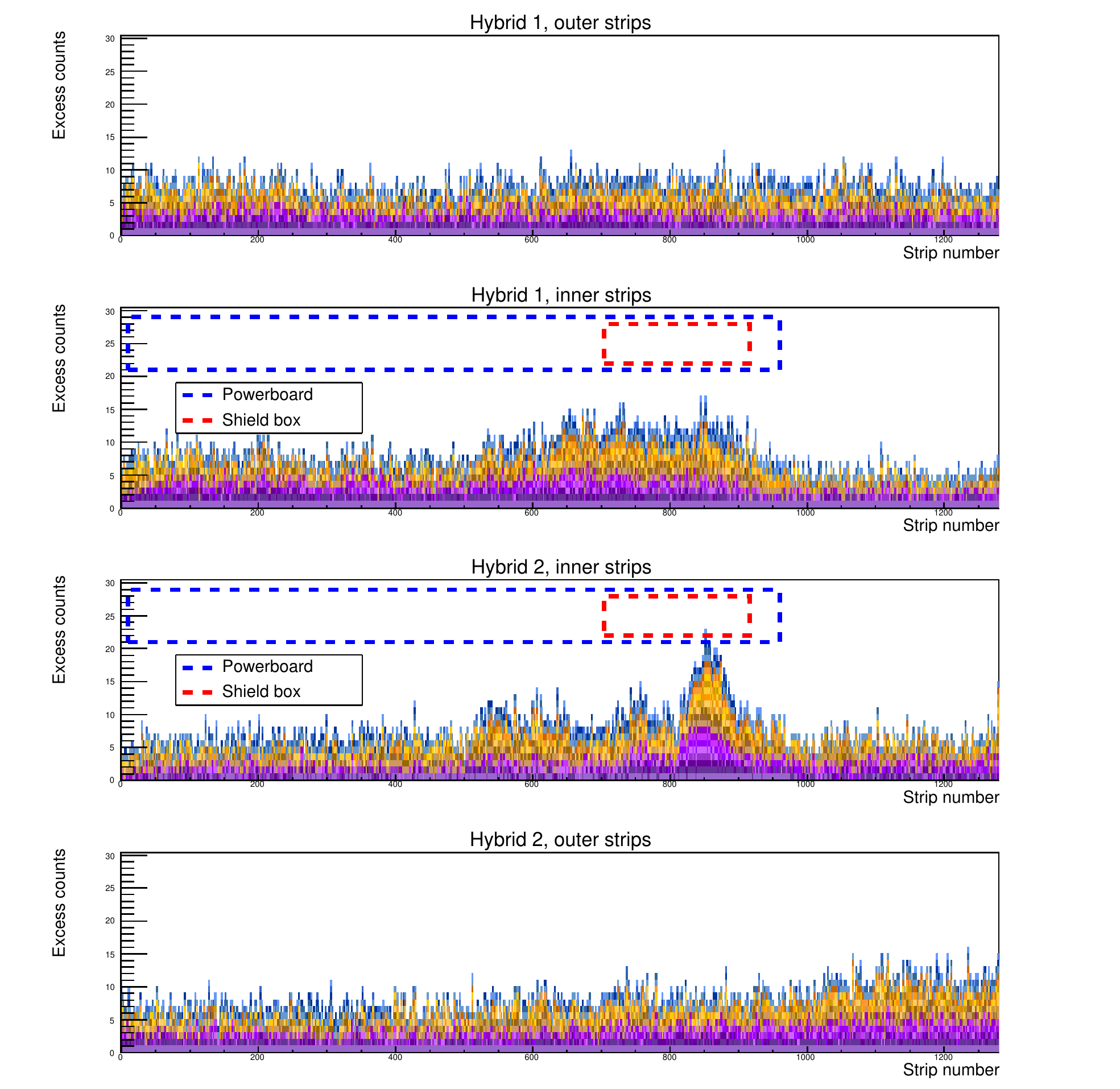}
\caption{Map of module channels which showed an excess $(\unit[> 2.5]{\%})$ compared to the average noise ratio on that strip segment. Blue, orange and violet indicate different module production sites, with each colour variation corresponding to contributions from one individual module. Combined results from different module production sites show a similar distribution of noise increase as an individual module (see figure~\ref{fig:PBnoise2}). Inner sensor segments show an increased noise around the sensor shield box, while outer rows show only randomly distributed noise.}
\label{fig:PBnoise2}
\end{figure}

In order to compare the noise increase after powerboard attachment, module test results from all produced modules were combined. Since the absolute ratio depends on the environmental conditions during testing, test results from different modules were not combined using the absolute calculated noise ratios. Instead, an average noise ratio was determined for each strip segment in each test using a linear fit. The calculated average was used to determine, on a strip-by-strip base, channels where the calculated noise ratio was more than \unit[2.5]{\%} higher than the average ratio for that strip segment. Each channel which exceeded this threshold was added to a map, which was then compared to the position of the powerboard on the modules (see figure~\ref{fig:PBnoise2}).
The combination of test results from different modules and construction sites confirmed that the observed noise increase was a systematic effect caused by the presence of the powerboard.

Based on these measurements, the powerboard design was modified to increase its shielding properties.

\subsection{Irradiated modules}

\label{subsec:irradModules}

\begin{figure}
 \centering

\begin{subfigure}{.48\textwidth}
 \centering
 \includegraphics[width=\linewidth]{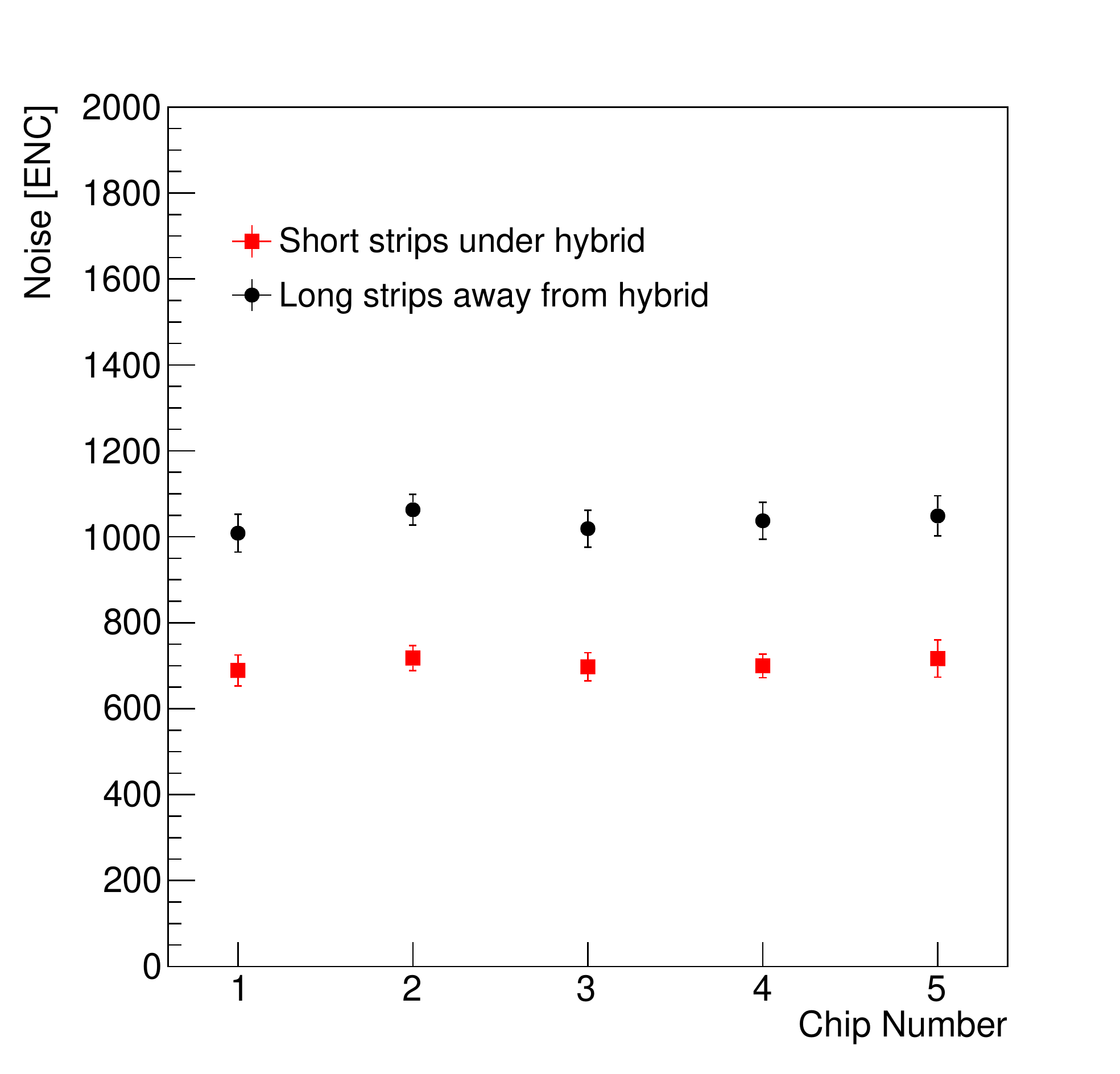}
 \caption{Noise before irradiation}
 \label{fig:irrad2a}
\end{subfigure}
\begin{subfigure}{.48\textwidth}
 \centering
 \includegraphics[width=\linewidth]{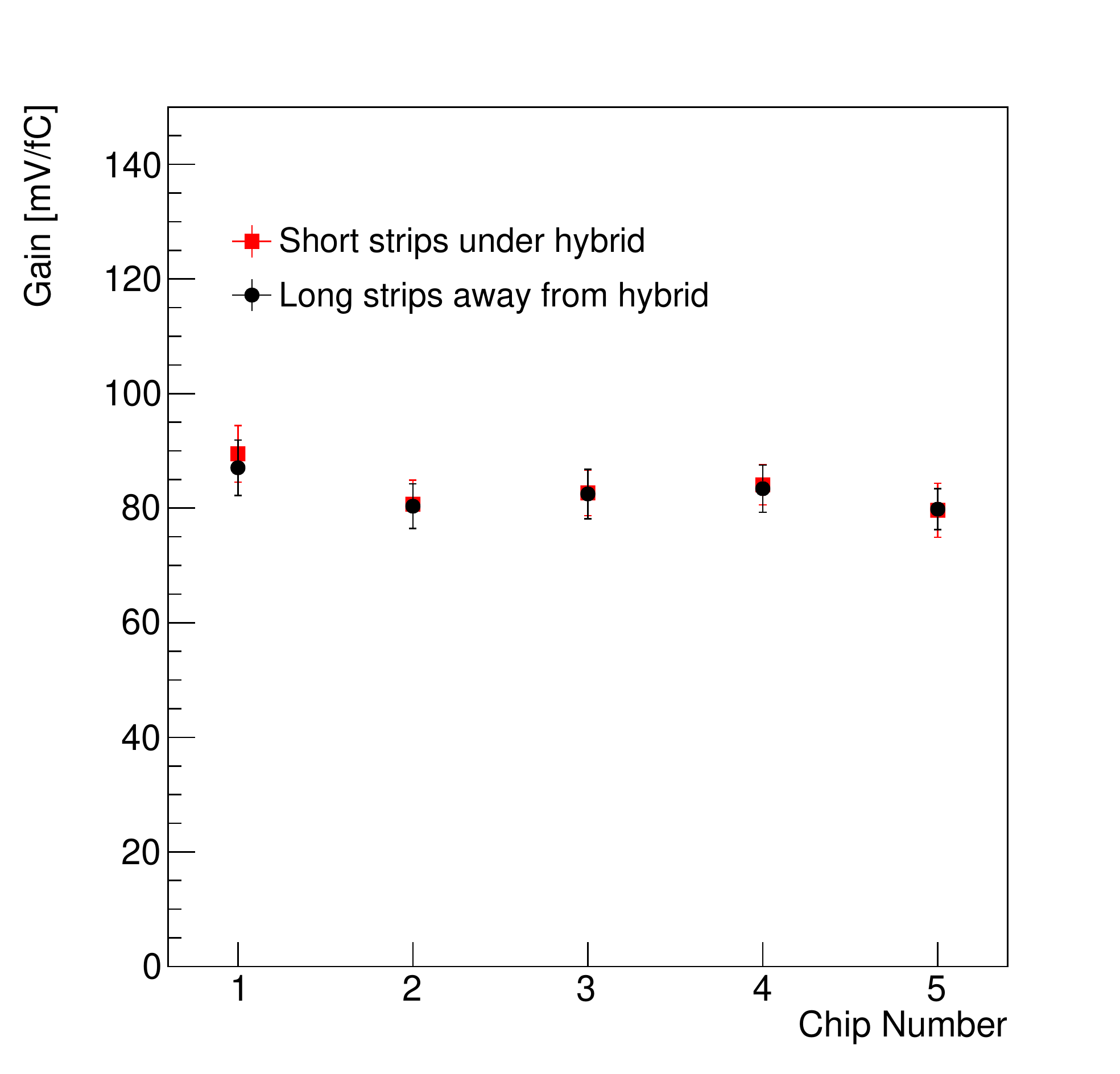}
 \caption{Gain before irradiation}
 \label{fig:irrad2b}
\end{subfigure}
\begin{subfigure}{.48\textwidth}
 \centering
 \includegraphics[width=\linewidth]{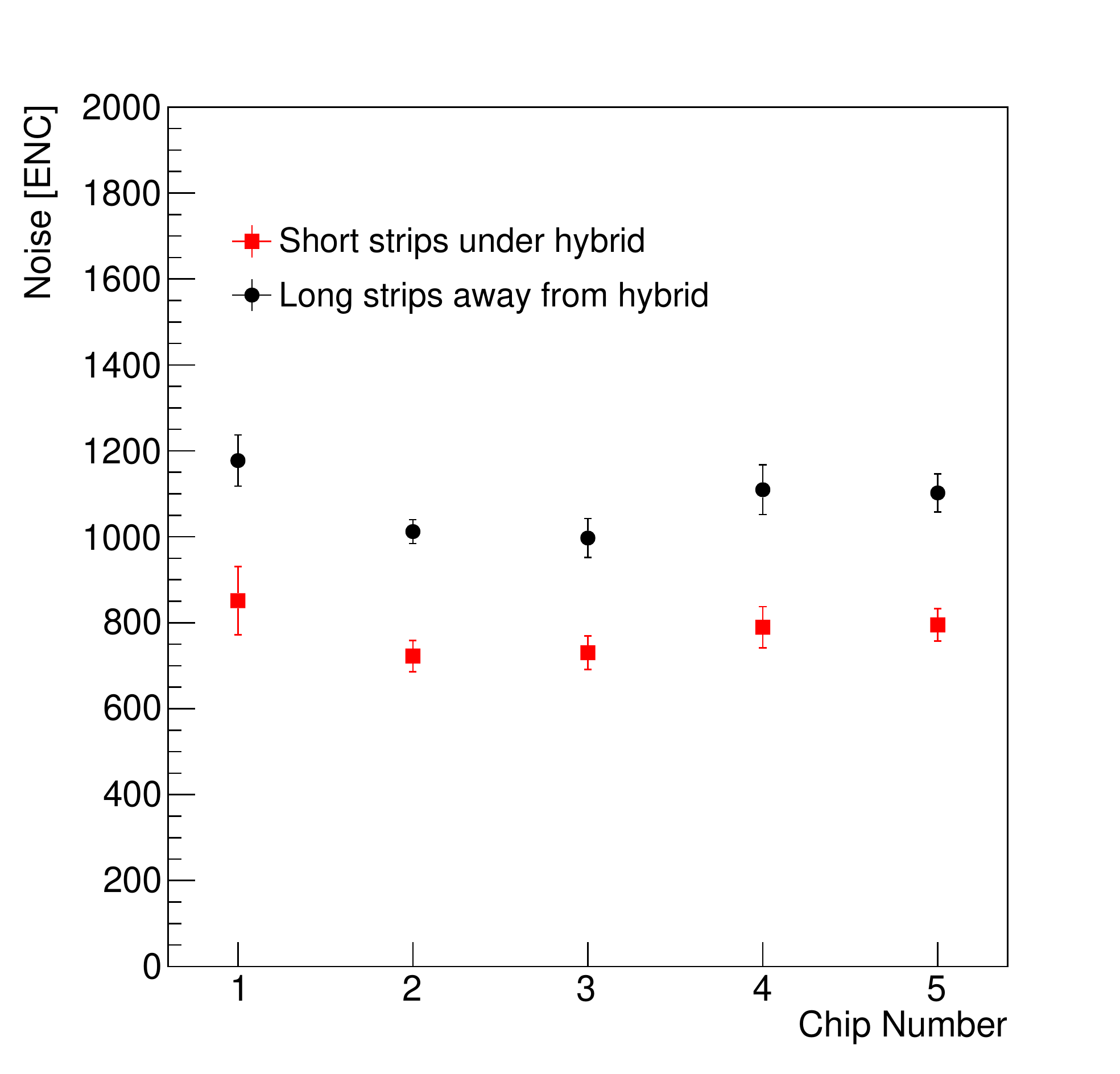}
 \caption{Noise after irradiation}
 \label{fig:irrad3a}
\end{subfigure}
\begin{subfigure}{.48\textwidth}
 \centering
 \includegraphics[width=\linewidth]{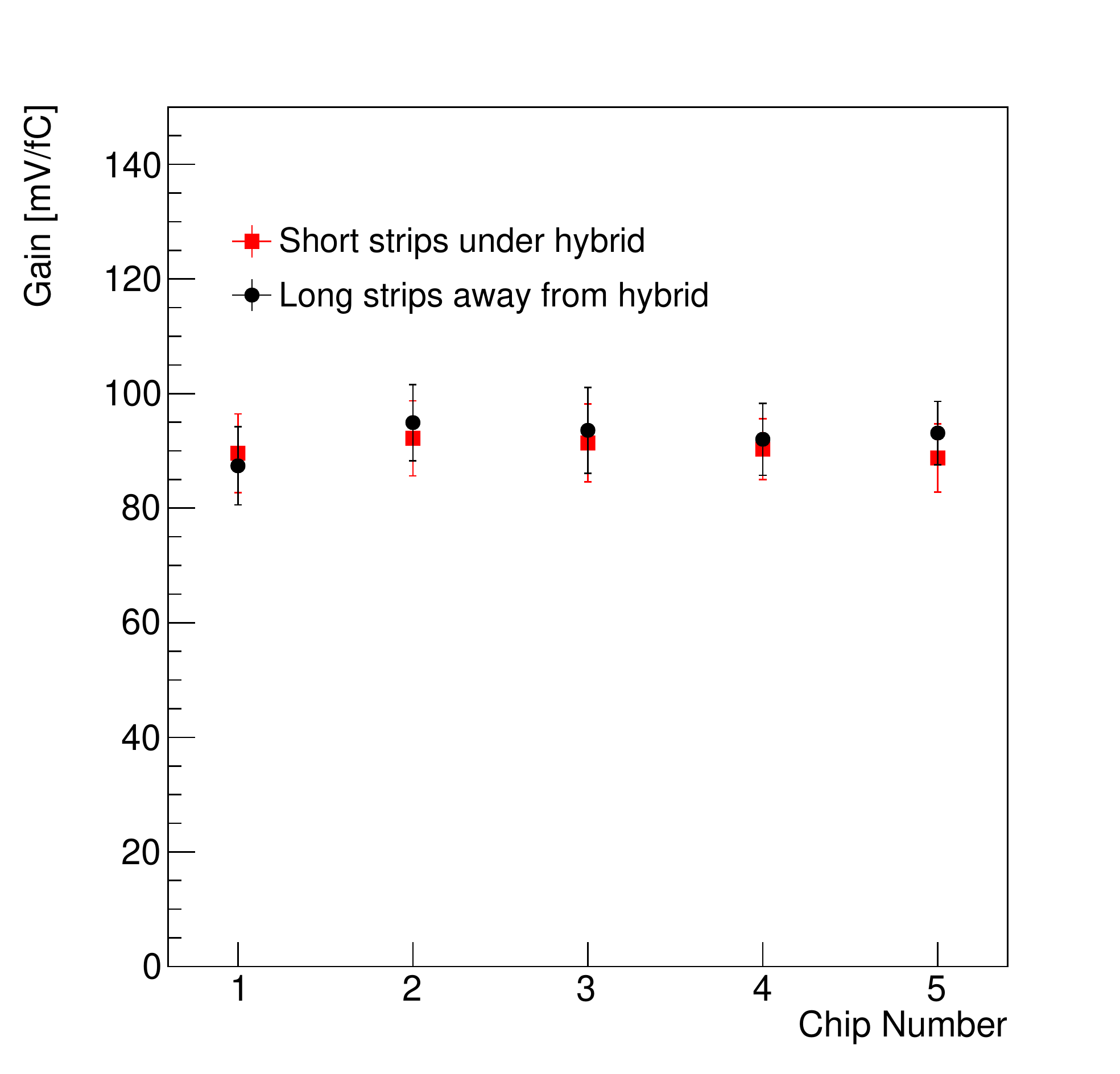}
 \caption{Gain after irradiation}
 \label{fig:irrad3b}
\end{subfigure}
\caption{Measured noise and gain on a mixed strip length ABC130 barrel module before and after irradiation with protons to $\unit[8\times10^{-14}]{n_{eq}/cm^2}$ at the CERN PS. The channels under the hybrid (red) are connected to short strips whilst those running away from the hybrid (black) are connected to long strips formed by ganging together short strips.  Measurements before irradiation were performed at room temperature whilst measurements after were performed at $\unit[-15]{^{\circ}C}$.}
\label{fig:irrad}
\end{figure}

In order to investigate the radiation tolerance of the ABC130 barrel module, a mixed strip length module was irradiated with protons at the CERN PS to a fluence of $\unit[8\times10^{14}]{n_{eq}/cm^2}$, corresponding to the maximum NIEL fluence expected at the end of life-time in the short strip region of the ITk strip barrel. The same charge injection based measurements were performed on this module and the results can be seen in figure~\ref{fig:irrad}, which shows the noise and gain of the module before and after irradiation. Only 5 chips are shown due to issues during the assembly process.

As can be seen, there is an increase in the measured noise on the module by about $\unit[20]{\%}$. It is specified that at end of life the detector must have a charged particle detection efficiency of at least $\unit[99]{\%}$ and an occupancy from noise hits of no more than $10^{-3}$. This requirement approximately equates to a requirement of a signal-to-noise ratio (SNR) of at least 10 throughout the life of the experiment. Given the conservative estimate of the charge collection at end of life of $\unit[1.6]{fC}$ ($\unit[10000]{e^-}$) and the measured noise of $\unit[950]{e^-}$ for the short strips on this module, an SNR of 10.5 is achieved and the end of life requirement is met.  The end-of-life requirement has been further studied in test-beam studies~\cite{testbeam}.

\subsection{Bad channel identification}
As described in section~\ref{chartests}, the Three Point Gain test can be used to determine electrical characteristics of module channels and assess their quality. This section investigates different types of channel defects and optimising their identification.

\subsubsection{Electrical Shorts} \label{elshorts}
Due to a lack of natural abundance of strip defects in the prototype sensors (see figure~\ref{fig:badchannels}), defects were intentionally added to a short-strip module in the form of additional wire bonds for some of the strips. Three types of defects were added corresponding to different scenarios that could arise:
\begin{itemize}
 \item bonds from strip to DC pad (added to three strips) to simulate shorts between the strip and implant (pinholes)
 \item bonds between neighbouring strips (added to four pairs of strips) to simulate shorts at the sensor
 \item bonds between non-neighbouring strips (added to four pairs of strips) to simulate shorts between ABC130 input channels or signal wirebonds
\end{itemize}
Figure~\ref{{defectimages}} shows an example of each type of defect.
\begin{figure}
	\centering
	\includegraphics[scale=0.375]{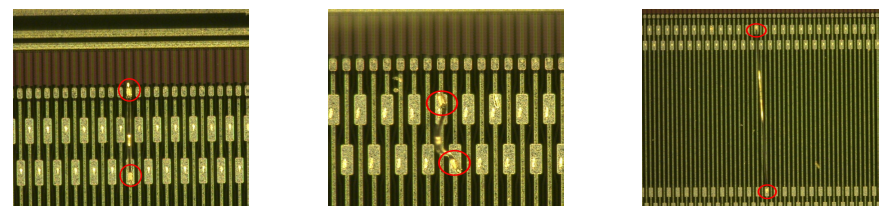}
	\caption{Examples of the three types of defects that were added to a short-strip module for the study in section~\ref{elshorts}. The image on the left shows a simulated pinhole with a wirebond between AC and DC pads. The images on the middle and right show simulated AC shorts for neighbouring and non-neighbouring channels respectively.}
	\label{{defectimages}}
\end{figure}

\begin{figure}
	\centering
	\includegraphics[scale=0.25]{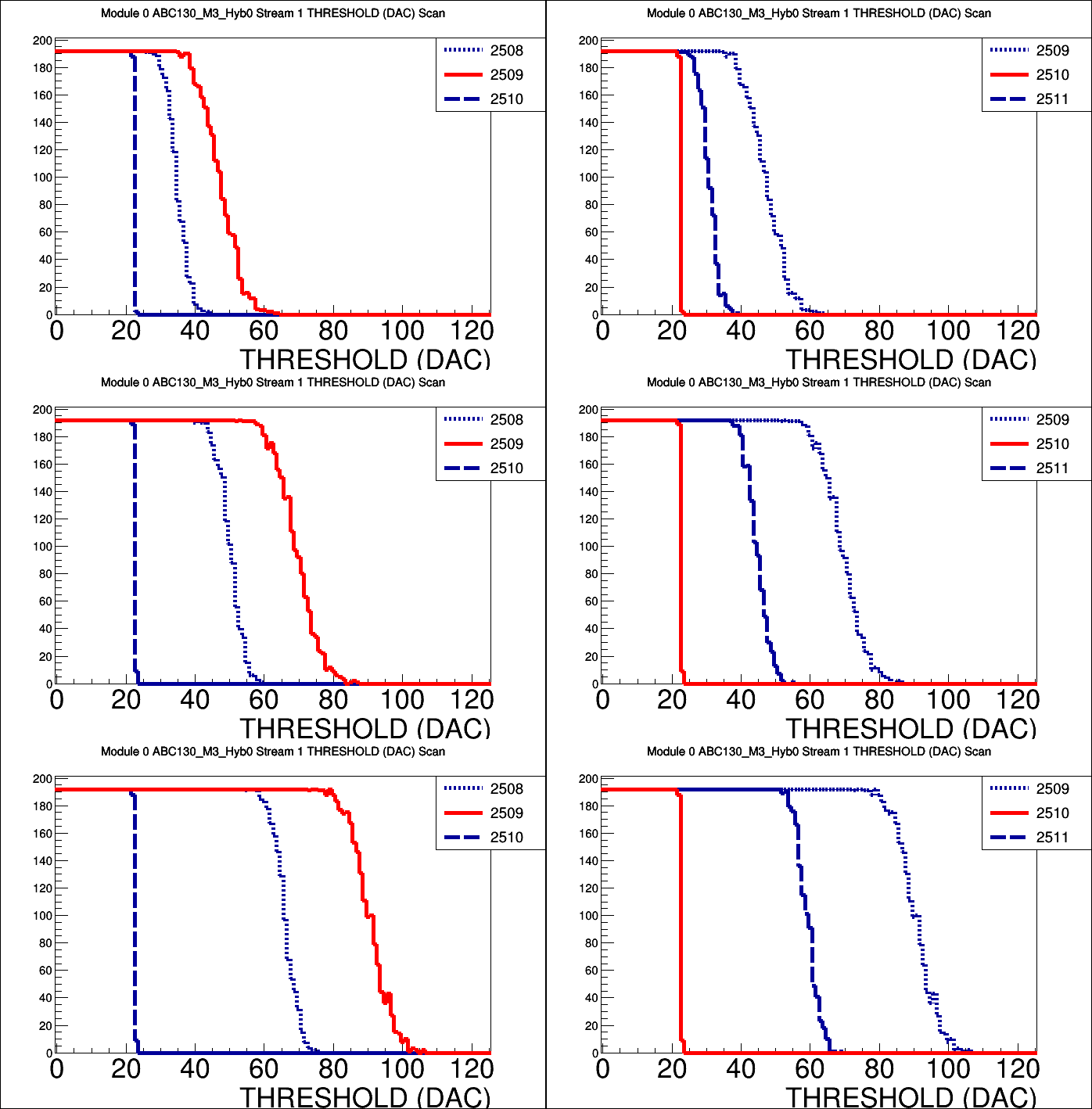}
	\caption{Example of threshold scans for each channel of bonded neighbouring strips are shown in the collection of plots in the left and right columns. The scans correspond to injected charges of {\unit[0.5]{fC}} (top), {\unit[1.0]{fC}} (middle), and {\unit[1.5]{fC}} (bottom). The red curves correspond to one of the channels from the pair. The neighbouring channels are shown in blue for reference.}
	\label{key19}
\end{figure}

The Three Point Gain test was run after these defects were added. The three strips bonded to simulate pinholes all had normal behaviour. All four pairs bonded to simulate shorts to neighbour strips showed similar behaviour where one channel has high gain and the other has low/nearly zero gain. The threshold scans in figure~\ref{key19} show an example of this behaviour. Three out of the four pairs of channels bonded to simulate shorts between non-neighbouring strips also showed a similar behaviour. The one pair that had different behaviour may have been due to an issue with wire bonding.

From these results it appears that pinholes cannot be identified from Three Point Gain tests while the shorts between strips can be identified from an obvious pattern. In regards to the inability to identify pinholes, it may be possible that after irradiation they may be visible from these tests and thus will require further investigation.

\subsubsection{Unbonded Classification and Bias Voltage}

The motivation for this study was to investigate the idea of running a Three Point Gain test at reverse bias voltage below full depletion for the purpose of identifying channels for which the signal bonds have become disconnected. Lowering the bias voltage of the sensor should increase the overall noise of the module and thus make it more distinct compared to hybrid-level noise, which helps to identify unbonded channels. 

\begin{figure}
	\centering
	\includegraphics[scale=0.5]{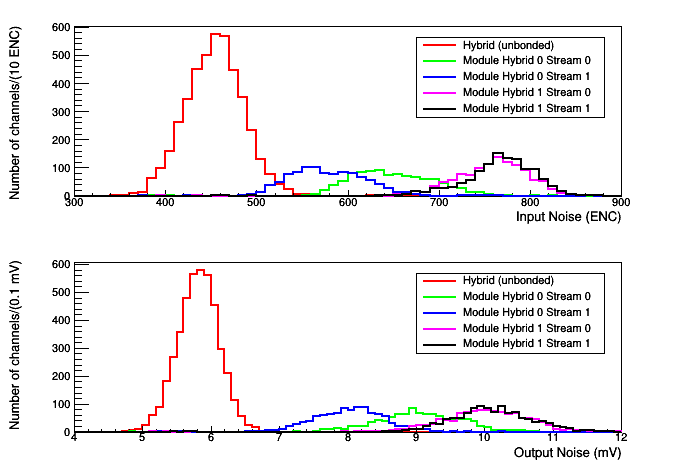}
	\caption{Histograms of the noise for channels from an R0 module with values from hybrid tests in red and the four other colours corresponding to different strip rows. Some overlap can be seen in the tails of the input noise distributions for the innermost strip row in blue and the hybrid values.}
	\label{r0histogram}
\end{figure}

The classification of channels based on noise measurements is anticipated to be more ambiguous for future ABCstar modules due to a smaller noise dependence on the input capacitance. For illustration purposes, this effect was studied for an R0 module with an ABC130 chip set (see figure~\ref{r0histogram}), which features one row of strips with short length (\unit[1.9]{cm})~\cite{R0ref}.

\begin{figure}
	\centering
	\includegraphics[scale=0.5]{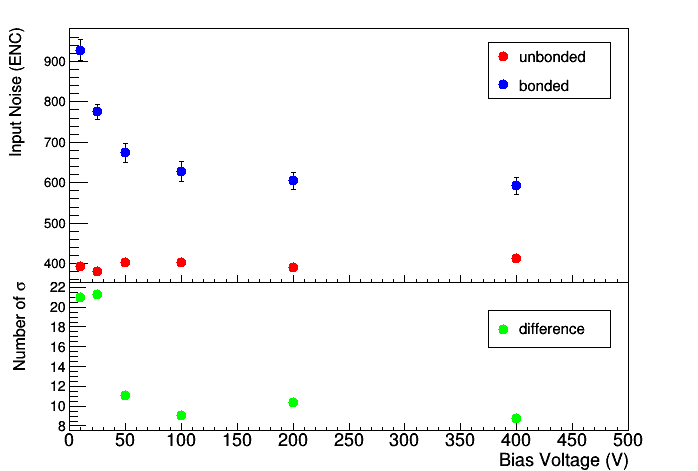}
	\caption{Comparison of the input noise between chip average (blue) and an unbonded channel (red) on the same chip as a function of the applied reverse bias voltage. The difference in terms of the number of standard deviations is shown below.}
	\label{key26}
\end{figure}

Three Point Gain tests were run at different reverse bias voltages between -1 and
\unit[-400]{V}. The tests were conducted using a short-strip module which had
a few channels that were intentionally unbonded to use as a comparison. An
example of the noise difference between bonded and unbonded channels is shown
in figure~\ref{key26}. The optimal value seems to be around \unit[-25]{V}
where it is possible to substantially increase the noise while still obtaining
reasonable results without large variations in the channel noise that appear
at lower bias voltages.

\subsection{Current increase with Total Ionising Dose (TID bump)}
\label{sec:TIDBUMP}

The \unit[130]{nm} process used to fabricate the ABC130 chip set is known to be
sensitive to certain radiation effects~\cite{TIDfaccio}. In particular, NMOS
transistors fabricated using this process show an increase in leakage current
when exposed to ionizing doses of radiation up to approximately \unit[1]{Mrad} (full lifetime dose is simulated to be \unit[53.2]{Mrad}).
Continued exposure to ionizing radiation (beyond \unit[1]{Mrad}) then gradually reduces
the leakage current back towards its nominal value. This phenomenon has been
named the ``TID bump". 

The size of the effect depends both on the
characteristics of the transistor and on the characteristics of the radiation.
The use of enclosed layout transistors in analog blocks of the chips, such as
those which perform the signal amplification and discrimination, completely
negates the impact of the TID bump for those transistors. However, the impact
is considerable for regions of the chips using digital functionality, where the
transistors do not use an enclosed layout. For these portions of the chips, it
is crucial to understand the impact of the TID bump, and its dependence on the
environmental conditions. Too much leakage current could result in an inability
to properly power the chips or to cool the detector.

\begin{figure}
	\centering
	\includegraphics[width=0.8\linewidth]{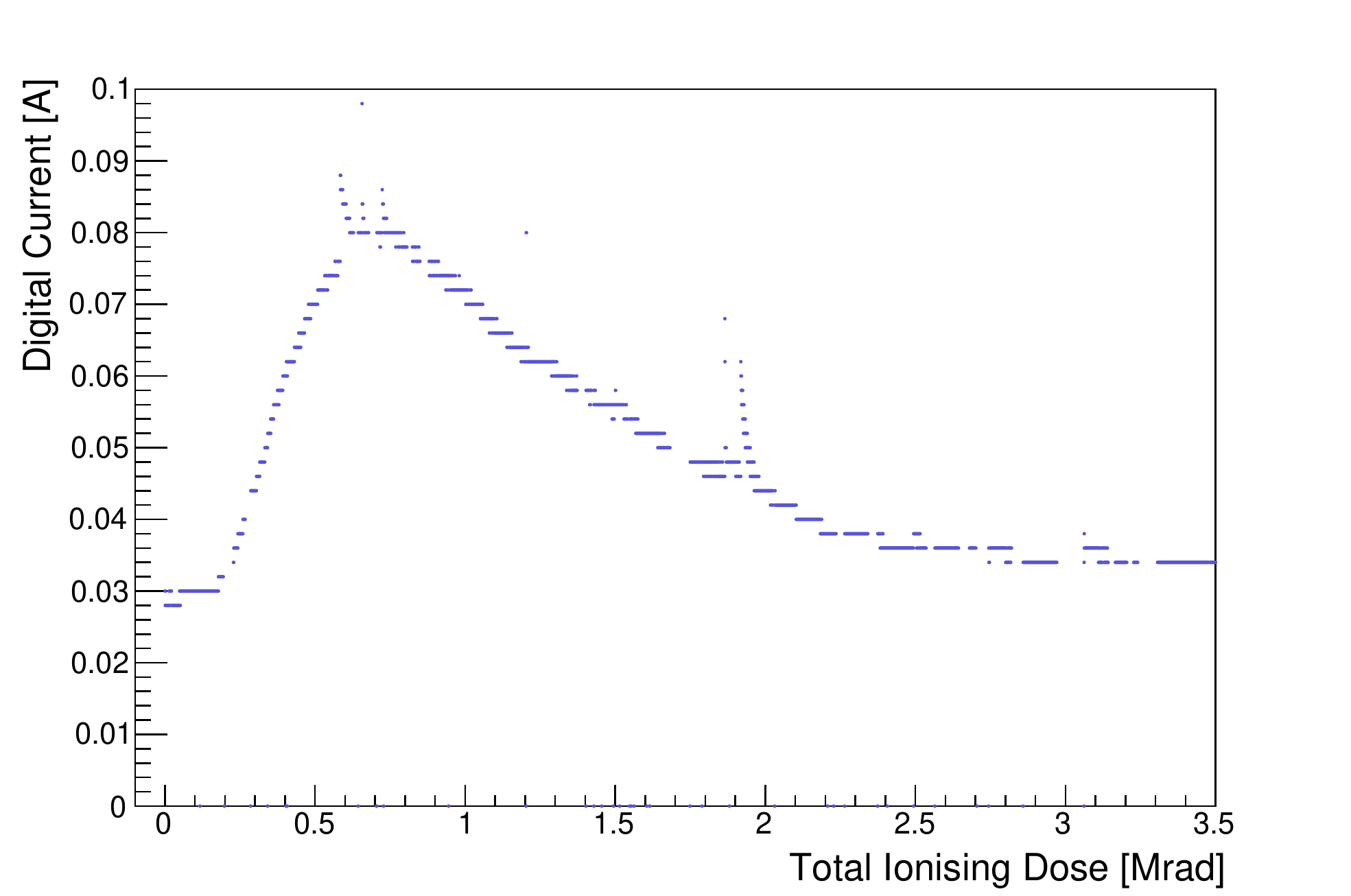}
	\caption{An example of the 'TID bump', an initial increase in the digital current supplied to an ABC130 chip when it is irradiated; the current reaches a maximum before {\unit[1]{Mrad}} of Total Ionising Dose then decreases again, approaching its initial value as the radiation continues. The plot shows the digital current of the chip as it is being irradiated at {\unit[2.5]{krad/h}} and cooled to {\unit[-10]{$^{\circ}$C}}, conditions chosen to be similar to those expected for the ITk strip tracker chips during HL-LHC operation. The radiation was performed using a Caesium-60 source.}
	\label{fig:TIDbump}
\end{figure}

Chips were tested while
being cooled and irradiated at dose rates compatible with those expected during
HL-LHC operation. Chips were cooled to between \unit[-25]{$^{\circ}$C} and \unit[0]{$^{\circ}$C}. They were irradiated at dose rates ranging from \unit[0.6]{krad/h} to
\unit[2.5]{krad/h}; these rates cover the expected range for chips in different
positions during nominal HL-LHC operation. The observed increase in leakage
currents for these tests ranged between \unit[$\sim30$]{\%} and \unit[$\sim160$]{\%}, with smaller increases at higher temperatures and lower dose rates~\cite{TDRs} (see figure~\ref{fig:TIDbump}). 
An empirical function was fit to the data in order to describe the dose rate and
temperature dependence of the bump for use in thermo-electrical models of
the detector.

Furthermore, tests of multiple chips revealed that not only does the size of the bump depend on the environmental conditions, but that there was significant
variance in the size of the bump measured across different chips tested in the
same conditions. Significant variations in the size of the TID bump were seen
both between different chips produced in the same batch, and between chips
produced in different batches.

\begin{figure}
	\centering
	\includegraphics[width=0.9\linewidth]{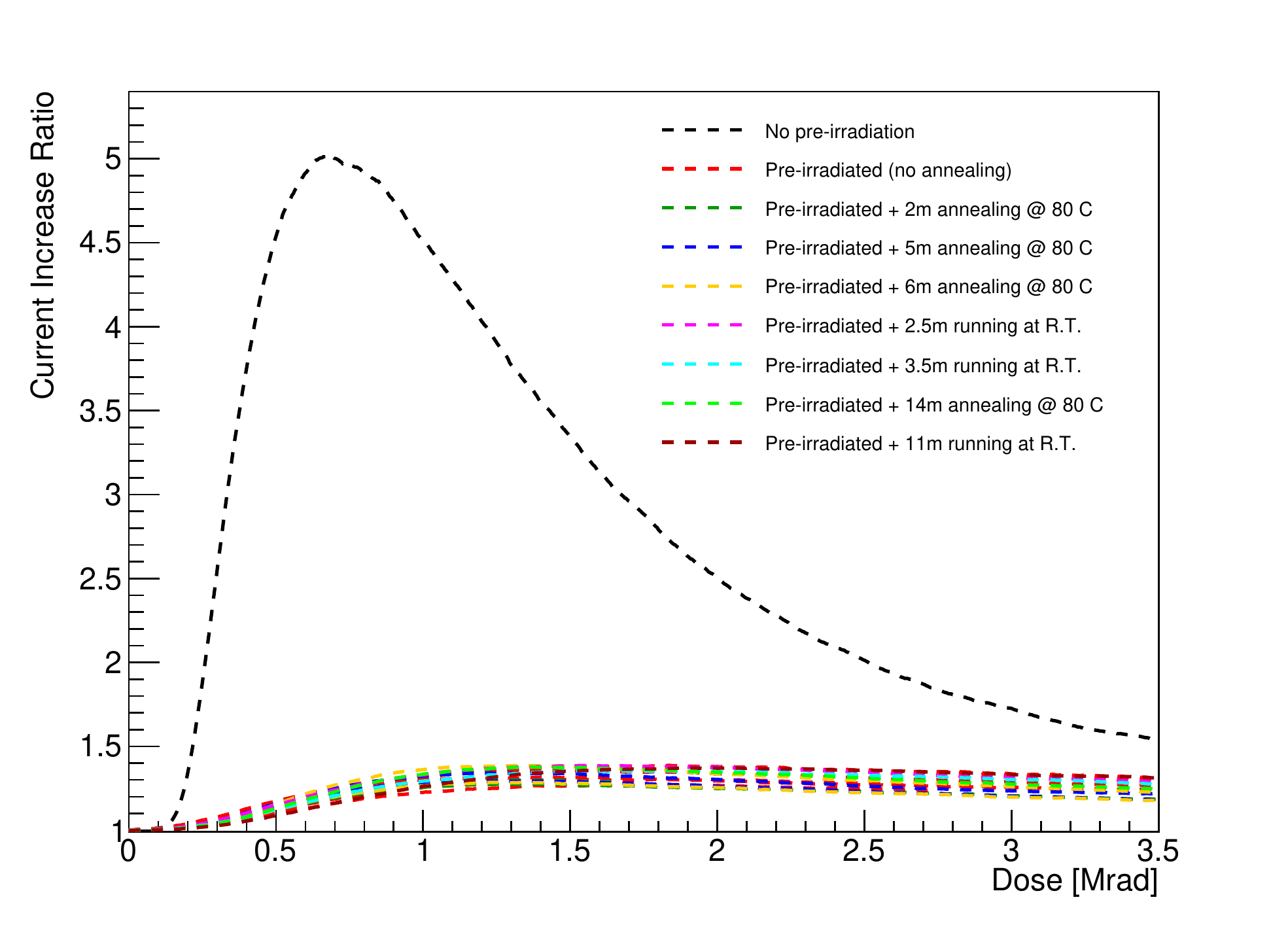}
	\caption{Comparison of current increase for un-pre-irradiated and pre-irradiated ASICs with various annealing processes carried out - chips were either left unpowered in an oven at {\unit[80]{$^{\circ}$C}} or powered and read-out constantly at room temperature.  Independent of the annealing applied, pre-irradiated ASICs showed a current increase of less than {\unit[50]{\%}} - significantly less than the primary TID bump.  Results are shown for multiple ASICs at each annealing option and all chips were taken from the same wafer.}
	\label{fig:TIDcomparison}
\end{figure}

Ultimately, in order to mitigate the current increase, a strategy of
pre-irradiation was chosen: chips undergo high dose-rate irradiation 
up to a dose of \unit[$\sim10$]{Mrad}, well beyond the TID bump. Using
this method, the chips have already passed the bump so that their leakage
current has returned to that before any irradiation. A number of tests performed on annealed chips confirmed that they would not undergo a secondary bump after long room temperature annealing periods between the pre-irradiation and operation at the HL-LHC.  These tests include up to 14 months with the chip stored at \unit[80]{$^{\circ}$C} and up to 11 months with the chip left powered and running at room temperature. These chips were irradiated to \unit[8]{Mrad} using Co-60 and re-irradiated at room temperature at \unit[0.7]{Mrad/hr} using a \unit[3]{kV} tungsten x-ray tube. The effect of this pre-irradiation procedure can be seen in figure~\ref{fig:TIDcomparison}.

\section{Conclusion and Outlook}

Within the scope of the ABC130 barrel module programme, about 100 modules were
constructed at ten institutes foreseen to assemble barrel modules for the
future ITk strip tracker. While a small number of LS modules was built, the
majority of assembled modules were SS modules, as their more complicated
assembly and readout was considered more challenging. Their construction 
allowed to develop assembly tools and test procedures for future module
construction.

The barrel module prototyping programme demonstrated that the established procedures for assembly and quality control ensured the production of modules within the required specifications. The assembly procedures and developed tooling were found to result in the required module geometries as evidenced by the outcome of metrological surveys. The electrical testing procedures confirmed that the Signal-to-Noise-ratio required by the \mbox{ATLAS} strip tracker for the subsequent module generation could be achieved. Areas for improvement, such as an increased noise under the powerboard shield box, could be identified and corrected in subsequent iterations.

The assembled modules were used to construct complete prototype versions of
higher-level detector structures called staves. Staves consist of a carbon fibre
core with integrated cooling, power, and data I/O infrastructures on both sides, onto which
13 or 14 modules per side are glued and electrically connected~\cite{stave}.
The availability of 100 barrel modules allowed the assembly of one fully loaded
stave, three half loaded staves (where only one side of the stave was fully
populated) and one partially loaded double-sided stave (where both stave sides
were partially populated with modules). The availability of several populated stave sides permitted tests of fully assembled structures as well as system tests, where potential interactions of neighbour staves could be studied. The extensive test program performed on staves~\cite{Staves} goes beyond the scope of this publication.

The scale and results of this prototyping program demonstrated a high degree of development for all involved components, as well as the module assembly and testing processes. The major driver for the next design evolution was an increase of the trigger rate requirement to MHz range. To address this challenge, a new readout architecture was necessary, instigating the development of the subsequent generation of readout chips called star chips. While ABC130 chips were read out in chains of five chips in series, the star chip set was developed to allow the direct readout of each individual ASIC to increase the bandwidth. A new generation of components based on the star chip design was designed based on the findings from the ABC130 barrel module programme and will be developed into modules for the ITk strip tracker.

\section*{Acknowledgements}

Individual authors were supported in part by the U.S. Department of Energy under Contract No. DE-AC02-05CH11231. This study was supported by National Key Programme for S\&T Research and Development (Grant No.: 2016YFA0400101). The work at SCIPP was supported by the Department of Energy, grant DE-SC0010107. This work was supported by the Science and Technology Facilities Council [grant number ST/R002592/1], the Polish Ministry of Science and Higher Education, Grant No.: DIR/WK/2018/04, the Canada Foundation for Innovation and the Natural Sciences and Engineering Research Council of Canada and the Australian Research Council.

\bibliographystyle{unsrt}
\bibliography{bibliography.bib}

\end{document}